\documentclass[twocolumn,times,tighten]{aastex631}
\usepackage{graphicx,multirow,hyperref,url,color}
\hypersetup{colorlinks=true, urlcolor=blue, citecolor=blue}

\usepackage{longtable}

\newcommand{\msun}{{M_\odot}}
\newcommand{\ergs}{{\rm erg\,s^{-1}}}

\graphicspath{{./}{figures/}}

\begin{document}
\shorttitle{Diffuse Radio Lobes in $\rm M\,87$}
\shortauthors{Wu et. al.}

\title{MWA and VLA Observations of Diffuse Radio Lobes in M\,87}


\author[0000-0003-3454-6522]{Linhui Wu}
\thanks{E-mail: wulinhui@shao.ac.cn}
\affil{Shanghai Astronomical Observatory, Chinese Academy of Sciences, 80 Nandan Road, Shanghai, 200030, P. R. China}

\author[0000-0001-9969-2091]{Fu-Guo Xie}
\thanks{E-mail: fgxie@shao.ac.cn}
\affil{Shanghai Astronomical Observatory, Chinese Academy of Sciences, 80 Nandan Road, Shanghai, 200030, P. R. China}
\affil{State Key Laboratory of Radio Astronomy and Technology, Chinese Academy of Sciences, A20 Datun Road, Chaoyang District, Beijing, 100101, P. R. China}

\author{Qian Zheng}
\thanks{E-mail: qzheng@shao.ac.cn}
\affil{Shanghai Astronomical Observatory, Chinese Academy of Sciences, 80 Nandan Road, Shanghai, 200030, P. R. China}
\affil{State Key Laboratory of Radio Astronomy and Technology, Chinese Academy of Sciences, A20 Datun Road, Chaoyang District, Beijing, 100101, P. R. China}

\author{Quan Guo}
\affil{Shanghai Astronomical Observatory, Chinese Academy of Sciences, 80 Nandan Road, Shanghai, 200030, P. R. China}
\affil{State Key Laboratory of Radio Astronomy and Technology, Chinese Academy of Sciences, A20 Datun Road, Chaoyang District, Beijing, 100101, P. R. China}

\author{Huanyuan Shan}
\affil{Shanghai Astronomical Observatory, Chinese Academy of Sciences, 80 Nandan Road, Shanghai, 200030, P. R. China}
\affil{State Key Laboratory of Radio Astronomy and Technology, Chinese Academy of Sciences, A20 Datun Road, Chaoyang District, Beijing, 100101, P. R. China}
\affil{University of Chinese Academy of Sciences, Beijing, 100049, P. R. China
}

\author{Dan Hu}
\affiliation{Department of Theoretical Physics and Astrophysics, Faculty of Science, Masaryk University, Kotl\'{a}\v{r}sk\'{a} 2, Brno, 611 37, Czech Republic}

\author[0000-0002-3846-0315]{Stefan W. Duchesne}
\affiliation{CSIRO Space and Astronomy, PO Box 1130, Bentley WA 6102, Australia}

\author{Nick Seymour}
\affiliation{International Centre for Radio Astronomy Research (ICRAR), Curtin University, Bentley, WA 6102, Australia}

\author[0000-0002-5598-2668]{Jingying Wang}
\affil{Shanghai Astronomical Observatory, Chinese Academy of Sciences, 80 Nandan Road, Shanghai, 200030, P. R. China}

\author[0000-0001-9765-6521]{Junhua Gu}
\affil{National Astronomical Observatories, Chinese Academy of Sciences, 20A Datun Road, Beijing, 100101, P. R. China}

\author{Qingwen Wu}
\affil{School of Physics, Huazhong University of Science and Technology, Wuhan 430074, P. R. China}

\author{Zhenghao Zhu}
\affil{Shanghai Astronomical Observatory, Chinese Academy of Sciences, 80 Nandan Road, Shanghai, 200030, P. R. China}

\author[0000-0003-2756-8301]{Melanie Johnston-Hollitt}
\affiliation{Curtin Institute for Computation, Curtin University, GPO Box U1987, Perth, WA 6845, Australia}

\author[0000-0002-3369-1085]{Christopher J. Riseley}
\affiliation{Astronomisches Institut der Ruhr-Universit\"{a}t Bochum (AIRUB), Universit\"{a}tsstra{\ss}e 150, 44801 Bochum, Germany}

\author{Xu-Liang Fan}
\affiliation{School of Mathematics, Physics and Statistics, Shanghai University of Engineering Science, Shanghai 201620, P. R. China}

\begin{abstract}
This study investigates the projected, quasi-symmetric $\sim\rm46\,kpc$-scale diffuse radio lobes surrounding the giant elliptical galaxy M\,87, utilizing well-sampled wideband ($\rm 60\,MHz-10.55\,GHz$) observations from MWA and VLA, supplemented by data from LOFAR and Effelsberg. The observed structures feature sharp edges and filaments, with nearly uniform and moderately steep spectral indices ($\alpha$, mostly within $-1.2\leq\alpha\leq-0.8$), indicating turbulence. Well-sampled radio spectra for the lobes’ diffuse region are derived using the continuous injection (CI) model (with $\alpha_{\rm inj}\simeq-0.86$ and $\nu_{\rm b}\simeq1.72\rm\,GHz$), and for its three localized regions using the impulsive injection model (e.g., JP model). From energy equipartition analysis, we estimate the typical magnetic field strength in the lobes’ diffuse region to be $B_{\rm eq}\simeq10\,\mu\rm G$. The age of the lobes is estimated as $\sim30-50\,\rm~Myr$, based on lifetimes derived from the CI and JP models and sound crossing time. Outflow powers of $\sim(0.2-2)\times10^{44}\,\ergs$ for the lobes’ diffuse components and $\sim(1-11)\times10^{44}\,\ergs$ for the whole source are calculated. With this power assessment, we conclude that the galactic stellar wind has a negligible effect, the active galactic nucleus (AGN)-driven jet can provide the necessary energy for the whole system. Furthermore, we argue that while the wind driven by current AGN activity is unlikely to power the lobes’ diffuse components, an average enhancement of AGN activity by a factor of $\sim 10^2$ over the past $\sim 30-50$ Myr remains plausible.

\end{abstract}


\section{Introduction} \label{sec:intro}

Approximately $10\%$ of active galaxies exhibit prominent radio structures (also called radio-active galactic nuclei, radio-AGNs), with some showing lobe-like structures extending from kiloparsecs (kpc) to megaparsecs (Mpc) (e.g., \citealt{Dre84,Car91,Colbert96,Ro96,Fabian00,McNamara00,Ow00,Kha06,Seb20,Oei24}). These large-scale lobes can uplift gas from the galaxy's center, interacting with the surrounding intra-cluster medium (ICM) via complex physical processes. This interaction redistributes the thermal state of the ICM, suppresses the cooling flow and influences star formation within the host galaxies (e.g., \citealt{McNamara12,Gilli19,Har20,Manuela22} and reference therein). Therefore, studying radio-AGN is crucial not only for understanding the physics behind lobe formation but also for revealing their role in AGN feedback processes.

Lobes are generated by outflows in the galaxy, which primarily involve three mechanisms: relativistic jets, AGN winds, and stellar-driven galactic winds (or superwinds) \citep{Kha06,Manuela22,Sil23}. The most-accepted one is through relativistic jets, which can propagate to hundreds of kpc (e..g, in ICM) and dissipate kinetic (and magnetic) energy there \citep{Guo08a,Guo10,Bl19,Duan20,Ehl21,Duan24}. This interpretation is widely adopted in the study of radio galaxies (\citealt{Lon73,Sch74,Bl74,Ehl18,Bl19}, and references therein). However, this interpretation is challenged by several observations (e.g., \citealt{Harwood20} and references therein). Alternative scenario, e.g., material entrainment from AGN winds and interstellar medium (ISM) can decelerate these outflows (jets) while simultaneously transporting energy (e.g., \citealt {Bow96,Wyk15,Har20,Silpa21b,Bl22}). In quasars, high-velocity AGN outflows can supply significant energy into ICM extending up to $\sim10$ kpc or further (e.g., \citealt{Liu13, Hil14, Wy16, Sil22, Sil23}). Methods have also been proposed to distinguish the radio emission in quasars between that originates from stellar activity and that originates from AGN activity \citep{Macfarlane21,Yue25}. In starburst galaxies, the lobe-like radio structures can be explained by the interaction of starburst superwinds with the ISM \citep{Hec93,Mar05,Rup05}. In Seyfert galaxies, both AGN and starburst activities are considered to explain the observed lobe-like radio structures \citep{Gen95,Colbert96,Mai98,Sebastian19,Silpa21}.

Currently, the spatial resolution and sensitivity of most telescopes are insufficient to directly observe the outflow's launching site. Consequently, identifying the mechanisms driving the outflow remains difficult. Several approaches, such as morphology analysis, spectrum examination, timescale considerations (e.g., dynamic and spectral aging), energetic assessments, power estimations, absorption/emission lines analysis, often combined with simulations, have been developed to infer the primary mechanism behind the outflow. However, even within our own Galaxy, the origin of Fermi and associated bubbles remains uncertain. It could stem from a jet or spherical outflow (AGN wind) from a supermassive black hole (SMBH) (\citealt{Zub11,Guo12a,Guo12b}), or from a superwind generated by supernova explosions (e.g., \citealt{Cro11}). 

M\,87 is the most prominent elliptical galaxy in the Virgo cluster. It is one of the closest radio galaxies. It features a notable one-sided jet that has a line-of-sight viewing angle of $\approx 17\degr$ (e.g., \citealt{Mertens16,Kim23}) and has been extensively studied across wavelengths, from radio to $\gamma$-rays (e.g., \citealt{EHTM21} and references therein). The nucleus of M\,87 is quite dim, with a bolometric luminosity of $L_{\rm bol}\approx 1.1\times10^{-6}\,L_{\rm Edd}$ \citep{EHTM21,Xie23}, classifying it as a low-luminosity AGN. Here, the Eddington luminosity for accretion onto a SMBH with mass $M_{\rm BH}$ is defined as $L_{\rm Edd}\approx 1.3\times 10^{47}\,\ergs\,(M_{\rm BH}/\rm 10^9\msun)$. Additionally, M\,87 displays giant radio lobes with a projected diameter of about 46 kpc (see Figures \ref{fig:region} and \ref{fig:projection} below), along with diffuse hot gas emitting X-rays \citep{Mat02,For07,Mil10}. 

Previous studies of M\,87's large-scale properties have mainly focused on the bright components and structures within its lobes. For instance, \cite{Hines89} estimated cooling processes (i.e., synchrotron, inverse Compton, adiabatic expansion) and shear acceleration in the bright structures of the inner lobes. They suggested that turbulence likely plays a role in sustaining the brightness of these structures, although the underlying physical mechanisms remained unclear. Using LOw-Frequency
ARray (LOFAR) observations, \cite{Gas12} traced the flows (zones) in lobes, noting that the low-frequency end of spectra is associated with the adiabatic expansion of the lobes, which were formed by continuous jet injection. However, detailed analysis based on observations from the Event Horizon Telescope (EHT) and Very Long Baseline Array (VLBA) suggests that there is an AGN wind near the SMBH of M\,87 \citep{Park19,Bl22,Yuan22,Lu23}, likely driven by black hole accretion. Wind from low-luminosity AGN is well-expected theoretically (e.g., \citealt{Yuan12a, Yuan12b, Yuan15, Yang21}, and \citealt{yuan14} for a review.), and may play a non-negligible role in feedback processes in low-luminosity AGNs (e.g., \citealt{Weinberger17,Yuan18}). These findings imply that AGN-driven wind, coupled to the jet, may contribute a non-negligible energy to the large-scale outflow as observed in low-luminosity AGNs like M\,87.

In this study, we investigate the diffuse radio emission from two giant radio lobes of M\,87, using observational data from the Murchison Widefield Array (MWA; \citealt{Tin13,Wayth18}) at $\sim70-230$ MHz and the Expanded Very Large Array (VLA; \citealt{Perley11}) at $\sim1-4$ GHz. We also include data from LOFAR at 60 MHz and 140 MHz (provided to us by Francesco de Gasperin), VLA at 325 MHz (provided by Frazer Owen) and Effelsberg at 10.55 GHz (provided by Helge Rottmann). Our findings suggest the need for continuous outflows injection to account for the large-scale diffuse radio emission in the lobes. We then explore potential mechanisms responsible for driving these outflows in M\,87, extending up to approximately $\sim 46\rm\,kpc$ scale, including galactic stellar winds, AGN jets, and AGN winds. This work is organized as follows: Section \ref{sec:observation} provides details on the MWA and VLA observations and data reduction. The results are presented in Section \ref{sec:results}, followed by a discussion in Section \ref{sec:discussions}. The final Section \ref{sec:conclusion} offers a summary. Throughout this work, the spectral index, $\alpha$, is defined as $S_{\nu}\propto \nu^{\alpha}$, where $S_\nu$ is the flux at frequency $\nu$. The distance and the SMBH mass of M\,87 are fixed at $d = 16.9$ Mpc and $M_{\rm BH}=6.2 \times 10^{9} \rm\,\msun$ \citep{Geb11,EHT19}.

\begin{table*}
\renewcommand{\thetable}{\arabic{table}}
\centering
\small
\caption{Details of our MWA and VLA observations.} \label{Tab:Obs_inf}
\renewcommand{\arraystretch}{1.2}
\begin{tabular}{ccccccccccc}
\hline
\hline
ObsID & Array &Conf. & $\nu$ & BW & Date & $t_{\rm duration}$ & $t_{\rm sampling}$& $\rm UV_{max}$ & $\rm UV_{min}$ & $\rm Ang_{max}$\\ 
  & &  & $\rm MHz$& $\rm MHz$ & &$\rm min$&$\rm s$& $\rm k\lambda$ & $\rm k\lambda$ & \arcmin \\
(1)  &(2) &(3)  &(4) & (5) & (6)&(7)& (8) & (9) & (10) & (11)\\
\hline\hline
1213610128& MWA& ext.  &88& 30.72&2018.06.21  & 2 &4 & 1.6 & $<0.01$ & $>340$ \\
1213610248& MWA& ext.  &118& 30.72&2018.06.21 & 2 &4 & 2.2 & $<0.01$ & $>340$ \\
1205430776 & MWA& ext.  &154& 30.72&2018.03.18 & 2&4 & 3.0 & $0.01-0.02$ & $170-340$ \\
1205430896 & MWA& ext.  &185& 30.72&2018.03.18 & 2&4 & 3.5 & $0.01-0.02$ & $170-340$ \\
1213697368 & MWA& ext.  &216& 30.72&2018.06.22 & 2&4 & 3.9 & $0.01-0.02$ & $170-340$ \\
TALG001& VLA& D  &1515& 1000&2010.05.23 & 360 &1& $3.8-7.0$ & $0.11-0.20$ & $17-31$ \\
13A-096& VLA & D & 3000 & 2000 &2013.04.08& 222&5& $6.8-12.4$ &$0.21-0.38$ & $9-16$ \\
\hline\hline
\end{tabular}

Notes.
The 1st column presents the observation (or project) ID. The 2nd and 3rd columns present the array type and corresponding configuration of the observation, respectively. The 4th and 5th columns present the central frequency and its bandwidth, respectively. The 6th column presents the observation date. The 7th and 8th columns present the total duration and the sampling time of each observation, respectively. The 9th and 10th columns present the maximum and minimum \textit{uv}-wavelength. The last column presents the maximum angular size (i.e., $\rm Ang_{max}\equiv3440/uv_{min}\,arcmin$) of the structures that can be recovered. 
\end{table*}

\section{Observations and data reduction}
\label{sec:observation}

\begin{table*}
\renewcommand{\thetable}{\arabic{table}}
\centering
\small
\caption{Image information of our MWA and VLA observations.} \label{Tab:Im_inf}
\begin{tabular}{cccccccc}
\hline
\hline
ObsID & Array & $\nu$ & weighting& PSF&Flux& $\sigma_{\rm rms}$    \\ 
  &  & $\rm MHz$& &$\arcsec\times\arcsec$ &Jy& $\rm mJy/beam$&\\
(1) & (2) & (3) & (4) & (5) & (6) & (7)\\
\hline\hline
1213610128 & MWA  &99 &0.5 &134$\times$102&$1705\pm170$ & 139 \\
1213610248 & MWA  &130&0.5 &102$\times$78&$1373\pm137$ &93 \\
1205430776 & MWA  &166&0.5 &78$\times$59&$1133\pm113$  &55 \\
1205430896 & MWA  &196&0.5 &66$\times$50&$991\pm99$   &60 \\
1213697368 & MWA  &227&0.5 &58$\times$44&$883\pm88$   &45 \\
TALG001    & VLA &1515&0.5 &37$\times$25&$196\pm10$  &2.3\\
13A-096    & VLA &3000&0.5 &23$\times$18&$113\pm6$  &1.8\\
\hline
\end{tabular}

Notes.
The 1st column presents the observation (or project) ID. The 2nd column presents the array type. The 3rd column presents the frequency of the image. The 4th column presents the Briggs weighting for imaging. The 5th presents the resolution of the image. The 6th column represents the total flux density of M\,87 at different frequencies. The last column presents the noise level ($\sigma_{\rm rms}$) of the intensity image. 
\end{table*}

\subsection{MWA observations}
The MWA is a low-frequency radio interferometer that operates at $70-300\rm\,MHz$ \citep{Tin13}. MWA Phase II comprises 256 `tiles', each containing a 4×4 array of 32 dipoles. Each tile measures two instrumental polarizations `X' (16 dipoles oriented East-West) and `Y' (16 dipoles oriented North-South). Full details of the Phase II array layouts are provided in \citet{Wayth18} and the key science of the MWA may be found in \citet{Bow13} and \citet{Bear19}.

In this work, we use MWA phase II datasets from MWA project G0047 (PI: N. Seymour), selecting observations where M\,87 is within 3 degrees from the center of field of view (FOV). The observations only use 128 of the MWA tiles for the limitation of the tiles number that can be correlated. These tiles were configured in the extended baseline configuration with the baseline length range from $<40\rm\,m$ to $3\rm\,km$. We conduct observations across five frequency bands, spanning from $70\rm \,MHz$ to $230\rm \,MHz$, with each band having a bandwidth of $30.72\rm \,MHz$. M\,87, located at (RA, Dec) = (12:30:49.42, +12:23:28.00), is one of the brightest radio sources in the sky at low frequencies and a 2-min snapshot in each band provides sufficient signal-to-noise (S/N) for our investigation. Basic information on these observations is displayed in Table \ref{Tab:Obs_inf}.

The data processing strategies essentially follow those detailed in \citet[][see also \citealt{Du20}]{Hur17}. We use the MWA All-Sky Virtual Observatory (ASVO)\footnote{\href{https://asvo.mwatelescope.org}{https://asvo.mwatelescope.org}} system to convert raw telescope products to the standard `MeasurementSet' format using \texttt{COTTER} \citep{Off15} and perform preliminary radio frequency interference (RFI) flagging using \texttt{AOFlagger} \citep{Off12}. Raw visibilities are averaged to have a frequency resolution of 40\,kHz, and a time resolution of 4\,s. Before calibration, bad tiles and channels are manually flagged. These averaged visibilities are then calibrated based on the \texttt{Mitchcal} algorithm \citep{Off16} using a global sky model as described in \citet{Du20}. Imaging for per snapshot is performed using \texttt{WSClean} \citep{Off14,Off17} with multi-scale and multi-frequency CLEANing.\footnote{In this work, the max scale size for the multi-scale cleaning method \citep[see][]{Cor08} is set about the size of the lobe of M87, i.e., $\sim 9\arcmin$, to recover the diffuse emission from the large-scale lobes.} After one round of phase and amplitude self-calibration, deep imaging is performed across the whole 30 MHz band---split into four subbands---using Briggs weighting ($\rm robust= 0.5$, see \citealt{Bri95}) to balance the sensitivity and resolution.\footnote{Unless specified, `resolution' in this work refers to spatial resolution.} In this work, we correct the MWA images of M\,87 for flux scale using the method described in Section \ref{sec:scaling} and for astrometry by aligning the brightest core position to maintain consistency with the following VLA images. Figure \ref{fig:mwa_obs} presents the fourth sub-band images for each band, where the frequency of each image is specified in each plot (see also a brief summary in Table \ref{Tab:Im_inf}).

\begin{figure*}
    \centering
    \includegraphics[width=0.6\textwidth]{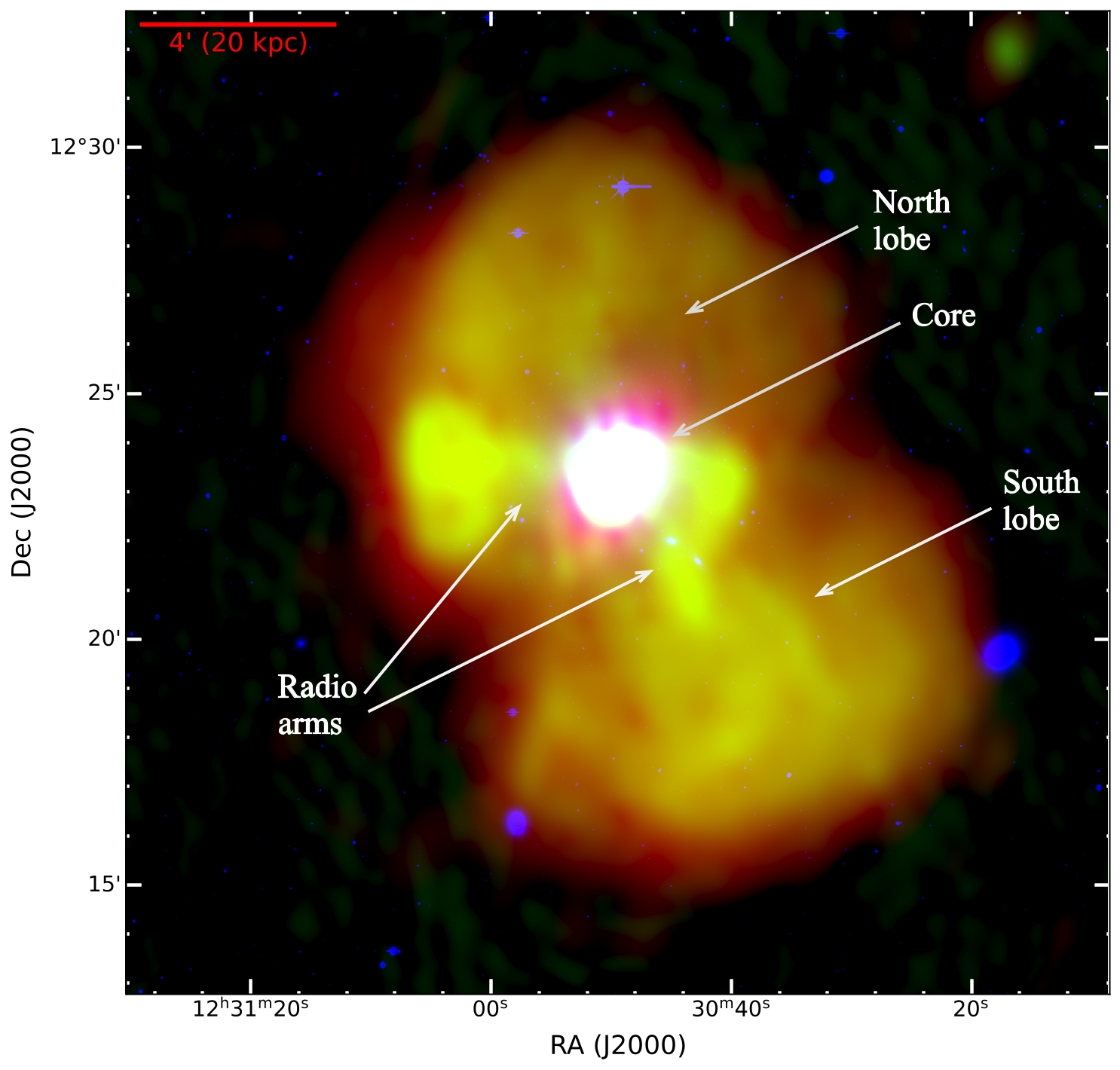}
    \caption{Composite Image of M\,87. This image combines radio and optical observations of M\,87, with the radio map from MWA at 166 MHz shown in red (see also Figure \ref{fig:mwa_obs}), the VLA map at 1.5 GHz in green (see also Figure \ref{fig:vla_obs}), and the optical map from the DESI Legacy Imaging Surveys (\citealt{Dey19}) in blue. Several most prominent features are labeled in this figure.}
    \label{fig:im_color}
\end{figure*}

\begin{figure*}
    \centering
    \includegraphics[width=0.32\linewidth]{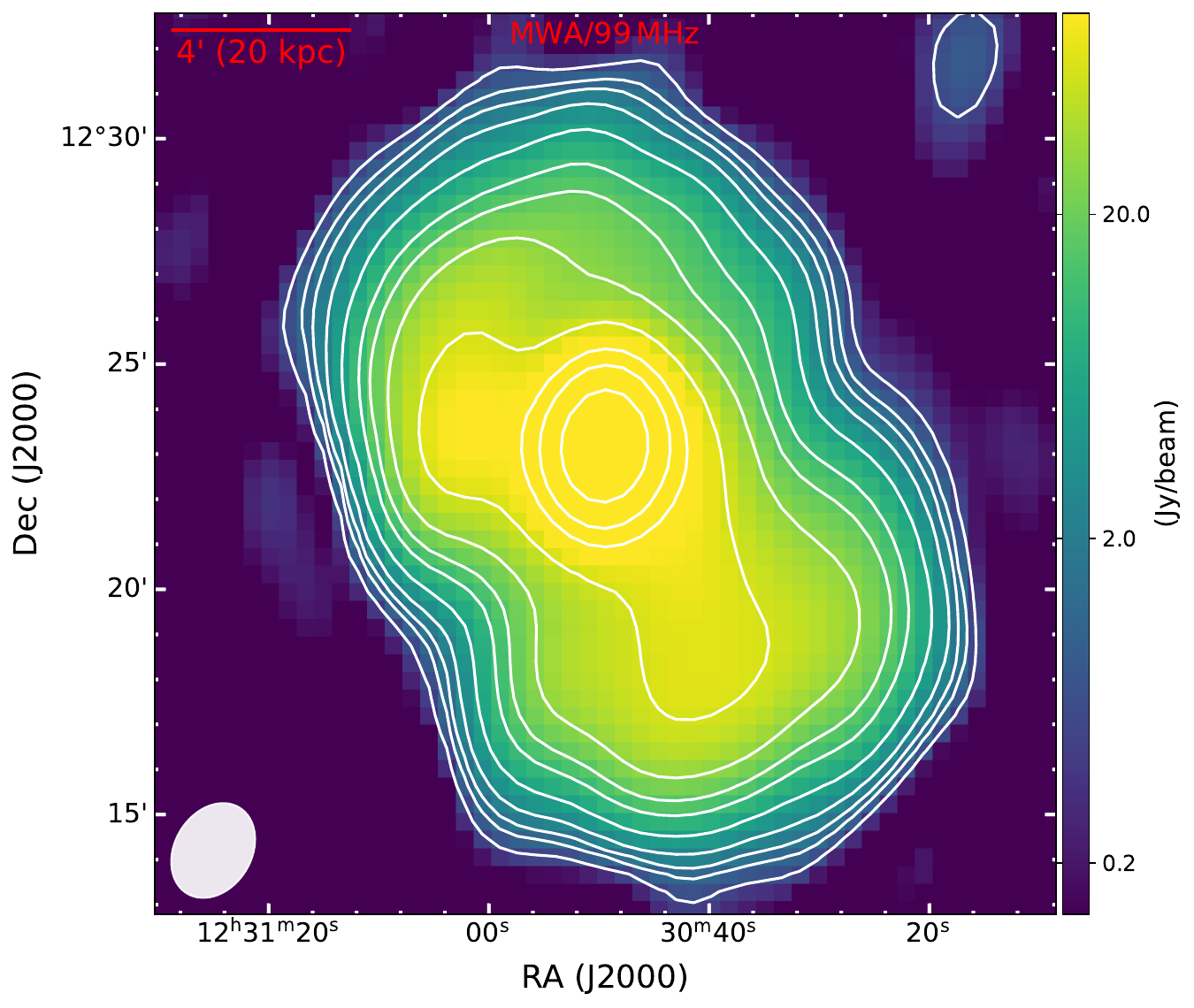}
    \includegraphics[width=0.32\linewidth]{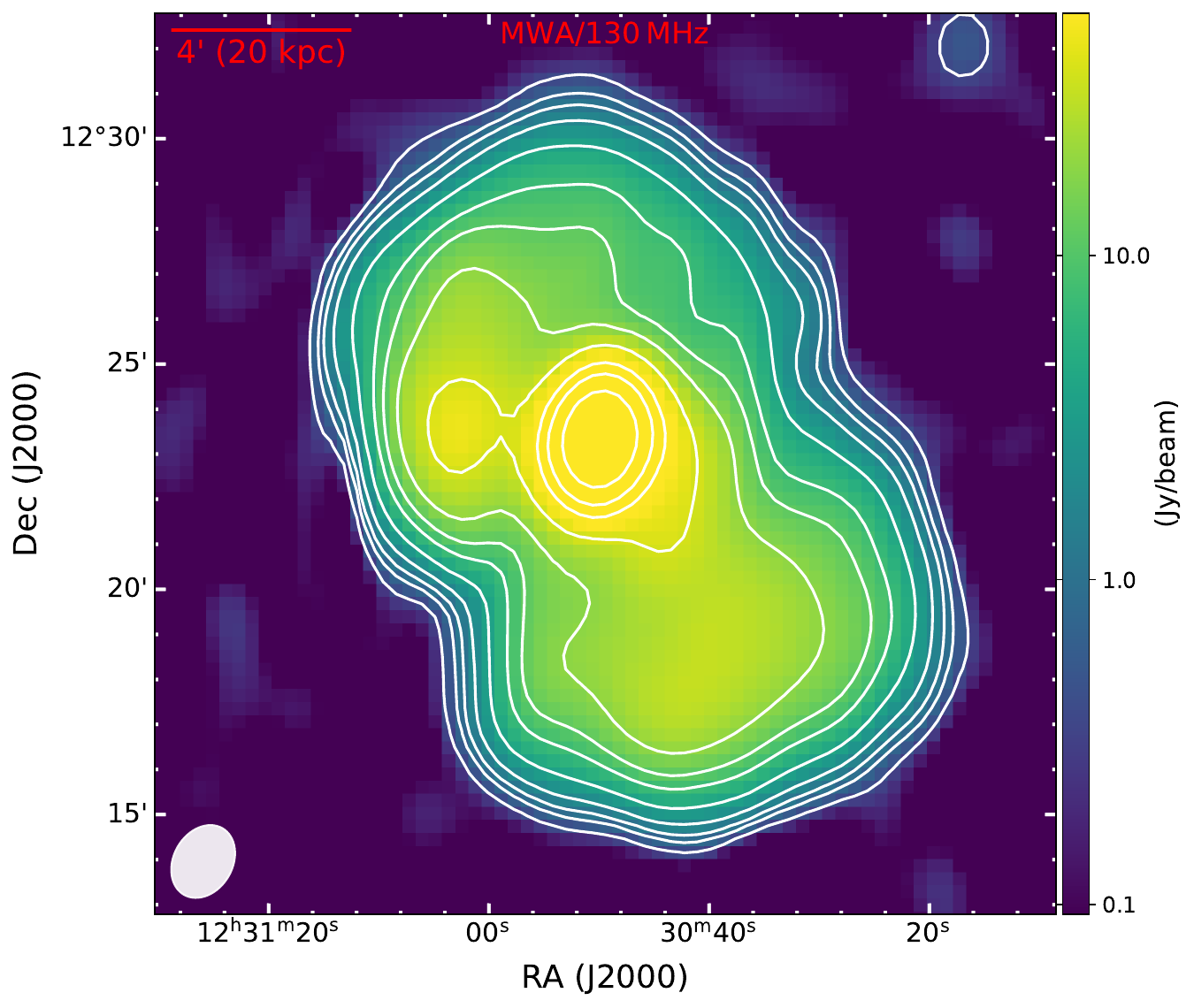}
    \includegraphics[width=0.32\linewidth]{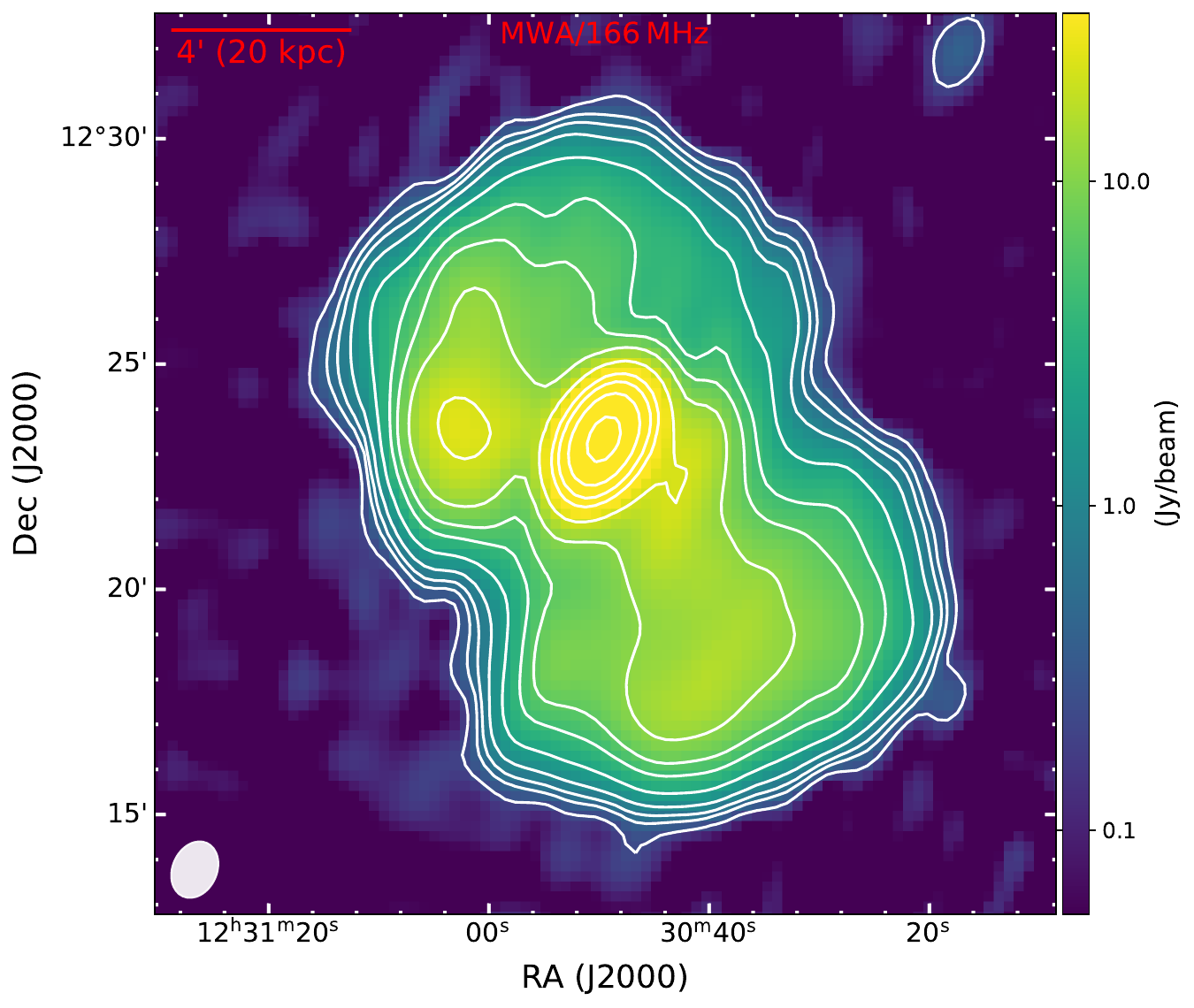}
    \includegraphics[width=0.32\linewidth]{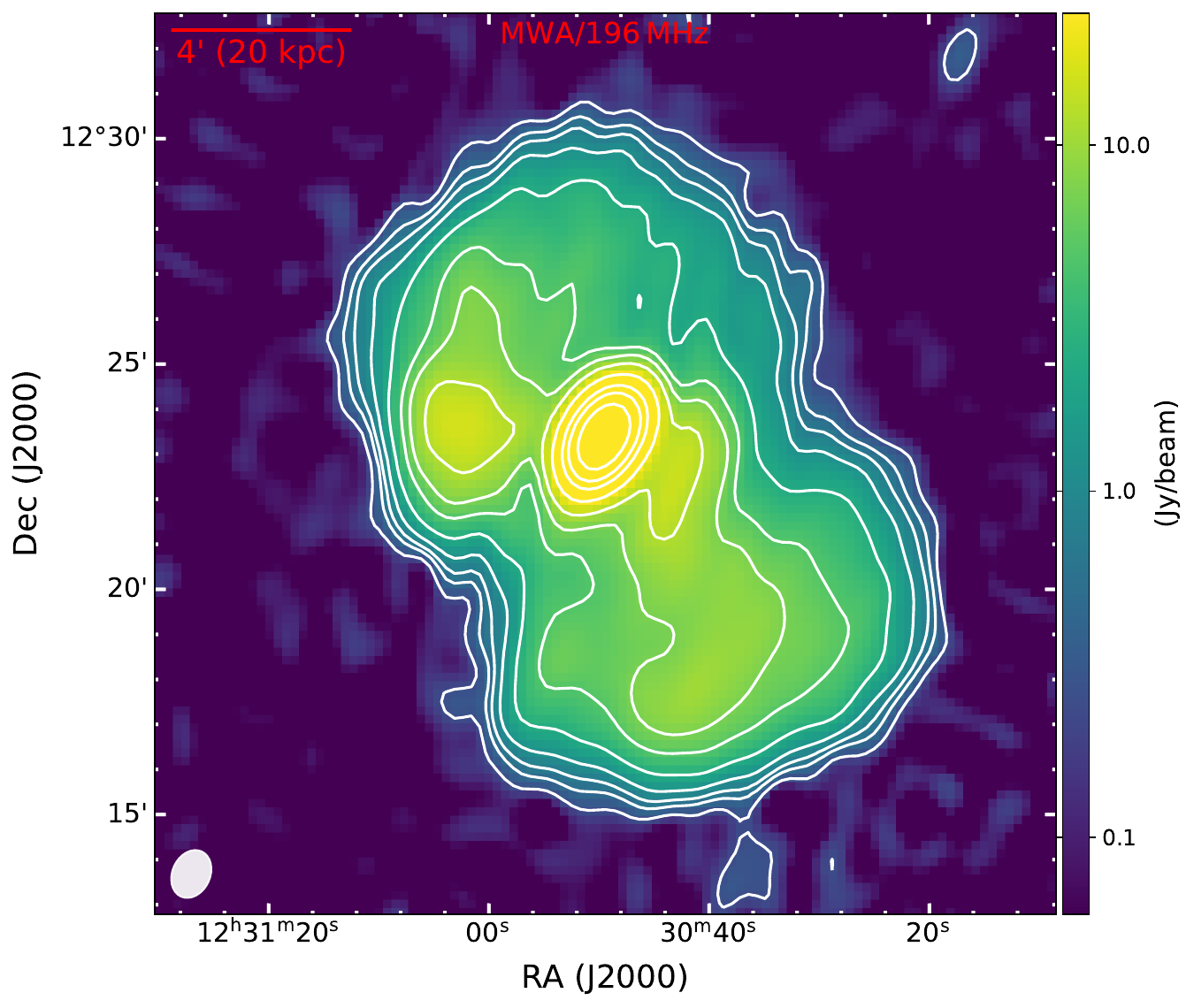}
    \includegraphics[width=0.32\linewidth]{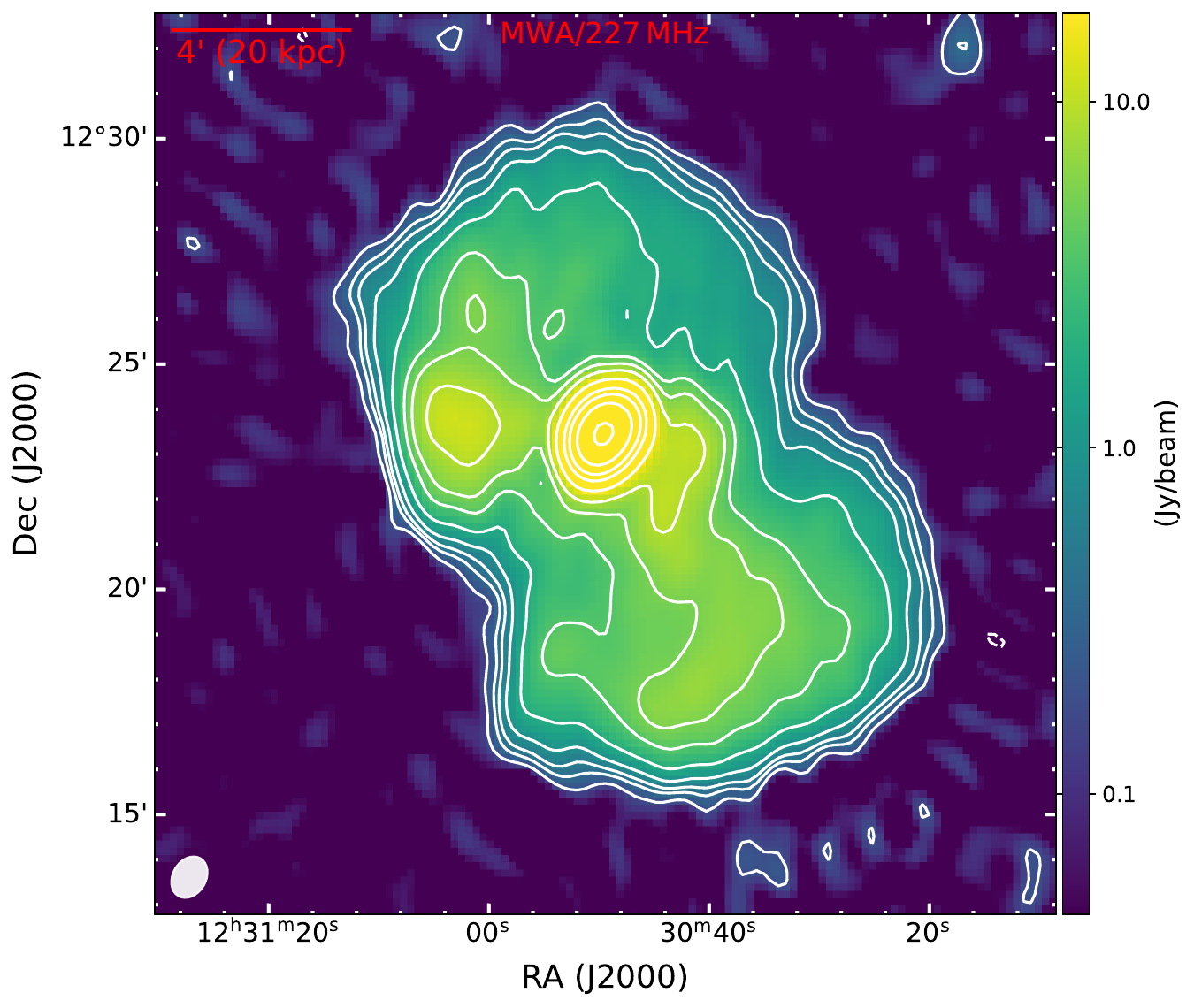}
    \hspace{5.74cm}
    \caption{The MWA intensity images of M\,87 at $\rm 99\,MHz$ (upper left), $\rm 130\,MHz$ (upper middle), $\rm 166\,MHz$ (upper right), $\rm 196\,MHz$ (lower left) and $\rm 227\,MHz$ (lower middle), respectively. The beam is shown at the bottom-left corner in each image. The contour levels are defined as $(-1,1,2,3,5,10,20,30,50,100,200,300,500,1000)\times 4\sigma_{\rm rms}$, where the negative contour is shown by the dotted curves. The $\sigma_{\rm rms}$ and beam size are of each panel are listed in Table \ref{Tab:Im_inf}. Note that 1\arcsec\ at the distance of M\,87 physically represents $\sim 85\rm\,pc$.}
    \label{fig:mwa_obs}
\end{figure*}

\begin{figure*}
    \centering
    \includegraphics[width=0.45\linewidth]{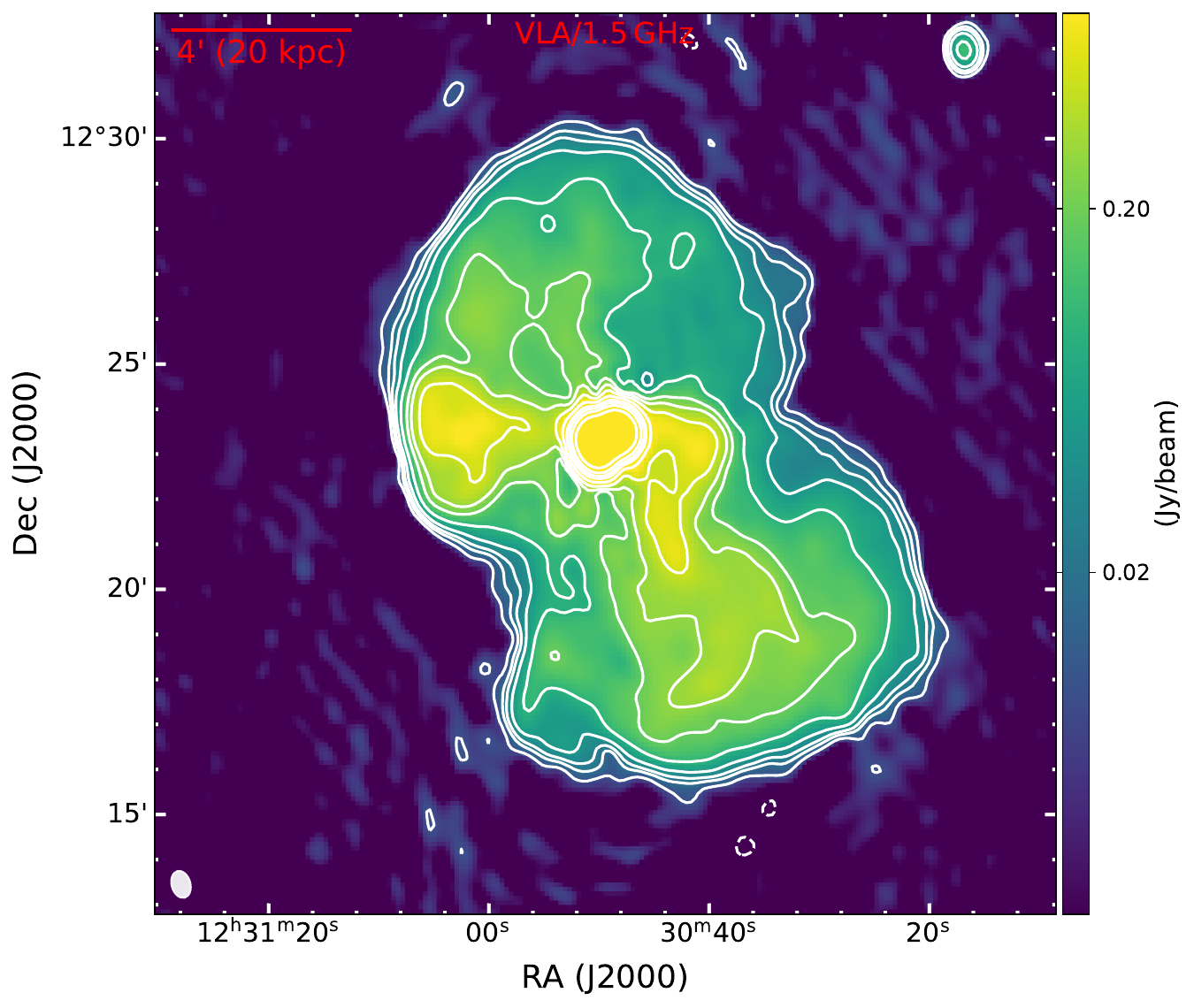}
    \includegraphics[width=0.45\linewidth]{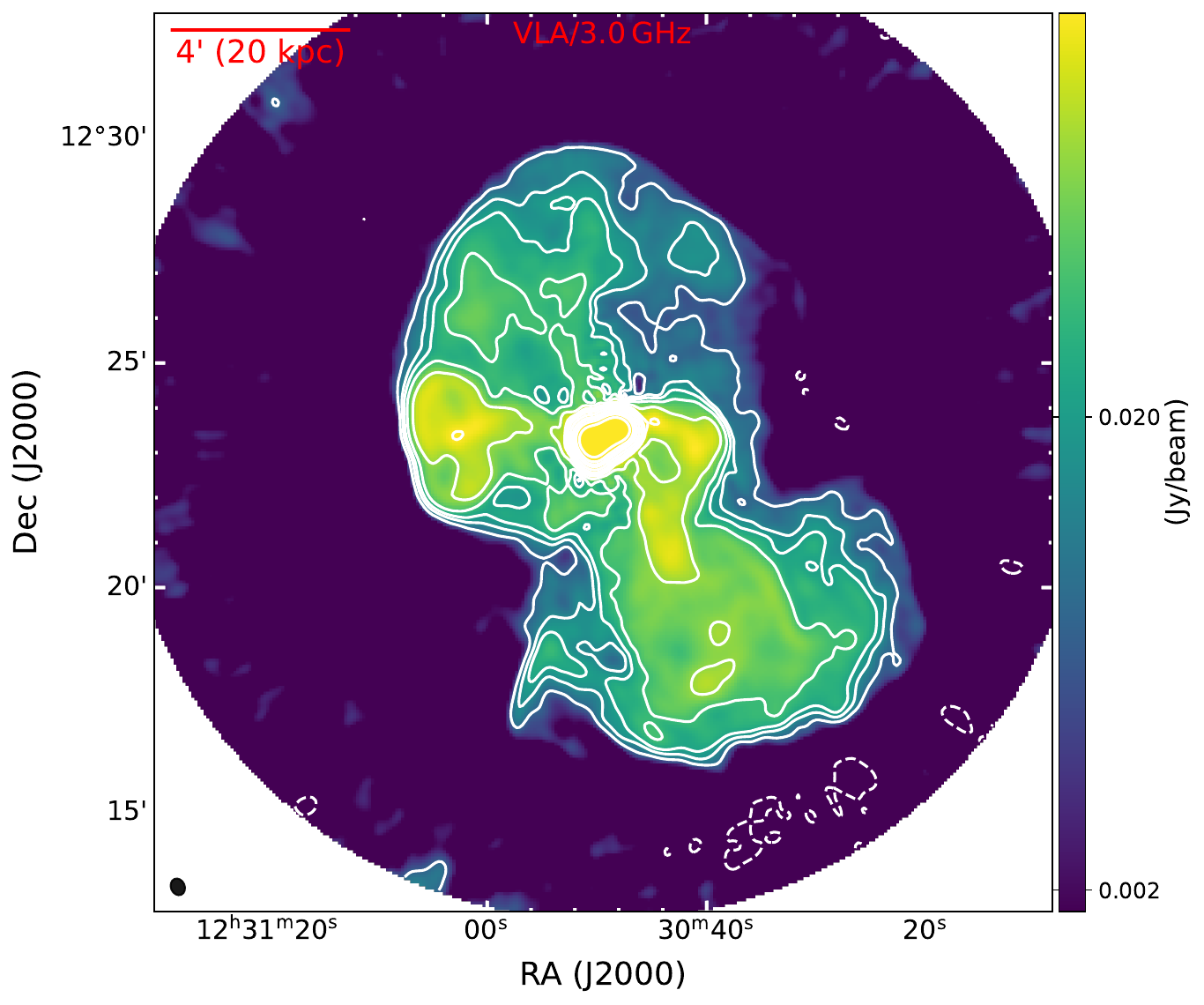}\\
    \caption{Same as Figure \ref{fig:mwa_obs}, but represent the VLA intensity images at $\rm 1.5\,GHz$ (left panel) and $\rm 3.0\,GHz$ (right panel).}
    \label{fig:vla_obs}
\end{figure*}

\subsection{VLA observations}
This study incorporates two VLA datasets, as listed in Table~\ref{Tab:Obs_inf}. Both datasets are obtained in the D configuration using 27 antennas. One dataset was observed with the L band receiver on May 23, 2010 (PI: Frazer Owen), the other was observed with the S band receiver on April 8, 2013 (PI: Francesco de Gasperin). The details of these observations can be found in the observer log file\footnote{\href{http://www.vla.nrao.edu/operators/logs}{http://www.vla.nrao.edu/operators/logs}}. To our knowledge, these datasets have not been previously reported.

The datasets are reduced using the Common Astronomy Software Applications package (\texttt{CASA}\footnote{\href{https://casa.nrao.edu}{https://casa.nrao.edu}}) with version 5.6.3, following standard tutorials\footnote{\href{https://casaguides.nrao.edu/index.php/Karl_G._Jansky_VLA_Tutorials}{https://casaguides.nrao.edu/index.php/Karl\_G.\_Jansky\_VLA\_Tutorial}, see also \citet{Mc07}.}. Below, we provide a detailed description of the calibration and imaging processes for the L band data and briefly highlight the differences in processing the S band data.

{\bf VLA L band data.} The total bandwidth of the $1-2\rm\,GHz$ L band is $1\rm\,GHz$, divided into 8 sub-bands, each consisting of 64 frequency channels and covering a bandwidth of $128\rm\,MHz$. The target, flux density calibrator ($\rm 3C\,286$) and phase calibrator ($\rm J\,1347+1217$) were observed for approximately 5 hours, 7 minutes and 52 minutes, respectively, with a sampling time of 1 second.

First, we examine and flag bad data, manually removing the 5th spectral window (spw) due to severe RFI effects. Subsequently, a priori calibration is performed, including antenna position correction and gain curve calibration. Next, delay calibration and bandpass calibration are conducted using the task \texttt{GAINCAL} and the task \texttt{BANDPASS}, respectively. The flux density scale for 3C 286 is set using the task \texttt{SETJY}, based on the Perley-Butler 2013 scale \citep{Per13}. Following this, complex gain calibration is performed with \texttt{GAINCAL}, and the flux density of J\,1347+1217 is scaled with the task \texttt{FLUXSCALE}, after which the solutions are applied using the task \texttt{APPLYCAL}. Finally, two rounds of phase self-calibration with solution times set to one scan time and sampling time, respectively, are performed, followed by two rounds of phase and amplitude self-calibration with the solution time set to the sampling time for the target source, M\,87. We note that bad data and RFI flagging are conducted throughout the entire calibration process.

Imaging of M\,87 is conducted using the task \texttt{TCLEAN} in `mtmfs' and `wproject' mode with ${\rm nterms} = 2$ for its broad frequency range. The Briggs weighting is set to ${\rm robust} = 0.5$. We perform primary beam correction using the task \texttt{WIDEBANDPBCOR}. This image has a reference frequency of $1515\rm\,MHz$ and a resolution of $37\arcsec\times25\arcsec$, shown in the left panel of Figure \ref{fig:vla_obs}. The off-source root-mean-square (rms) of the image is $\sigma_{\rm rms}\sim2.3\rm\,mJy/beam$, resulting in a dynamic range in intensity of about 25 000.

{\bf VLA S band data.} The total bandwidth of the $2-4\rm\,GHz$ S band is $2\rm\,GHz$, divided into 16 sub-bands, each consisting of 64 frequency channels and covering a bandwidth of $128\rm\,MHz$. The target, flux density calibrator ($\rm 3C\,286$) and phase calibrator ($\rm J\,1254+1141$) were observed for $\sim3.2$ hours, $\sim12$ minutes and $\sim18$ minutes, respectively, with a sampling time of 5 seconds. We note that the second and third spws are affected by RFI, and the 14th, 15th and 16th spws show poor calibration solutions. These spws are flagged and excluded from further analysis. 

M\,87 was observed with five pointing centers. The calibration process follows a similar procedure to that of the L band data. We now use $\rm J\,1254+1141$ to derive the complex gain calibration, and set the `mtmfs' and `mosaic' mode in the task \texttt{TCLEAN} to combine all the five-pointings data for imaging during the self-calibration process.

Imaging of M\,87 is conducted using the task \texttt{TCLEAN} in `mtmfs' and `mosaic' mode with ${\rm nterms} = 2$ for its broad frequency range and five pointing centers. The Briggs weighting is set to ${\rm robust} = 0.5$. We perform primary beam correction using the task \texttt{WIDEBANDPBCOR}. This image has a reference frequency of $3000\rm\,MHz$ and a resolution of $23\arcsec\times18\arcsec$, shown in the right panel of Figure \ref{fig:vla_obs}. The rms of the image is $\sigma_{\rm rms}\sim1.8\rm\,mJy/beam$, resulting in a dynamic range in intensity of about 14 000.

\subsection{Flux scaling}\label{sec:scaling}
To conduct a reliable spectral analysis of M\,87 using data from LOFAR, MWA, VLA and Effelsberg, it is crucial to have a reliable flux calibration/scaling across all observations, so we opt to align them to the RCB scale \citep{Ro73}. We note from previous studies that the total flux of M\,87 across different frequencies generally follows a power-law distribution \citep{Ke96,Ro73,Baars77,Gas12}.

First, the MWA's low frequency and widefield observations. They can be affected by direction-dependent effects (e.g., ionospheric distortions and primary beam variations), leading to deviations in source flux and position. Since these effects are not accounted for during the calibration process, post-processing corrections for images are required using hundreds of sources in the FOV \citep[e.g.,][]{Hin16,Hur17,Du20}. In this work, particularly for M\,87, we follow the flux scaling method outlined by \citet{Gas12}, where the LOFAR data are analyzed.

Within MWA’s frequency range, the total flux of M\,87 can be scaled to the matching with the RCB scale, using a frequency-dependent linear model:
\begin{equation}
         S_{\nu}[\rm Jy]=A_{\rm 0}\sum\limits_{i=1}^{N}10^{A_{i}\rm log^{i}[\nu/150 \rm\,MHz]}.
\end{equation}
\citet[][see also \citealt{Ro73}]{Gas12} tested polynomial fits up to fourth order and found that a first-order polynomial provides the best fit, with ${\rm A}_0 = 1226 \pm 17$ and ${\rm A}_1 = -0.79 \pm 0.008$. Using this linear model, we estimate the expected total flux of M\,87 at MWA’s observed frequencies. We estimate a correction factor at each frequency with the ratio of the expected total flux to the observed value, which is used to rescale MWA’s flux measurements. For our MWA observations, we find these correction factors range from $\sim1.3$ to $\sim2.0$, which are consistent with similar post-imaging flux scale corrections found by \cite{Du20} for MWA snapshot images (see their figure 10 and figure 11). This method preserves the flux ratios among different components of the radio morphology.

Second, the VLA data at L and S bands. For these data, the RCB scale is consistent with the KPW scale \citep{Ke96}. By fitting a power-law model, $S_{\nu}[\rm Jy]=A_{\rm GHz}(\nu/GHz)^{\alpha_{t}}$, to four data points between $0.75-5$ GHz from Table 3 of \cite{Ke96}, we obtain $A_{\rm GHz}=155.8 \pm 2.4$ and $\alpha_{t}=-0.83\pm0.02$. Using this result, we scale the total flux of M\,87 observed by VLA at L and S bands in a manner similar to the MWA observations. For our VLA observations, we find these correction factors range from $\sim0.98$ to $\sim1.09$, which are comparable to the ratios between different flux density scales at different frequencies for M\,87 as presented in Table 7 of \cite{Baars77}.

Third, the LOFAR data, VLA data at $325\rm\,MHz$, and Effelsberg data at $10.55\rm\,GHz$. We supplement our analysis with these datasets from the literature, and the data are directly provided to us by Francesco de Gasperin and Frazer Owen. Their analysis follows \citet{Gas12}, thus ensures the same flux scale, i.e., the RCB scale.

After flux scaling, we adopt a 10\% flux error (plus statistic error combined in quadrature) for the LOFAR, MWA, and Effelsberg data \citep{Gas12,Hur17} and 5\% flux error (plus statistic error combined in quadrature) for the VLA data \citep{Per13}.

\begin{figure*}
    \centering
    \includegraphics[width=0.38\textwidth]{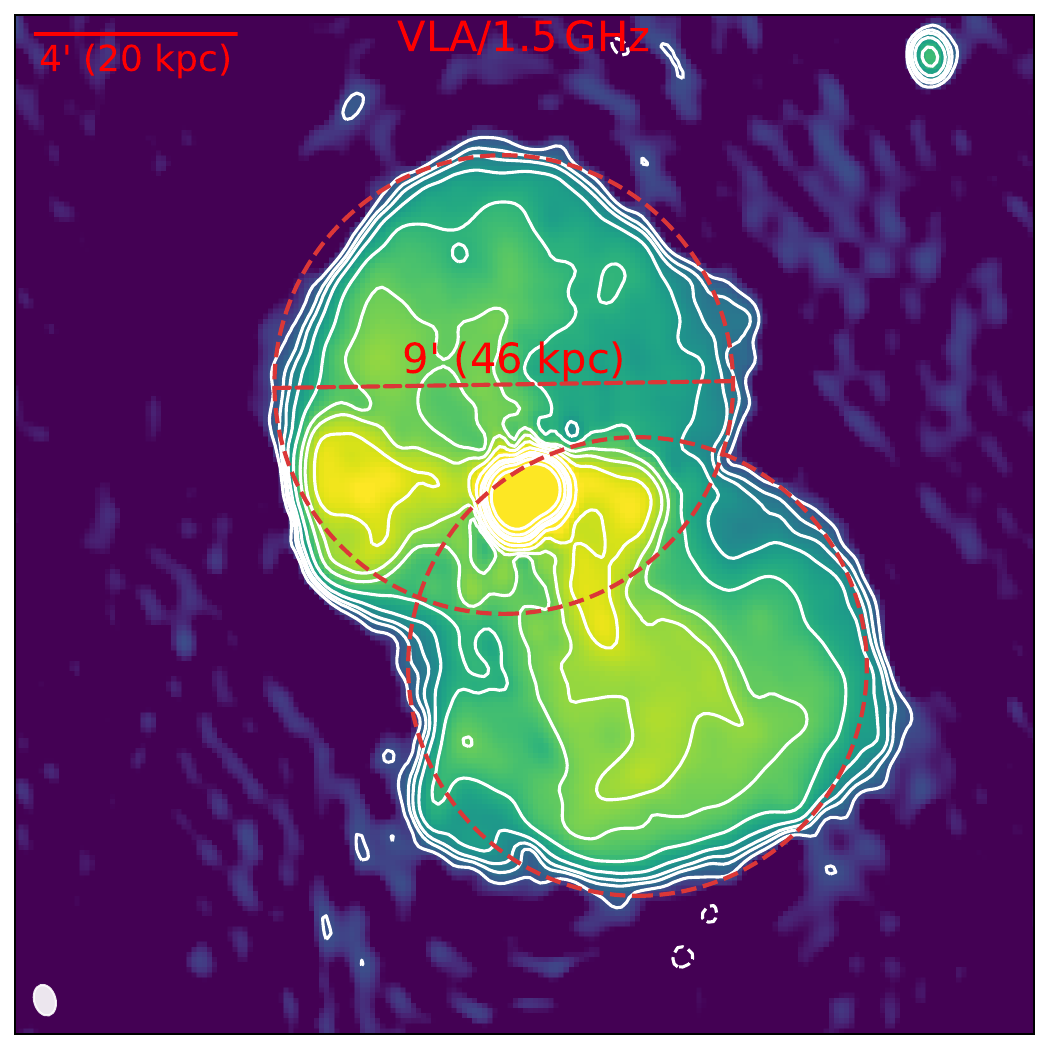}
    \includegraphics[width=0.38\textwidth]{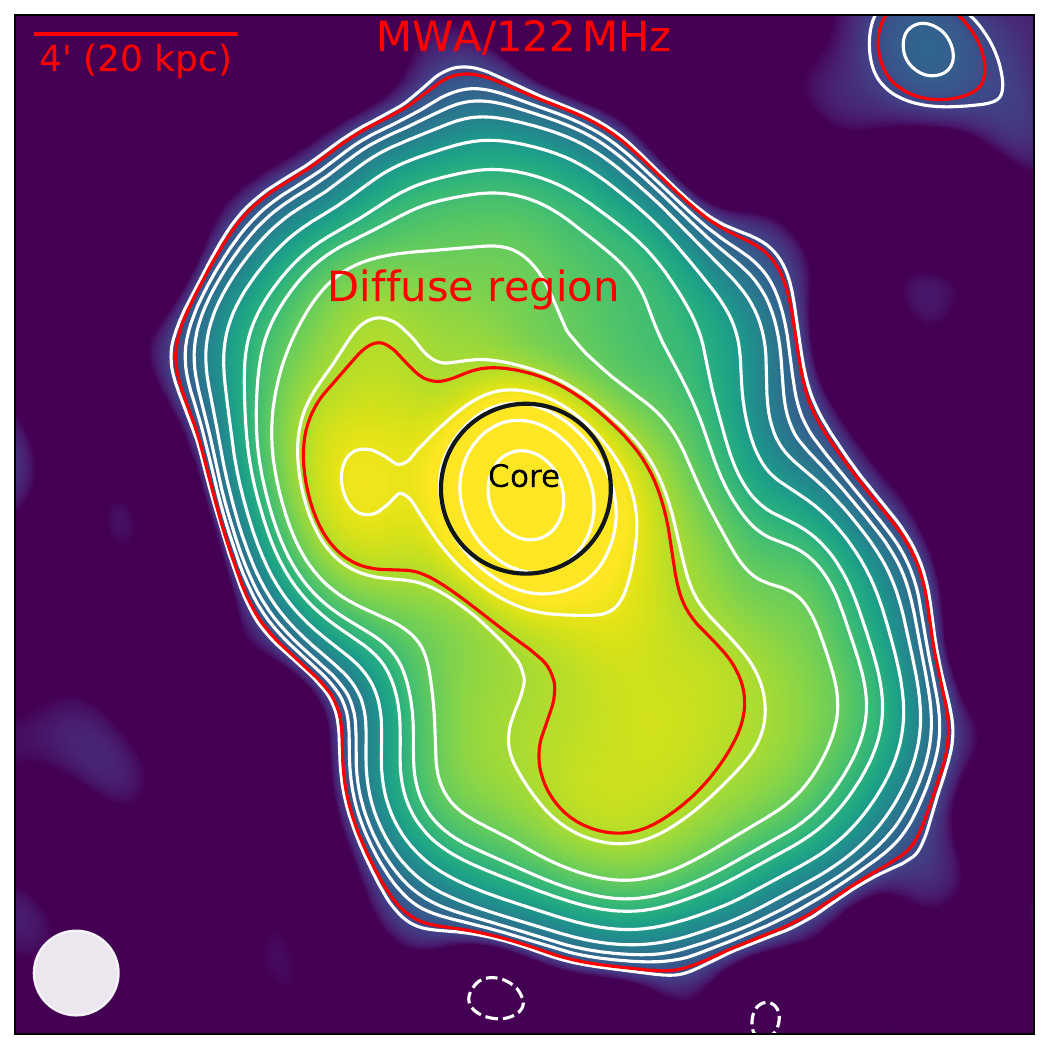}
    
    \caption{Left panel: The $\rm 1.5\,GHz$ high-resolution VLA image of M\,87 (left panel of Figure \ref{fig:vla_obs}). Two dashed circles, each with a diameter of 9\arcmin{} ($\sim 46$ kpc), represent the northern and southern lobe. The black circle marks the nuclei of M\,87, which is also labelled as `core'. Right panel: The low-resolution (100\arcsec) $\rm 122\,MHz$ MWA image of M\,87 with $\sigma_{\rm rms}=60\rm\,mJy/beam$. Contour levels are the same as Figure \ref{fig:mwa_obs}. Red line region, having intensity range from 5 to 500 $\sigma_{\rm rms}$, as labelled `Diffuse region', is overlapped on the image.}
    \label{fig:region}
\end{figure*}

\begin{table*}[htp!]
\renewcommand{\thetable}{\arabic{table}}
\begin{centering}
\small
\caption{Basic Properties of the lobes' diffuse components of M\,87.} \label{Tab:Im_measure}
\setlength{\tabcolsep}{2pt}
\begin{tabular}{c|c|cccccc|ccc}
\hline
\hline
Component &$N_{\rm B}$ &\multicolumn{6}{c|}{Flux (Jy)} &$\bar{\alpha}$& \,$\bar{B}_{\rm e}$& $\bar{P}_{\rm min}$    \\ 
  & & 122 &150 & 227 & 1033 & 1741 &2593& & $\mu$G & $10^{-12}\rm\,dyn\,cm^{-2}$\\
(1) & (2) & (3) & (4) & (5) & (6) & (7) & (8) & (9) & (10) & (11) \\
\hline
Diffuse region & 64.0 &$415\pm41$&$350\pm35$ &$217\pm22$& $47\pm3$&$25\pm2$&$15\pm1.4$&$-0.99\pm0.21$ & $10\pm2$ & $9\pm3$\\
Diffuse components$^\star$ &87.5&$568\pm57$&$479\pm48$ &$296\pm30$& $64\pm4$&$34\pm3$& $20\pm1.9$&$-0.99\pm0.21$ & $10\pm2$ & $9\pm3$\\
\hline
\end{tabular}
\end{centering}

{\bf Notes.} The 1st column presents diffuse region in the lobes (see Figure. \ref{fig:region}). The 2nd column presents the number of beams ($N_{\rm B}$) of each component. The 3rd-8th columns present the flux density at 122, 150, 227, 1033, 1741, $2593\rm\,MHz$, in units of Jy. The 9th column shows the mean spectral index $\bar{\alpha}$. The 10th column presents the mean equipartition magnetic field strength in unit of $\mu$G. The last column presents the mean minimum pressure in unit of $10^{-12}\rm\,dyn\,cm^{-2}$. The uncertainty of $\bar{\alpha}$ is derived as the mean value of the spectral index error image (cf. upper right panel of Figure \ref{fig:spec}); other uncertainties are at the 1$\sigma$ level.\\
$^\star$: Properties of the whole diffuse components in the lobes (cf. Figure \ref{fig:region}), as derived based on the properties of lobes' diffuse region. See Section \ref{sec:images} for details.
\end{table*}

\section{Results}\label{sec:results}

\subsection{MWA and VLA images of M\,87}\label{sec:images}

Figure \ref{fig:im_color} represents a composite multi-wavelength view of M\,87. It combines the radio observations of MWA at $166\rm\,MHz$ and VLA at $1.5\rm\,GHz$, and the optical images from the DESI Legacy Imaging Surveys \citep{Dey19}. For clarity, several key features are annotated in this plot. Figure \ref{fig:mwa_obs} and \ref{fig:vla_obs} display the intensity images of M\,87 from the MWA and VLA observations, while Table \ref{Tab:Im_inf} provides basic image properties (i.e., frequency, resolution, flux, noise level) of these observations. 

In the low-resolution MWA images, the central bright cocoon signifies the core region, with two radio `arms' extending eastward, and westward with a southern extension. Additionally, two fainter lobes (or `bubbles') appear symmetrically distributed above and below the core, each spanning over 40 kpc in projected diameter. Previous studies suggest the southern lobe is inclined toward our line of sight, while the northern lobe recedes, as inferred from kpc-scale jet kinematics \citep{Ava16} and the southern lobe’s higher polarization \citep{Ro96}. The lobes exhibit sharp edges and maintain consistent sizes across frequencies, as seen in high-resolution VLA image (Figure \ref{fig:im_color}). These properties align with earlier works \citep{Ka93,Ro96,Ow00,Gas12,Gas20,Gas25}, suggesting that the emitting plasma is effectively confined potentially due to the substantial external pressure from the ICM or lobe-boundary magnetic fields (e.g., \citealt{Gas12}). Intriguingly, the axis connecting the lobes’ centers is nearly perpendicular to the pc-scale jet direction in projection, implying distinct formation mechanisms and/or temporal evolution between the relativistic jet and the large-scale lobes.

Figure \ref{fig:vla_obs} (left/right panels) displays high-resolution images at 1.5 and 3.0 GHz, respectively, providing clearer insights into the aforementioned structures. Additionally, we can marginally see some filamentary structures permeate the lobes, also evident in images at $50\rm\,MHz$, $140\rm\,MHz$ and $325\rm\,MHz$ \citep{Ow00,Gas20,Gas25}, though their origins remain ambiguous. Possible explanations include turbulence and/or magnetic flux tubes \citep{Ow00,Seb19}.

\subsection{Flux density of the lobes' diffuse components}\label{sec:flux_diffuse}

The two large-scale lobes, each exhibits a roughly circular morphology in projection spanning about 9\arcmin{} ($\sim 46\rm\,kpc$) in diameter, as delineated by dashed circles in the 1.5 GHz high-resolution image (Figure \ref{fig:region}, left panel). To avoid potential contamination from bright components (e.g., the central cocoon and `radio arms'), our analysis targets the diffuse emission in the lobes, defined by intensities of $5-500~\sigma_{\rm rms}$ ($\sigma_{\rm rms}\sim60~\rm mJy/beam$) in the $\rm 122\,MHz$ image (see Figure \ref{fig:region})\footnote{The lobes' diffuse region selected here is arbitrary, but a smaller upper limit intensity doesn't influence our conclusion.}. This image is the lowest frequency of the MWA image that can be re-convolved to 100\arcsec (see Section \ref{sec:spec_model}). Table \ref{Tab:Im_measure} summarizes the size, flux density at six typical frequencies and derived physical properties of the diffuse region. For each specific frequency, we compute the mean intensity ($F_\nu/N_{\rm B}$, where $F_\nu$ is the flux density and $N_{\rm B}$ is the region size in beams) across the lobes' diffuse region ($\sim64.0$ beams). Under the assumption that diffuse radio emission uniformly fills both large-scale lobes, we derive the total integrated flux density of the (lobes') diffuse components (spanning $\sim87.5$ beams). For instance, at $\rm 150\,MHz$, the mean intensity is approximately 5.5 Jy/beam for the lobes' diffuse region, yielding the total flux of the lobes' diffuse components of about 480 Jy. Similar calculations for other typical frequencies are tabulated in Table \ref{Tab:Im_measure}.


\begin{figure*}
    \centering
    \includegraphics[width=0.4\textwidth]{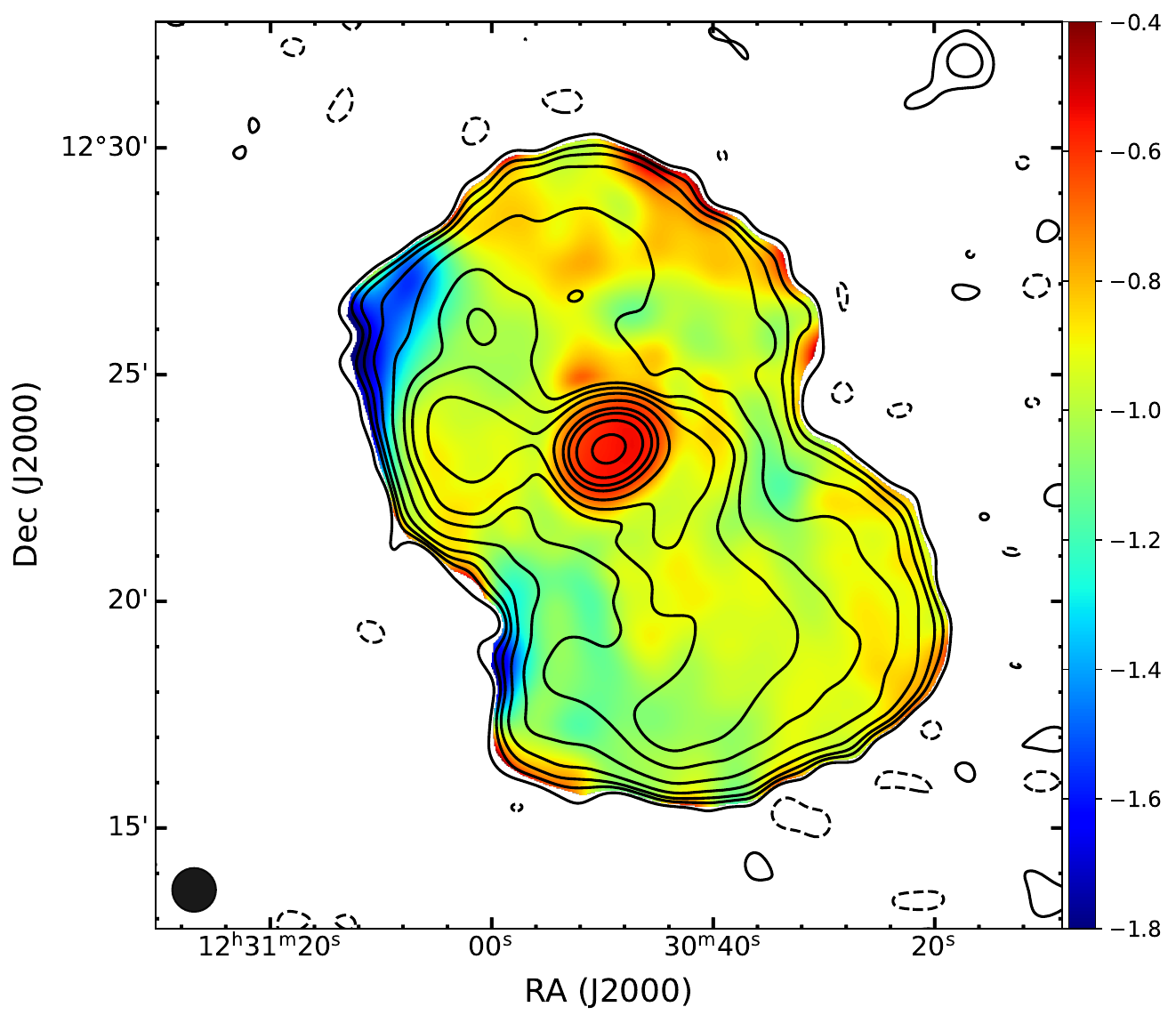}
    \hspace{0.5cm}
    \includegraphics[width=0.4\textwidth]{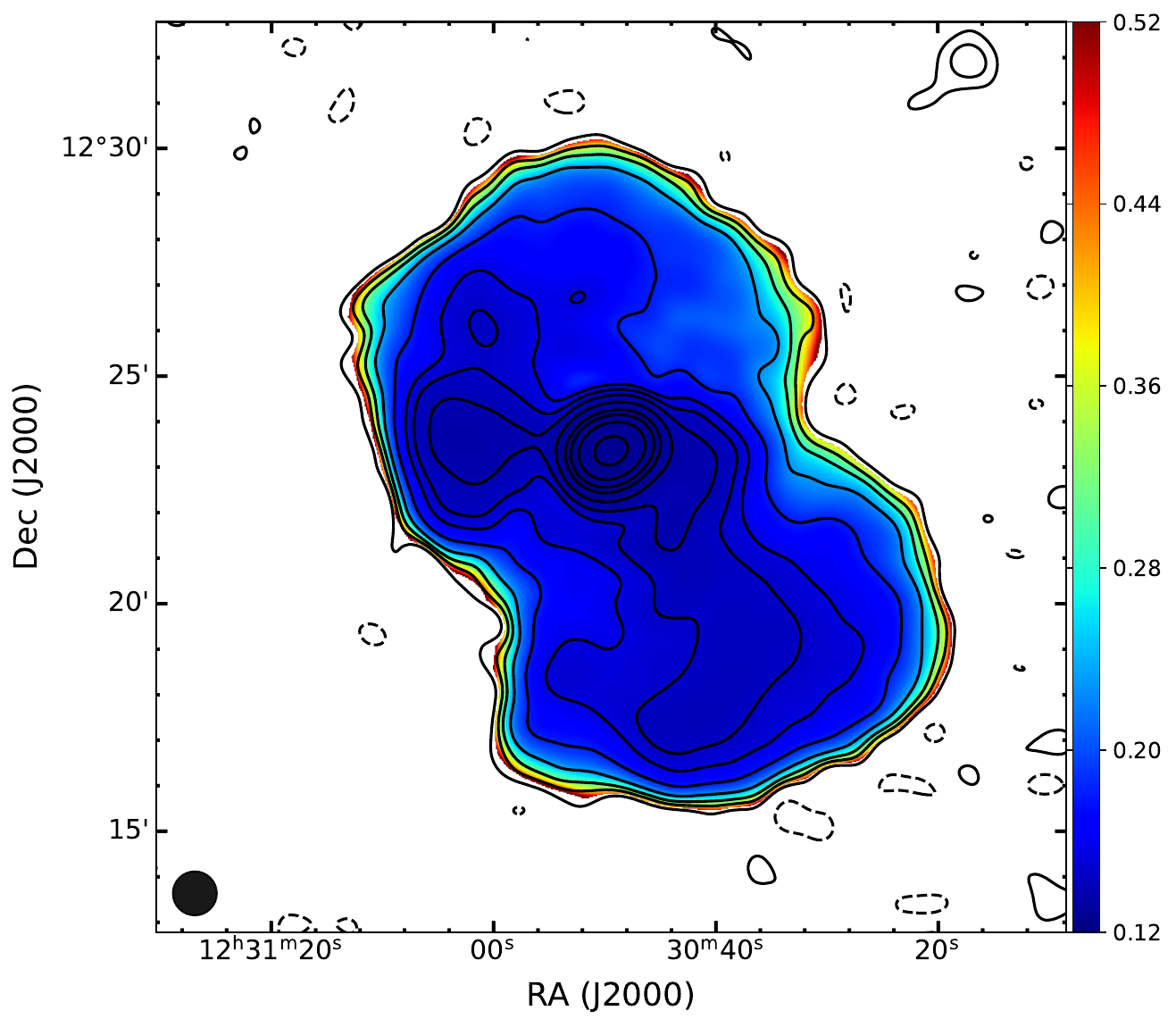}\\
    \vspace{0.2cm}
    \includegraphics[width=0.4\textwidth]{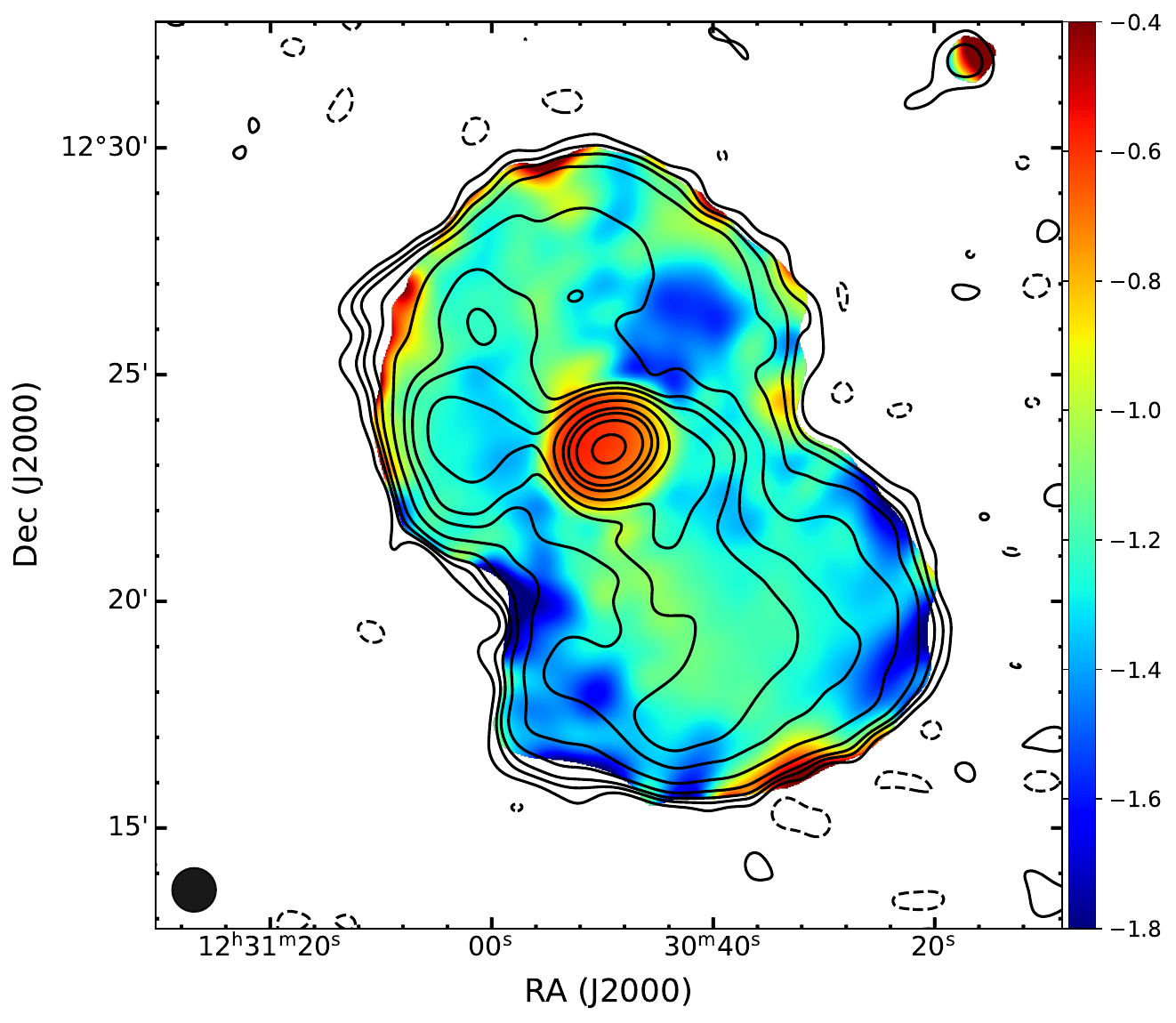}
    \hspace{0.5cm}
    \includegraphics[width=0.4\textwidth]{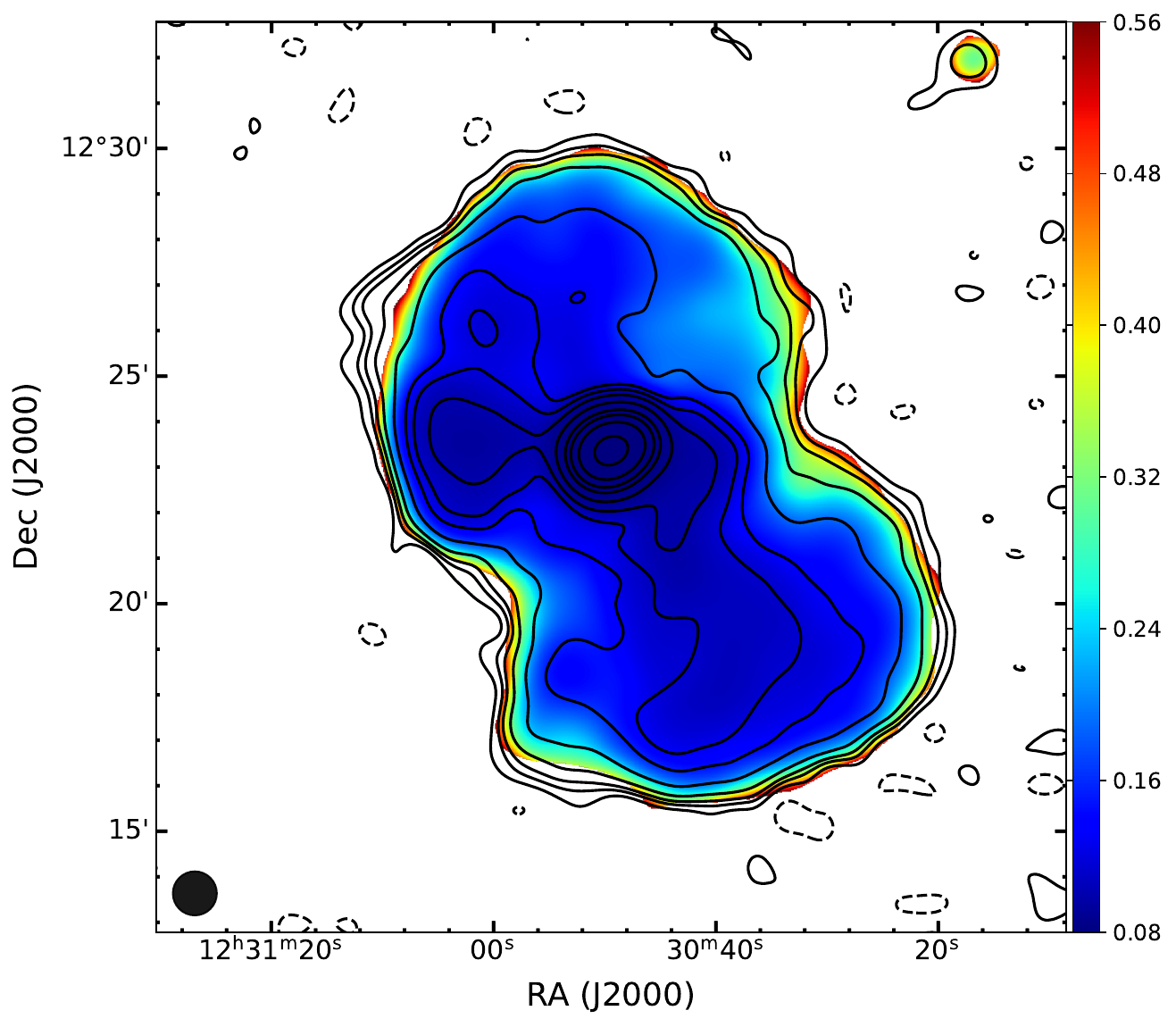}
    \caption{The spectral index (left panels, shown in color) and its $1\sigma$ uncertainty (right panels, shown in color) images of M\,87. The upper panels are derived based on the $\rm 140\,MHz$, $\rm 227\,MHz$ and $\rm 325\,MHz$ images. The lower panels are derived based on the $\rm 1063\,MHz$, $\rm 1711\,MHz$ and $\rm 2563\,MHz$ images. The spatial resolution of these images is 58\arcsec. Contour levels are represented by black lines at $(-1,1,2,3,5,10,20,30,50,100,200,300,500,1000)\times 3\sigma_{\rm rms}$ ($\sigma_{\rm rms}=74\rm\,mJy/beam$) with negative contour is shown with dotted lines.}
    \label{fig:spec}
\end{figure*}

\subsection{Spectral properties}

\subsubsection{Spectral index image}
\label{sec:spec_im}
In this section, we generate higher resolution images using the highest frequency band observed by MWA, as well as the first and sixth spws at L band and the fifth spw at S band observed by VLA. As discussed in \ref{sec:spec_model}, these observations effectively recover the large-scale structures of $\rm M\,87$. During the imaging process, we use a multi-scale deconvolution algorithm, apply uniform Briggs weighting and consider a specific \textit{uv}-range of $0.2-3.9\rm\,k\lambda$, where $0.2\rm\,k\lambda$ corresponds to the max angular scale of M\,87 (see section \ref{sec:spec_model}), and $3.9\rm\,k\lambda$ corresponds to the longest baseline length of the MWA data. In addition, for the MWA data, deep imaging is performed across the full 30 MHz band---split into four subbands---and the fourth subband, centered at 227\,MHz and offering the highest resolution, is used to derive the spectral index image. For the VLA data, we image each selected spw separately—first and sixth spws at L band and the fifth spw at S band—followed by primary beam correction using the CASA task \texttt{IMPBCOR}. This results in central frequencies of 1063 MHz, 1711 MHz, and 2563 MHz, respectively. Finally, we re-convolve these images, along with the 140\,MHz and 325\,MHz images, to a common resolution of 58\arcsec{} and align them based on the brightest core position. We note that, except for the LOFAR observations, flux scaling is conducted using the method described in Section \ref{sec:scaling}.

We derive the spectral index image based on a pixel-by-pixel power-law fitting. First, we use three low frequency images at $\rm 140\,MHz$\footnote{The quality of this LOFAR image is better \citep{Gas12}.}, $\rm 227\,MHz$ and $\rm 325\,MHz$. The derived spectral index and uncertainty images are shown in the upper panels of Figure \ref{fig:spec}. Next, we examine three higher-frequency images at $\rm 1063\,MHz$, $\rm 1711\,MHz$ and $\rm 2563\,MHz$, with the corresponding spectral index and uncertainty images displayed in the lower panels of Figure \ref{fig:spec}. Compared to the spectral index image in the upper panel, the lobe region exhibits a lower value of $\alpha$ (i.e., a steeper spectrum). This suggests the presence of a curved spectrum with a break frequency in the lobe region, as discussed in Section \ref{sec:spec_model}. 

We now focus on the spectral index results displayed in the upper panels of Figure \ref{fig:spec}. The brightest core region displays the flattest spectral index, with a mean value of $\sim-0.61$ and a mean error of 0.13. The spectral index between the core and the lobes' region presents a sharp decrease, indicating an adiabatic expansion of the lobes \citep{Gas12}. Within the lobes, the spectral index are range from $\sim-0.8$ to $\leq-1.8$, with the steepest values observed at the northeastern edge. Notably, the lobe region north of the core has a mean spectral index of $\sim -0.9$, which is typical for lobe regions at low frequencies \citep[e.g.,][]{Mor21}. In the LOFAR results presented by \citet{Gas12}, this region may have been affected by deconvolution errors, as speculated in their study, leading to some artificially flattened spectral index values.

\begin{figure}
    \centering
    \includegraphics[width=0.45\textwidth]{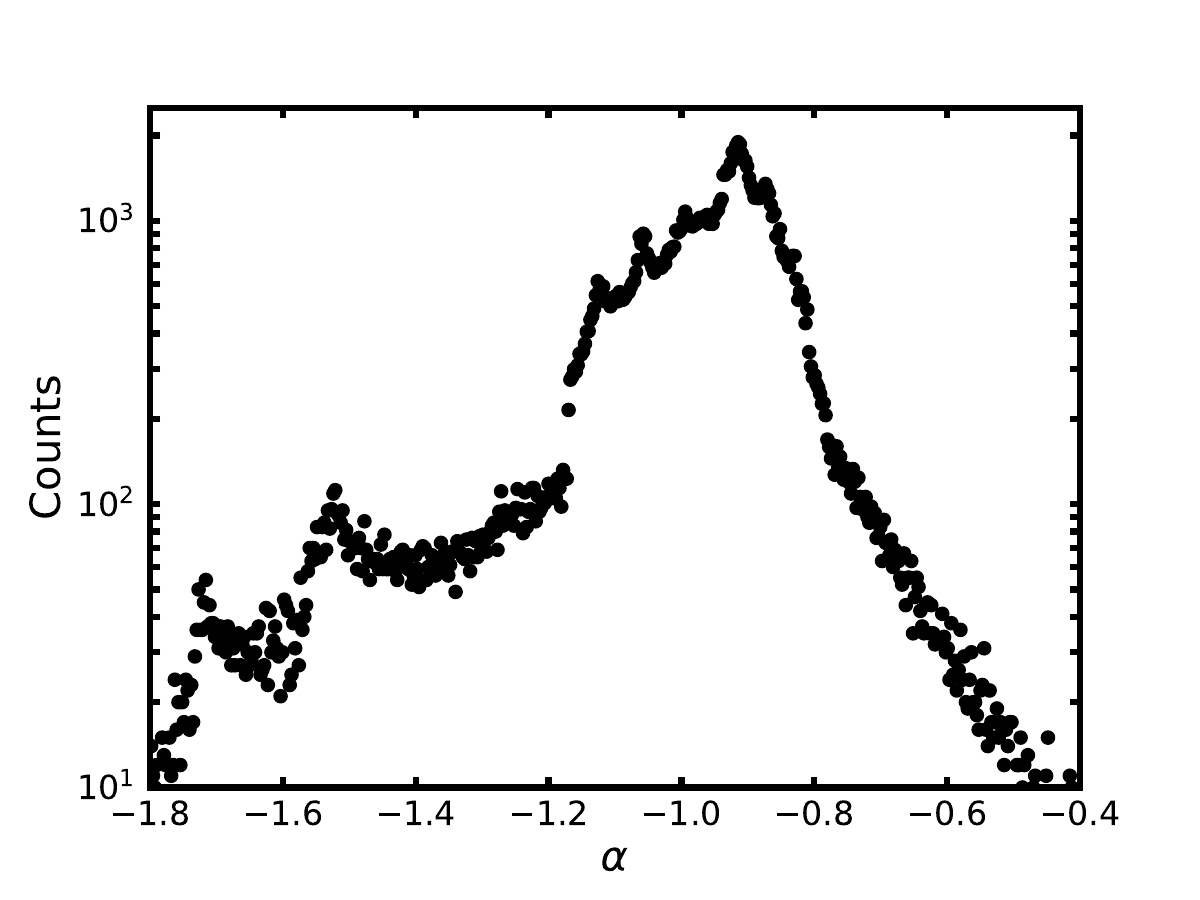}
    \caption{The spectral index statistical distribution of the lobes’ diffuse region.}
    \label{fig:lobe_spec} 
\end{figure}

Overall, the spectral index in the lobes appears relatively uniform, mostly ranging between $-1.2$ and $-0.8$, with no significant correlation to intensity. Figure \ref{fig:lobe_spec} presents the statistical distribution of the spectral index in the lobes' diffuse regions, showing a peak around $-0.9$ and a mean value of $-1.0\pm0.2$. Only a few pixels exhibit spectral indices that transition from steep to ultra-steep spectra (with a critical threshold of $\alpha\sim-1.2$, see \citealt{Pa07,Bri18}).

We conclude that the spectral index distribution in the two giant lobes is nearly uniform, moderately steep (a mean value of $\sim-1.0$), narrowly ranged (mostly within $-1.2\leq\alpha\leq-0.8$). Notably, the distribution shows no significant difference between bright and faint diffuse regions. These characteristics suggest that the lobes are turbulent.

\begin{figure*}
    \centering
    \includegraphics[width=0.48\textwidth]{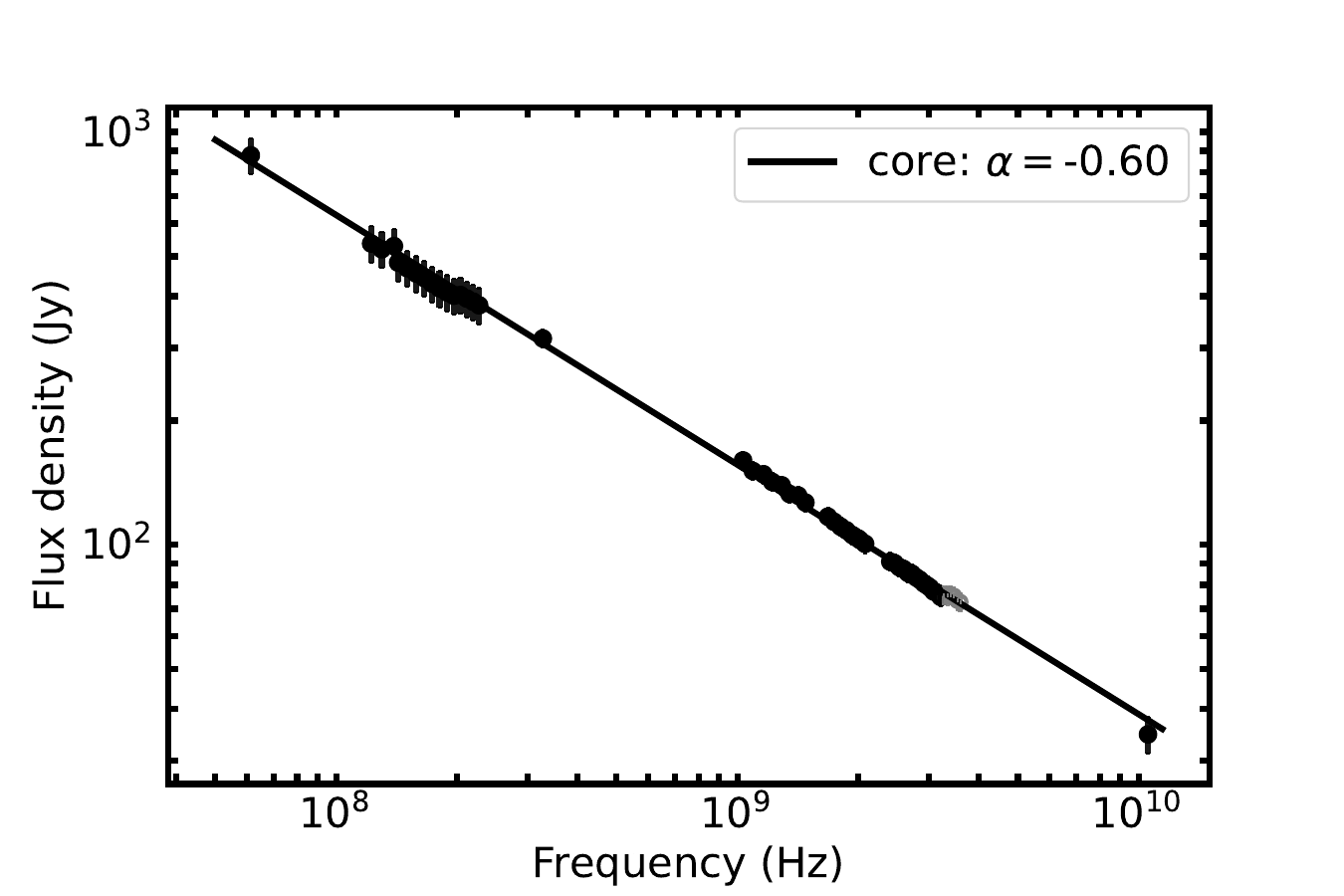}
    \includegraphics[width=0.445\textwidth]{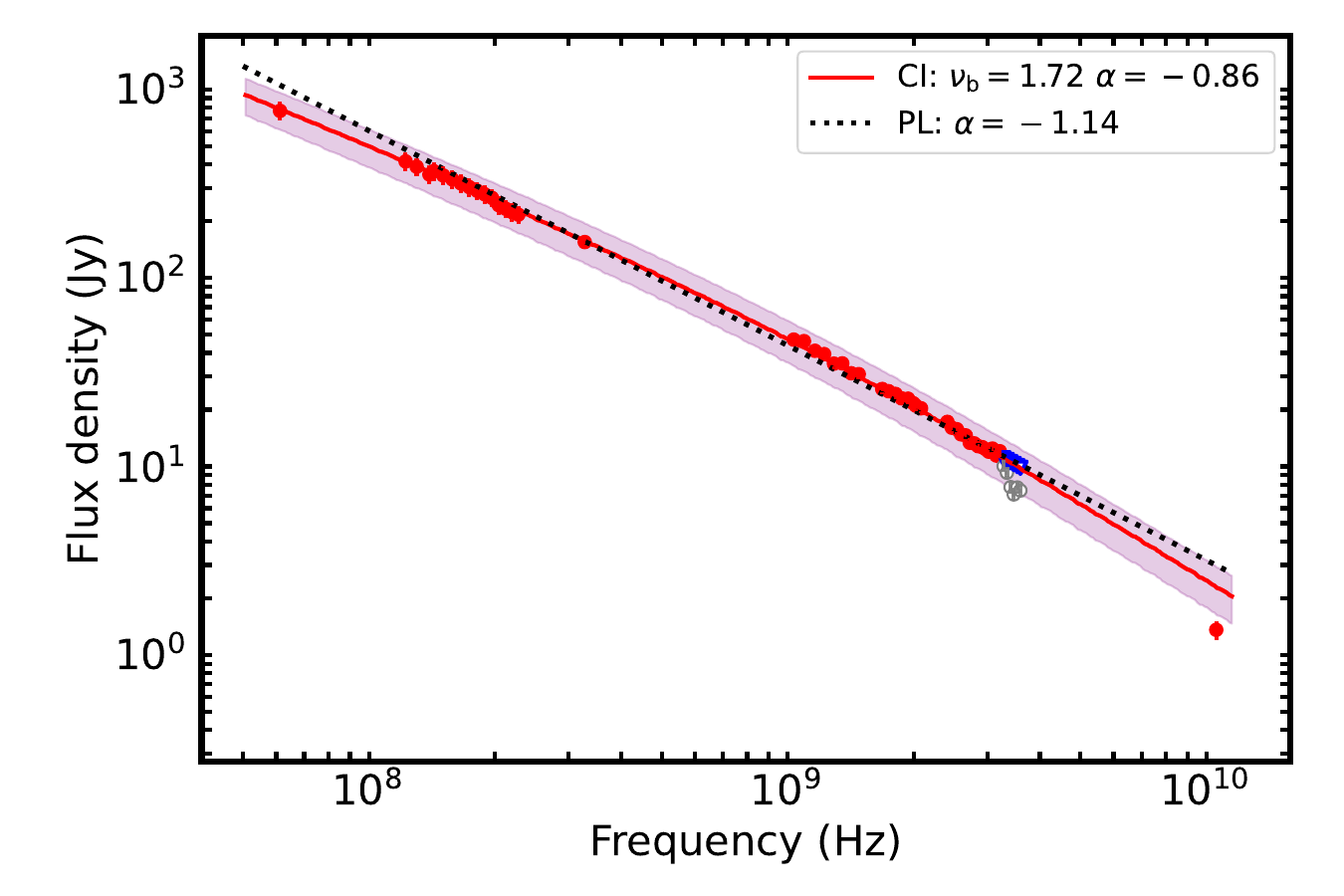}
    \caption{Left panel: The power-law modeling for the spectra of the core region of M\,87 with effective data points ( solid circles) observed by LOFAR, MWA, VLA and Effelsberg, where the best modeling presented with the black solid line. The last 6 data points (empty circles) at S band are dropped from the modeling. Right panel: Spectral modeling of the lobes' diffuse region, as depicted in Figure \ref{fig:region}. The meaning of the lines is labelled. The stripe presents the 2$\sigma$ errors for the CI model. The red-filled circles represent the effective data points adopted in the modeling, the grey circles represent those excluded, and the blue triangles represent the derived flux for the excluded data with the mean spectral index of the lobes' diffuse region. $\nu_{\rm b}$ is in GHz.}
    \label{fig:core_total_spec}
\end{figure*}

\subsubsection{Spectral modeling}
\label{sec:spec_model}

To generate well-sampled wideband spectra, we image the MWA data and VLA data at L and S bands using the multi-scale deconvolution algorithm, setting a uniform Briggs weighting and considering a specific \textit{uv}-range of $0.2-1.6\rm\,k\lambda$ (noting that $1.6\rm\,k\lambda$ is close to the longest \textit{uv}-range for the lowest frequency band of the MWA data). In addition, for the MWA data, deep imaging is performed across the full 30 MHz band---split into four subbands---resulting in a total of 20 sub-band images. Since lower-frequency images tend to have lower resolution, we exclude the very low-frequency MWA observations ($<$122$\rm\,MHz$) to maintain an angular resolution of 100\arcsec. For the VLA data, we perform imaging for every 32 channels, followed by primary beam correction using the CASA task \texttt{IMPBCOR}, yielding 14 images at L band and 22 images at S band. Finally, all available images---including those at the 60\,MHz, 140\,MHz, 325\,MHz and 10.55\,GHz---are then re-convolved to a common resolution of 100\arcsec and aligned based on the brightest core position. We note that, except for the LOFAR observations, flux scaling follows the method described in Section \ref{sec:scaling}.

\begin{figure*}
    \centering
    \hspace{0.6cm} \hspace{1cm}\includegraphics[width=0.30\textwidth]{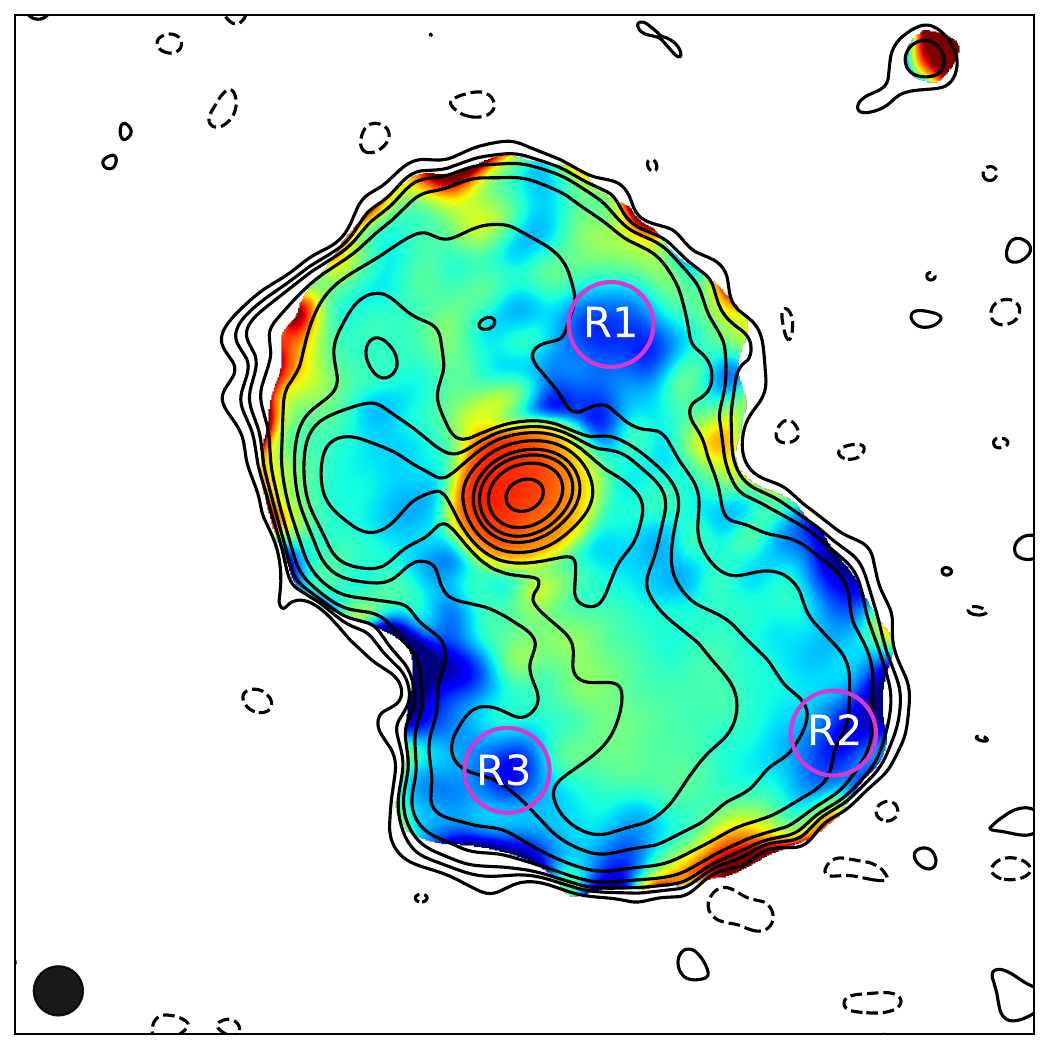}\hspace{1.cm}
    \includegraphics[width=0.45\textwidth]{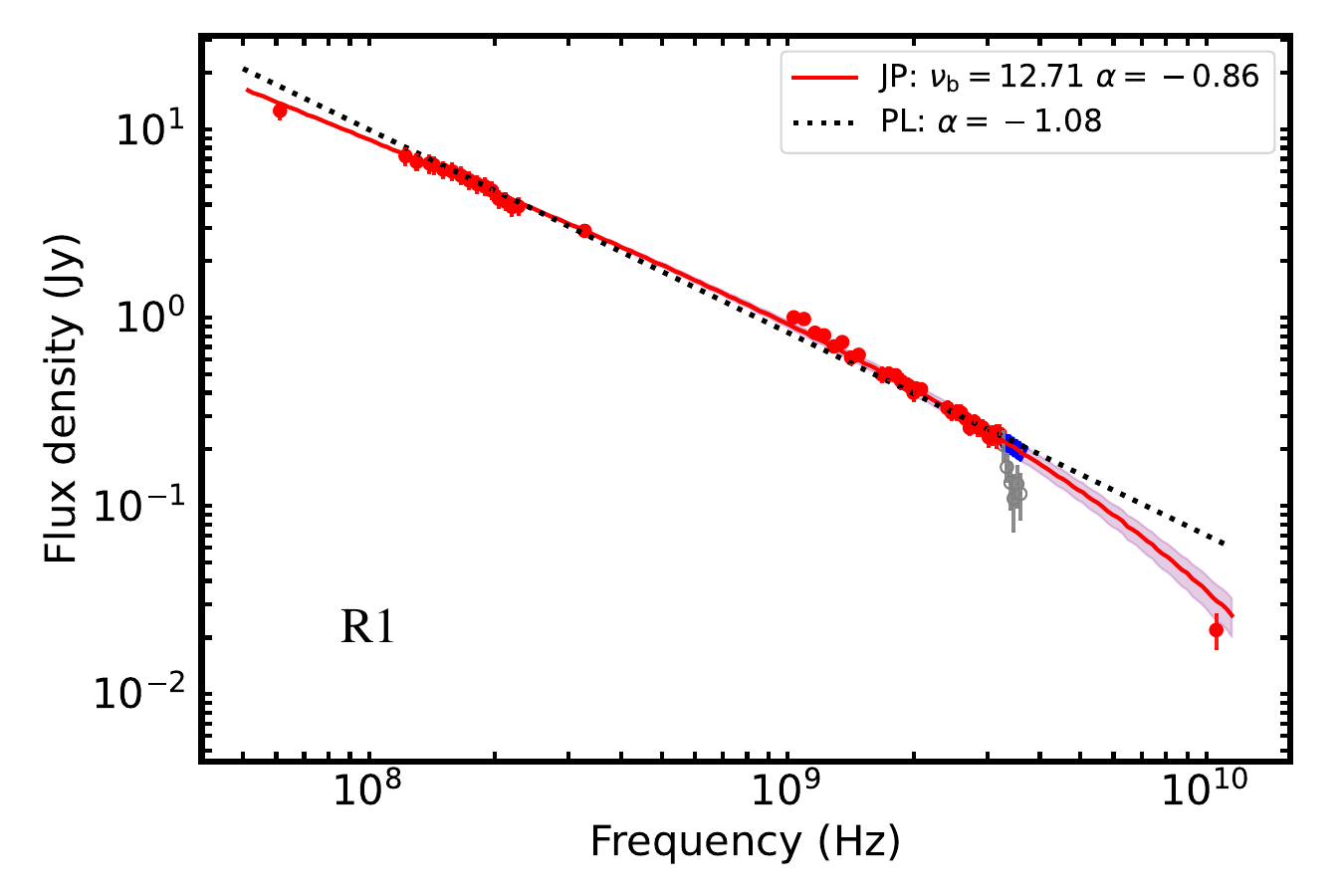}
    \includegraphics[width=0.45\textwidth]{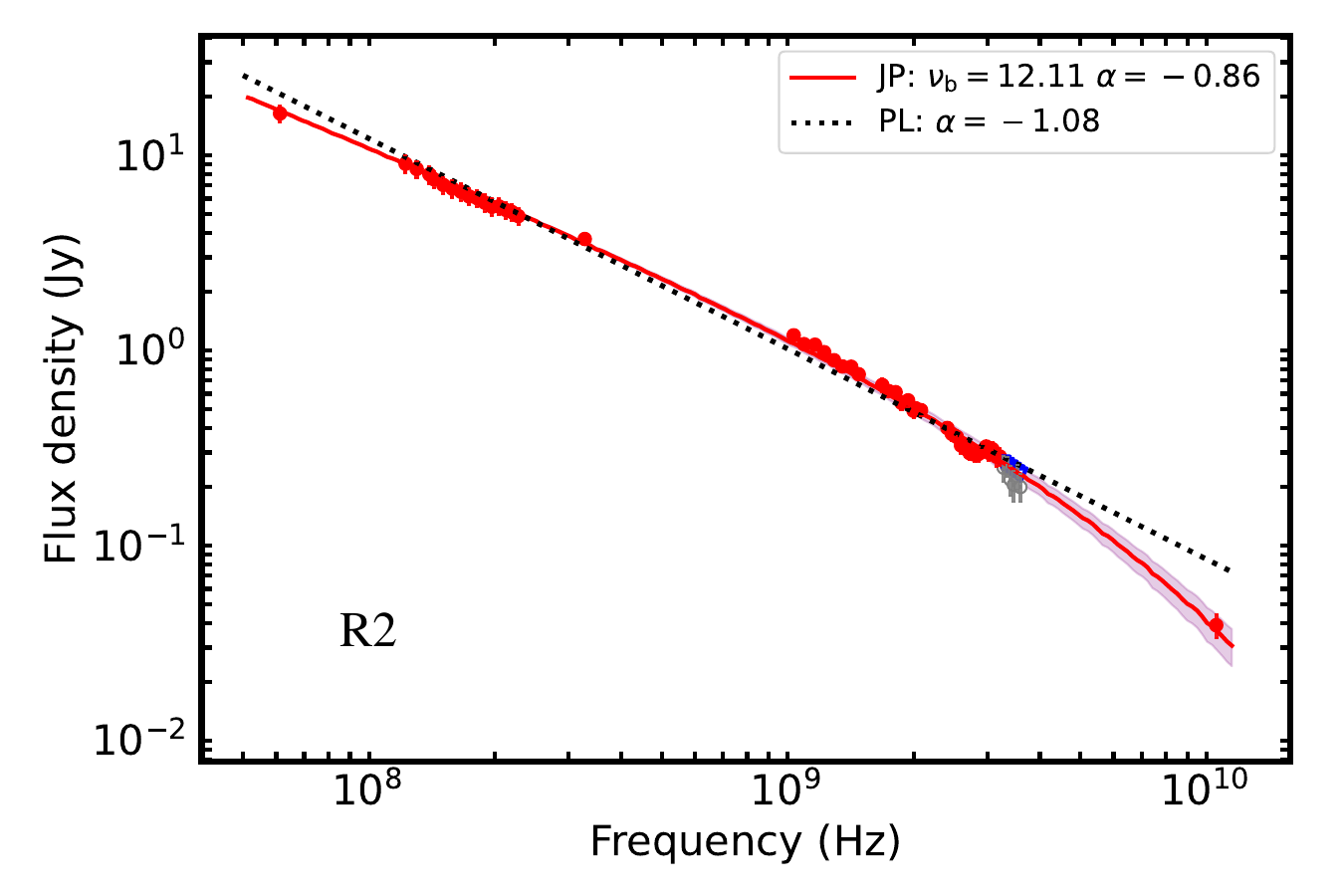}
    \includegraphics[width=0.45\textwidth]{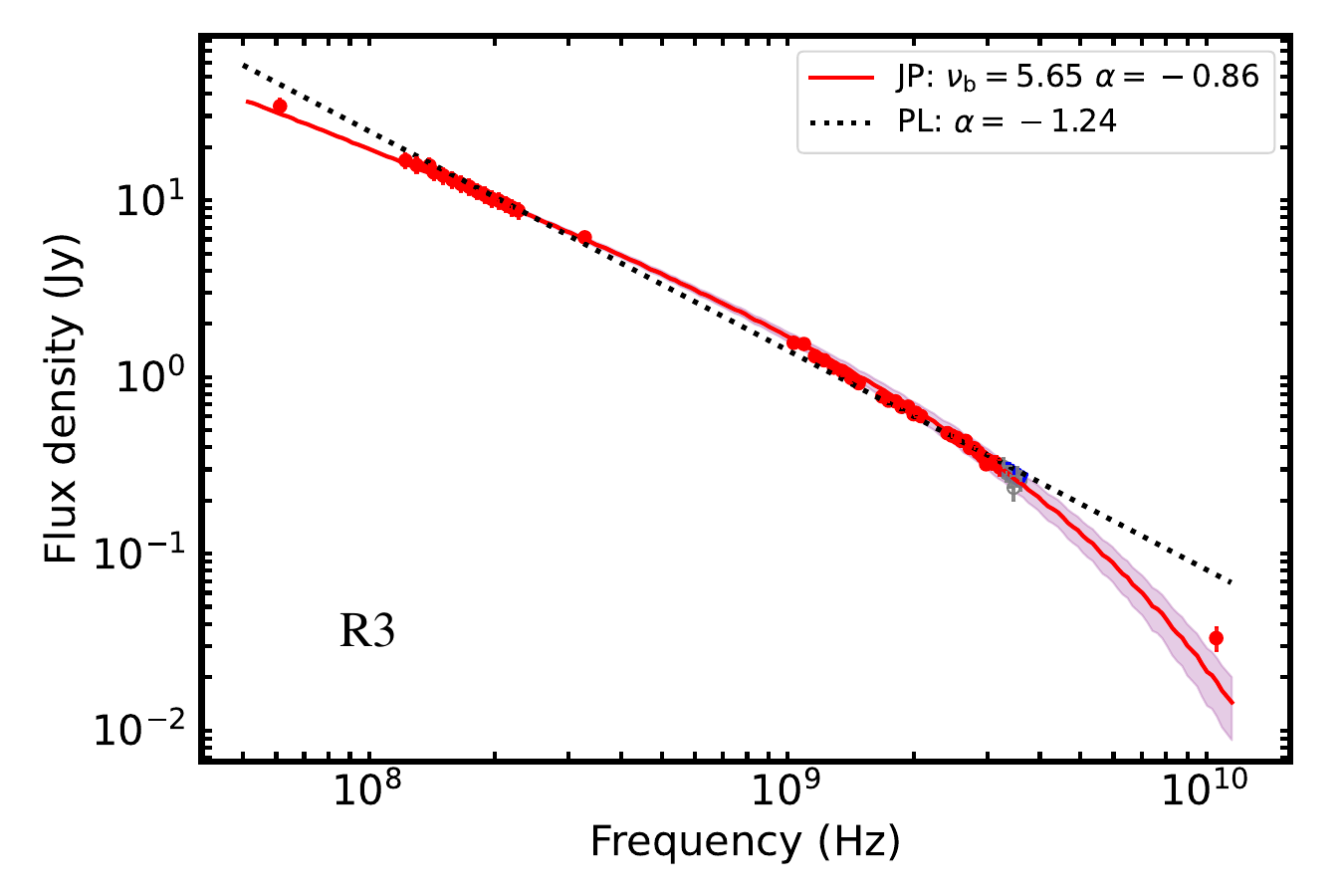}
    \caption{Upper left panel displays the three selected local regions (R1, R2, and R3) within the lobes’ diffuse region, where steeper spectral indices are observed (see also lower panel of Figure \ref{fig:spec}). Other panels present the JP model with fixed $\alpha$ for the spectra of these regions, as labeled at the lower left corner (i.e., R1, R2 and R3). The meanings of the lines are labelled. The stripes present the corresponding 2$\sigma$ errors for the JP model. The meanings of the data points are the same as the right panel of Figure \ref{fig:core_total_spec}}, but the blue triangles represent the derived from the mean spectral index of the corresponding regions. $\nu_{\rm b}$ is in GHz.
    \label{fig:reg_spec_fit_3} 
\end{figure*}

In total, we generated 54 low-resolution (100\arcsec) images spanning various frequencies observed by LOFAR, MWA, VLA, and Effelsberg. However, the six highest-frequency images at S band must be excluded,\footnote{M\,87 has a maximum angular scale of approximately 15\arcmin, corresponding to a baseline length of roughly $\sim0.23\,k\lambda$ in the \textit{uv}-plane. The lobes, with a size of $\sim9$\arcmin, correspond to a baseline length of roughly $\sim0.38\,k\lambda$. Observations with a shortest baseline length exceeding $\sim0.38\,k\lambda$ might fail to fully capture the lobe structures of M\,87. As shown in Table \ref{Tab:Obs_inf}, the shortest baseline length at higher S band frequencies is around $0.38\,k\lambda$, suggesting that these observations may miss emission from the larger-scale lobes.} as their total flux density exhibits a sudden drop before flux scaling (or an anomalously high core flux density after flux scaling, as shown in Figure \ref{fig:core_total_spec}). This issue arises from insufficient shortest baselines, which fail to recover the emission from the extended lobes. Ultimately, we retain 48 effective images, which we use to construct well-sampled wideband ($\rm 60\,MHz-10.55\,GHz$) spectra for the core region and the lobes' diffuse regions (as illustrated in Figure \ref{fig:region} and \ref{fig:reg_spec_fit_3}).

The left panel of Figure \ref{fig:core_total_spec} presents the constructed spectra for the core region of M\,87. Fitting a power-law model to the effective data points yields a spectral index of $\alpha=-0.60\pm0.01$ for the core region, consistent with \cite{Gas12}.

\begin{table*}[htp!]
\renewcommand{\thetable}{\arabic{table}}
\begin{centering}
\small
\caption{Basic properties of the three local regions.} \label{Tab:Im_measure_R123}
\setlength{\tabcolsep}{2pt}
\begin{tabular}{c|c|cccccc|ccc}
\hline
\hline
Component &$N_{\rm B}$ & \multicolumn{6}{c|}{Flux} &$\bar{\alpha}$& \,$\bar{B}_{\rm e}$& $\bar{P}_{\rm min}$    \\ 
  & & 122 &150 & 227 & 1033 & 1741 &2593& & &\\
(1) & (2) & (3) & (4) & (5) & (6) & (7) & (8) & (9) & (10) & (11)\\
\hline
R1 &1    &$7.2\pm0.7$&$6.1\pm0.6$&$3.9\pm0.4$&$1.0\pm0.07$&$0.5\pm0.04$&$0.3\pm0.03$&$-0.97\pm0.19$ & $10\pm0.4$ & $8\pm0.6$\\
R2 &1   &$9.0\pm0.9$&$7.1\pm0.7$&$4.9\pm0.5$&$1.2\pm0.08$&$0.6\pm0.05$&$0.3\pm0.03$&$-0.87\pm0.18$ & $11\pm1$ & $10\pm2$\\
R3 &1   &$16.8\pm1.7$&$13.8\pm1.4$&$8.7\pm0.9$&$1.6\pm0.09$&$0.7\pm0.05$&$0.4\pm0.03$&$-1.11\pm0.15$ & $12\pm0.5$ & $12\pm1$\\
\hline
\end{tabular}
\end{centering}

{\bf Notes.} Each column has the same meaning as in Table \ref{Tab:Im_measure}, but the first column presents three local regions in lobes' diffuse region (see Figure \ref{fig:reg_spec_fit_3}).\\
\end{table*}

\begin{table}[htp!]
\renewcommand{\thetable}{\arabic{table}}
\begin{centering}
\tiny
\caption{Spectral modeling results.} \label{Tab:spectral_model}
\setlength{\tabcolsep}{2pt}
\begin{tabular}{c|ccc|cc}
\hline
\hline
Component & \multicolumn{3}{c|}{CI or JP} &  \multicolumn{2}{c}{Power-law}\\
& $\chi^{2}$ & $\alpha$ & $\nu_{\rm b}$& $\chi^{2}$ & $\alpha$   \\ 
(1) & (2) & (3) & (4)& (5) & (6)  \\
\hline
Diffuse region & 1.26 &$-0.86\pm0.01$&$1.72\pm0.28$& 4.81 & $-1.14\pm0.01$ \\ 
R1 & 0.51 &$-0.86$&$12.7\pm2.3$& 3.58 & $-1.08\pm0.01$ \\
R2 & 0.62 &$-0.86$&$12.1\pm1.9$& 3.08 & $-1.08\pm0.01$ \\
R3 & 0.71 &$-0.86$&$5.7\pm0.9$& 2.78 & $-1.24\pm0.01$ \\
\hline
\end{tabular}
\end{centering}

{\bf Notes.} The assumed errors in the flux densities may overestimate the real random, thus resulting in artificially low $\chi^{2}$ values. $\nu_{\rm b}$ is in GHz. \\

\end{table}

Next, we analyze the spectra of the lobes' diffuse region, considering three models: a phenomenological power-law model, the interrupted or uninterrupted `continuous' injection (CI) model \citep{Pac70, Kom94}, and the `impulsive' injection (\citealt{Jaf73}, JP) model. It's worth to point out that the interrupted CI model was ruled out in \cite{Gas12}. The interrupted CI model also failed to explain the uniform spectral distribution in the lobes and the flatter spectra observed in the southwestern and northern lobe regions. Furthermore, the synchrotron aging time predicted by the interrupted CI model is significantly longer than the estimated dynamical time (see Section \ref{sec:dynamics}), making it an unreasonable interpretation.

The power-law modeling is performed in logarithmic space using Python's \texttt{curve\_fit} package. The JP model accounts for spectral aging due to synchrotron and inverse Compton losses, based on the assumption that the synchrotron-emitting electron population has a constant age. In this model, the pitch angles of these electrons are considered isotropic on timescales shorter than the radiative timescale, a reasonable assumption given that scattering and interactions with magnetic fields tend to isotropize pitch angle distributions. The CI model, in contrast, assumes that the central source continuously injects fresh relativistic particles. These high-energy particles age following the JP model's description. The CI model is particularly applicable in regions where injected particles remain trapped. Both models are implemented in the \texttt{synchrofit}\footnote{synchrofit: \href{https://github.com/synchrofit/synchrofit}{https://github.com/synchrofit/synchrofit}} (cf. \citealt{Tur18a,Tur18b,Quici2022}) and include three free parameters: the injection index ($s$), the break frequency ($\nu_{\rm b}$) and the normalization factor. The energy distribution of injected relativistic particles follows a power-law form, in a form of $N(E)dE\propto E^{-s}dE$, where $s$ denotes the injection index. For synchrotron emission, the corresponding injection spectral index, $\alpha_{\rm inj}$, can be expressed as $\alpha_{\rm inj} = (s-1)/2$. The break frequency ($\nu_{\rm b}$) marks the point where the spectrum steepens due to energy loss mechanisms.

Firstly, our analysis focuses on the spectra of the lobes' diffuse region, as shown in Figure \ref{fig:region}. The power-law modeling shown in the right panel of Figure \ref{fig:core_total_spec}, yields a spectral index of $\alpha=-1.14\pm0.01$. However, the model overestimates flux at both high and low frequencies and results in a higher $\chi^{2}$ value compared to the following CI model. This suggests spectral curvature at both ends (as discussed in Section \ref{sec:spec_im}), motivating the adoption of a physically motivated model---the CI model---that can trace the continuous injection process in the large-scale lobes’ diffuse regions.

The right panel of Figure \ref{fig:core_total_spec} presents the CI modeling results, while Table \ref{Tab:spectral_model} lists the corresponding best-fit parameters. We obtain $\nu_{\rm b}=1.72\pm0.28\rm\,GHz$ and $\alpha_{\rm inj}=-0.86\pm0.01$ (with $2\sigma$ errors), consistent with the CI model results of \cite{Gas12}, which modeled the `halo' region excluding the central cocoon and flows. Compared to their study, our analysis provides a slightly higher and more tightly constrained estimate of $\nu_{\rm b}$, likely due to the richer observational data available around GHz frequencies.

Secondly, we focus on the spectra of three local regions (i.e., R1, R2 and R3, see the upper left panel of Figure \ref{fig:reg_spec_fit_3}), which represent the steeper spectral regions within the lobes' diffuse region (also see their basic properties in Table \ref{Tab:Im_measure_R123}). Similar to the lobes' diffuse region, power-law modeling for these regions overestimates the flux at both high and low frequencies and yields a larger $\chi^{2}$ value compared to the following JP model. To better characterize these regions, we apply a JP model with a fixed injection index, assuming a constant-age synchrotron-emitting electron population. Using $\alpha_{\rm inj}=-0.86$ from our CI model, we fit the spectra with the JP model to account for spectral aging effects. The corresponding best-fit break frequency $\nu_{\rm b}$ is presented in Table \ref{Tab:spectral_model}. The break frequency $\nu_{\rm b}$ for these regions ranges from 
$5.7\pm0.9$ to $12.7\pm2.3$\,GHz, with an overall range of approximately $6-13$\,GHz. Compared to the results for the selected diffuse `halo' zones from \cite{Gas12}, where a fixed $\alpha_{\rm inj}=-0.85$ was used, our JP modeling yields slightly lower $\nu_{\rm b}$. These steeper spectral regions contain particles with longer cooling timescales, potentially tracing the longer evolution time of the lobes' diffuse region.

\subsection{Equipartition analysis and synchrotron ageing} \label{sec:equal}
We follow the methodology of \citet{Bec05} to perform a revised equipartition analysis, assuming that relativistic protons and electrons follow a power-law energy distribution with an injection spectral index of $\gamma_{0}=-2\alpha+1$. This analysis covers the entire M\,87 region, excluding areas where $\alpha \ga-0.5$. We derive the equipartition magnetic field strength ($B_{\rm eq}$), following the formulae \citep{Bec05}:
\begin{equation} \label{eq:eq_b}
    B_{\rm eq}=\lbrace\frac{4\pi(2\alpha-1)(K_{0}+1)I_{\nu}E^{1+2\alpha}_{\rm p}(\nu/2c_{1})^{-\alpha}}{(2\alpha+1)c_{2}(\alpha)flc_{4}}\rbrace^{\frac{1}{3-\alpha}},
\end{equation}
$E_{\rm p}$ is the proton rest energy. $I_{\nu}$ is the synchrotron intensity at frequency $\nu$, we utilize the $\rm 227\,MHz$ intensity image at a resolution of 58\arcsec, along with the spectral index ($\alpha$) image shown in the upper panel of Figure \ref{fig:spec}. We adopt a constant proton-to-electron number density ratio of $K_{0}=100$, assuming that relativistic protons and electrons have the same total number above an energy threshold of $E_{1}\simeq10\rm\,keV$. The depth through the emitting medium, $l$, is set to $\sim46\rm\,kpc$, corresponding to the lobe width, with a volume filling factor of $f=1.0$. $c_{1}$, $c_{2}$ and $c_{4}$ are constants. $c_{2}$ and $c_{4}$ depend on $\alpha$ and the magnetic ﬁeld’s inclination, respectively (see \citealt{Bec05} for details).


Based on the relationship between minimum-energy and equipartition field strengths, $B_{\rm min}/B_{\rm eq}=(\frac{1-\alpha}{2})^{1/(3-\alpha)}$, and the ratio of minimum-energy magnetic and particle energy densities, i.e, $\epsilon_{B_{\rm min}}/\epsilon_{\rm CR}=(1-\alpha)/2$ (see \citealt{Bec05}), we also derive the minimum total energy density---equivalent to the minimum pressure---as $P_{\rm min}=\epsilon_{B_{\rm min}}+\epsilon_{\rm CR}$. 

\begin{figure*}
    \centering
    \includegraphics[width=0.45\textwidth]{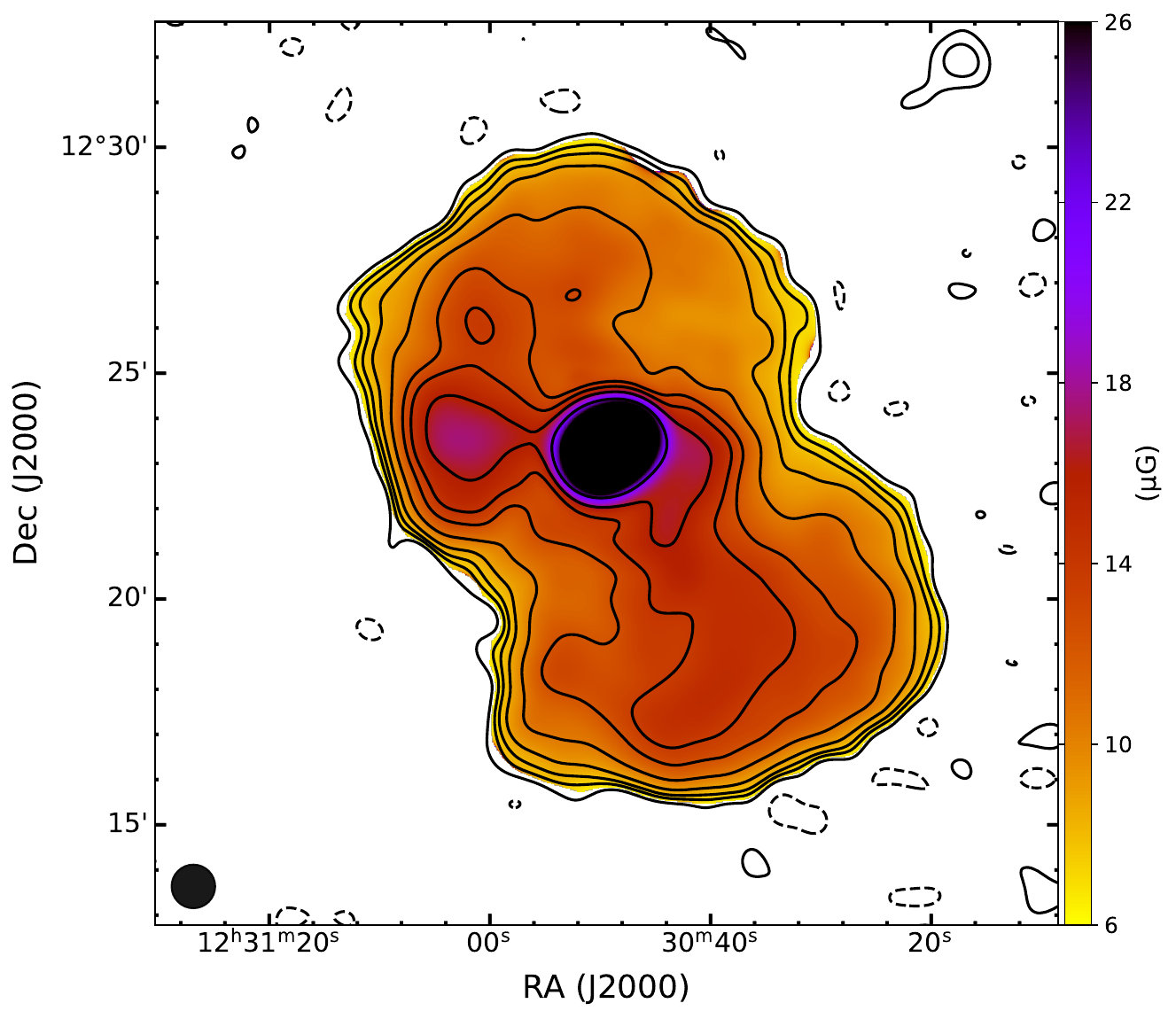}
    \hspace{1cm}
    \includegraphics[width=0.45\textwidth]{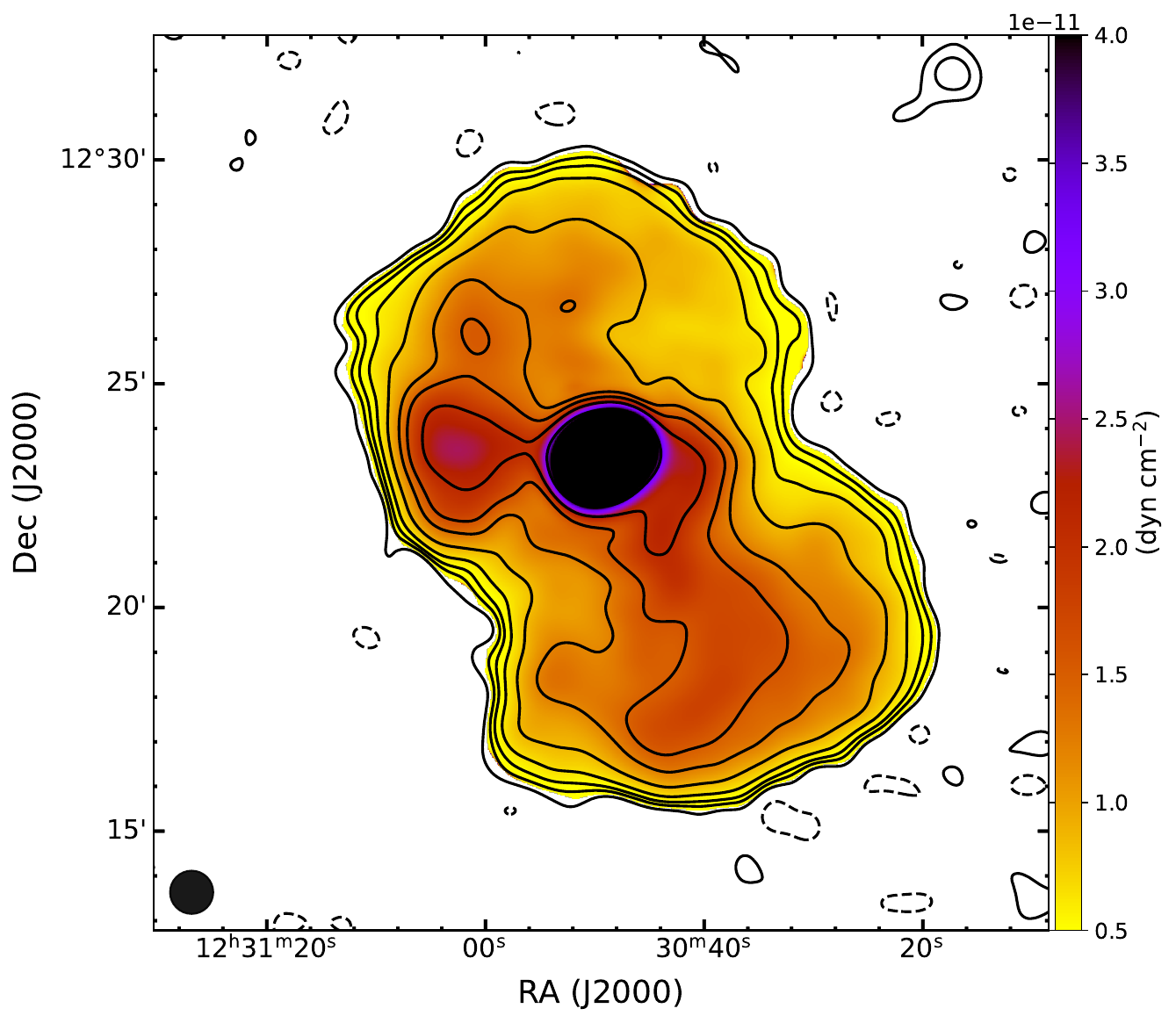}
    \caption{The left panel shows the distribution of the equipartition magnetic field strength ($B_{\rm eq}$), while the right panel presents the minimum pressure ($P_{\rm min}$) in M\,87. These calculations are based on an equipartition analysis. The contours are the same as Figure \ref{fig:spec}.}
    \label{fig:eq_cal} 
\end{figure*}

Figure \ref{fig:eq_cal} presents the results of our equipartition analysis, showing the spatial distribution of $B_{\rm eq}$ and $P_{\rm min}$. These images reveal a consistent trend: regions with greater intensity typically exhibit higher values of both $B_{\rm eq}$ and $P_{\rm min}$. The core region exhibit $B_{\rm eq}$ values larger than $20\rm\,\mu G$, reaching up to about $60\rm\,\mu G$. Within the lobes, the majority $B_{\rm eq}$ values ranging from 7 to $16\rm\,\mu G$, while $P_{\rm min}$ varies between $0.5\times10^{-11}$ and $2.0\times10^{-11}\rm\,dyn\,cm^{-2}$. The averaged $B_{\rm eq}$ and $P_{\rm min}$ values in the lobes' diffuse region, listed in Table \ref{Tab:Im_measure}, are $B_{\rm eq}\approx10\pm2\,\rm \mu G$ and $P_{\rm min}\approx(9\pm3)\times10^{-12}$ dyn\,cm$^{-2}$, which we take as representative for the lobes' diffuse components. The similar averaged $B_{\rm eq}$ and $P_{\rm min}$ values for R1, R2 and R3 regions are also listed in Table \ref{Tab:Im_measure_R123}. Our calculated $B_{\rm eq}$ is consistent with the results of \citet{Gas12} under the assumption of $k=1$ and $\gamma_{\rm min} = 100$, where $k$ denotes the energy ratio between protons and electrons.

We estimate the synchrotron ageing (or lifetime) $t_{\rm syn}$ using the break frequency and the (equipartition) magnetic field strength, following the formulae (e.g., \citealt{Mur11}): 
\begin{equation} \label{eq:eq_cal}
    t_{\rm syn}=1590\,{\rm Myr}\,\frac{B^{0.5}}{(B^{2}+B_{\rm CMB}^{2})\left[(1+z)\nu_{\rm b}\right]^{0.5}},
\end{equation}
where the magnetic field strength, $B$, is in $\mu$G, and the break frequency $\nu_{\rm b}$ is in $\rm GHz$. The inverse Compton equivalent magnetic field strength, $B_{\rm CMB}$, is given by $B_{\rm CMB}=3.25(1+z)^{2}\mu$G, which corresponds to the energy density of the cosmic microwave background (\citealt{Sle01}).

From the CI modeling result, we estimate the typical lifetime of the lobes' diffuse region to be $35^{+15}_{-10}\,\rm Myr$. This suggests that a continuous injection of relativistic particles for at least $\sim35\,\rm Myr$ is required to form the lobes’ diffuse region. Our improved spectral coverage at L and S bands corroborates the estimate of $t_{\rm syn}\sim 40 \rm\,Myr$ by \citet{Gas12}.
  
Since the lobes’ diffuse region contains particles of varying ages, we also analyze three local regions (R1, R2, and R3) that trace the older particles within the lobes. From our JP modeling results, we estimate the lifetime of these regions to be $13^{+2}_{-2}\,\rm Myr$ for R1, $11^{+3}_{-2}\,\rm Myr$ for R2 and $15^{+2}_{-2}\,\rm Myr$ for R3. Overall, our JP modeling suggests that the longest synchrotron lifetime of these regions, which contain the oldest particles, is approximately $15\,\rm Myr$.

Systemically, the CI modeling yields a longer typical lifetime compared to the longest lifetime derived from the JP model. This discrepancy arises because the JP model assumes a electron population with a constant age, while the distribution of particle ages within even small regions of the lobe. As a result, the JP model is known to underestimate the dynamical age of active radio lobes by a factor of $2-3$ \citep[e.g.,][]{Tur18c,Mah19}. Applying this correction, we estimate the true source age (longest lifetime) to be in the range of $30-45\rm\,Myr$ (or $25-50\rm\,Myr$, including uncertainties), is consistent with our CI modeling results. We caution that the synchrotron ageing depends on the estimated $B_{\rm eq}$. $B_{\rm eq}$ is determined based on several assumptions, as presented in Section \ref{sec:equal}, i.e., $B_{\rm eq}\propto((K_{0}+1)/fl)^{1/(3-\alpha)}$ (\citealt{Bec05}). Since the diffuse lobe regions have $\alpha\leq-0.8$, $B_{\rm eq}$ weakly depends on the adopted parameters of $K_{0}$, $f$ and $l$. For instance, an order-of-magnitude increase (decrease) in $K_{0}$ would result in an $\sim80\%$ ($\sim60\%$) increase (decrease) in $B_{\rm eq}$. According to Equation \ref{eq:eq_cal}, increasing (decreasing) the current $B_{\rm eq}$ of $\sim10\,\mu$G by a factor of 2 would result in a decrease (an increase) of the $t_{\rm syn}$ by a factor of $\sim2.6$ ($\sim2.2$).

\begin{figure*}
    \centering
    \hspace{1.6cm}\includegraphics[width=0.70\textwidth]{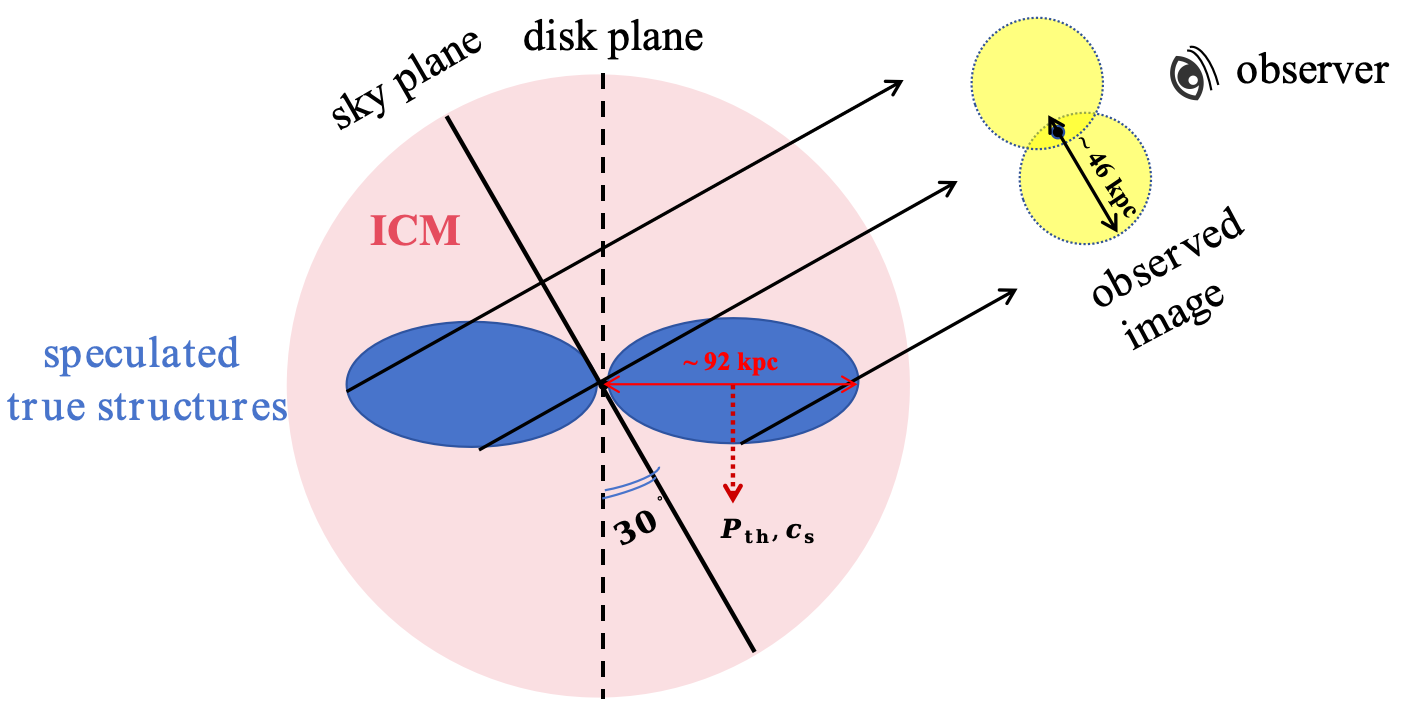}
    
    \caption{The projection effect for the two large-scale ellipsoidal lobes. The blue regions present the speculated true structures, i.e., the de-projected ellipsoidal lobes, and the yellow regions present the observed image, i.e., the projected lobes as seen with a projection angle about $30\degr$ \citep{Wer10}. The pink regions present the the isotropic ICM surrounding the ellipsoidal lobes. We also mark the position (i.e., the de-projected lobes' center) to derive the $P_{\rm th}$ and $c_{\rm s}$ (see Section \ref{sec:power} and \ref{sec:dynamics}).}
    \label{fig:projection}
\end{figure*}

\subsection{Energy and power}\label{sec:power}
In this section, based on our radio observation results, we estimate the total energy and outflow power to produce the lobes' diffuse components and the whole source, respectively. 

Before our estimation, it is important to determine the line-of-sight viewing angle of the two lobes. The viewing angle of the relativistic jet at pc scales is constrained to be $\sim17-18\degr$ \citep{Mertens16,Kim23}. Based on deprojected profiles of temperature and electron density, along with emission measures, \cite{Wer10} suggested that the large-scale `radio arms' are oriented at $\sim15–30\degr$ to our line of sight, implying a larger viewing angle for the more extended structures. \citet{Wer10} further constrained the viewing angle of the large-scale lobes to be $\theta_{\rm lobe}<35\degr$. We hereby adopt $\theta_{\rm lobe}=30\degr$ for the two large-scale lobes of M87, and approximate their intrinsic geometry as ellipsoids in Figure \ref{fig:projection} accounting for projection effects.

Using the typical $P_{\rm min}$ value of $9\times10^{-12}$ dyn\,cm$^{-2}$, we calculate the total energy of the lobes' diffuse components as $E_{\rm min}={\gamma\over(\gamma-1)}P_{\rm min}V\simeq 2.1\times10^{59}\,\rm erg$, where $\gamma=4/3$ is the adiabatic index for relativistic particles, and the volume for two lobes is given by $V=\pi d^3/(3\rm\,sin\theta_{\rm lobe})$. This provides a lower limit constraint for the total energy within the lobes' diffuse components.

Moreover, we calculate the total energy of the whole source using gas thermal pressure. This method is widely adopted to derive the total energy of X-ray cavities (e.g., \citealt{All06}). Based on X-ray observations, \cite{Mat02} derived the temperature and electron density profiles for the spherically symmetric hot (thermal) gas in M\,87. At a distance of $\sim9^\prime$ (i.e., $L_{\rm th}\sim46\rm\,kpc$) from the SMBH to the deprojected center of the lobes, the gas thermal pressure is $P_{\rm th}=n_{\rm e}kT/\mu_{\rm e}\approx4.4\times10^{-11}$dyn\,cm$^{-2}$, where $\mu_{\rm e}\simeq0.517$ (see \citealt{Pfr13}). This value is approximately five times larger than the typical minimum pressure in the lobes' diffuse region. By adopting this thermal pressure, the total energy or entropy of the whole source can be estimated as $E_{\rm th}={\gamma\over(\gamma-1)}P_{\rm th}V$. For an internally relativistic plasma ($\gamma=4/3$), we obtain$E_{\rm th}\simeq 10.5\times10^{59}\,\rm erg$, whereas for an internally non-relativistic plasma ($\gamma=5/3$), we obtain $E_{\rm th}\simeq 6.6\times10^{59}\,\rm erg$.

Assuming that the radio spectrum typically extends from 10\,MHz to 100\,GHz \citep{Pac70,Gas12}, we proceed with the following estimates. First, for the lobes' diffuse components, using the CI modeling result for the lobes' diffuse region (see the right panel of Figure \ref{fig:core_total_spec}) and scaling it by the size ratio of the lobes’ diffuse components to the lobes’ diffuse region (87.5/64.0), we estimate the total radio luminosity of the lobes' diffuse components as $\sim1.7\times10^{41}\,\ergs$. Second, for the whole source, based on power-law modeling with a spectral index of 
$\alpha\sim-0.80$, we estimate the total radio luminosity of the whole source as $\sim1.0\times10^{42}\,\ergs$. Assuming that 1\% of the outflow power is converted into radio emission \citep{Odea85,Kha06}, the corresponding outflow power estimates are $\sim1.7\times10^{43}\,\ergs$ for the lobes' diffuse components and $\sim1.0\times10^{44}\,\ergs$ for the whole source.

Alternatively, assuming the outflow is driven by the kinetic jet, we estimate the jet power based on the $P_{\rm kin}$–$L_{\rm151\,MHz}$ relation \citep{God13,Fan18}, where the radio luminosity at $151\rm\,MHz$ can be derived from the radio flux at $150\rm\,MHz$. For the lobes’ diffuse components, having a flux of 480\,Jy at 150\,MHz, the outflow (or jet) power is estimated to be $\sim8.1\times10^{43}\,\ergs$. For the whole source, having a flux of 1223\,Jy at 150\,MHz, the outflow (or jet) power is estimated to be $P_{\rm out}\sim1.5\times10^{44}\,\ergs$.

In summary, we estimate the minimum total energy of $E_{\rm min}\simeq 2.1\times10^{59}\,\rm erg$ for the lobes' diffuse components and the total energy of $E_{\rm th}\simeq(6.6-10.5)\times10^{59}\,\rm erg$ for the whole source. While our outflow power constraints are outlined in Table \ref{Tab:Contraints}.

\begin{table*}
\renewcommand{\thetable}{\arabic{table}}
\centering
\small
\caption{Power constraints.} \label{Tab:Contraints}
\begin{tabular}{c|c|c|c|c}
\hline
\hline
Region &Distance (kpc)&$P_{\rm out}$ ($10^{44} \rm \,erg\,s^{-1}$)& $p_{\rm out}$ ($10^{-5}$) &References \\
(1) & (2) & (3) & (4)&(5) \\
\hline\hline
Diffuse& $>20$ &0.17; 0.81; $1.3-2.2$ & $2-27$ &this work  \\
Total & $\sim20$ &1.0; 1.5; $4-11$ & $12-140$ & this work\\
Total & $\sim20$ &$6-10$  &$74-120$&1,2 \\
Inner & $\sim5$ &0.2 &2.5 &3,4 \\
Knot D& $\sim1$ &$>0.8$ &$>10$&2\\
Knot A& $\sim0.3$ &$>1-5$ &$>12-60$&2,5\\
HST-1 & $\sim0.07$ &$0.1-1$ &$1.2-12$&6 \\
SMBH  & 0          &$0.01-0.6$ &$0.1-7$&$7-14$\\
Bondi$^\star$ &$\sim0.12$  &$<5-10$ &$<60-120$&$15-16$ \\
\hline\hline
\end{tabular}

References: (1) -- \cite{Gas12}; (2) -- \cite{Ow00}; (3) -- \cite{Rey96}; (4) -- \cite{Pas21}; (5) -- \cite{Bic96}; (6) -- \cite{Sta06};  (7) --
\cite{EHTM21}; 
(8) -- \cite{EHT19}; (9) -- \cite{Yang24}; (10) -- \cite{EHT21}; (11) -- \cite{Kuo14}; (12) -- \cite{Feng16}; (13) -- \cite{Cruz22}; (14) -- \cite{Pri16};
(15) -- \cite{Di03}; (16) -- \cite{Rus15}

Notes.
The 1st column presents the structures/components in M\,87 that are used to constrain the outflow/jet power. Here `Diffuse' denotes the lobes' diffuse component, `Total' denotes the whole large-scale structures of M\,87 and `Inner' denotes the inner $5\rm\,kpc$ lobe. Others are the typical structures/components observed in M\,87. The 2nd column presents the the projected distance of the structures/components from the SMBH. The 3rd column presents the constrained outflow power (details see section \ref{sec:power}, \ref{sec:dynamics} and \ref{sec:AGN-driven}). The 4th column presents the dimensionless outflow power, defined as $p_{\rm out}=P_{\rm out}/L_{\rm Edd}$. The last column presents the references of the constraints. \\
$^\star$: The upper limit outflow power predicted from the Bondi accretion rate ($\dot{M}_{\rm B}$), i.e., about 10\% of $\dot{M}_{\rm B}c^{2}$. 
\end{table*}

\section{Discussions}\label{sec:discussions}

\subsection{Dynamics and timescales} \label{sec:dynamics}

Following the standard AGN picture, we assume that the two giant radio lobes of M\,87 are formed by outflows from the core/nuclear region (the outflow origin awaits to be specified, see Section \ref{sec:mechanisms}) and interacts with the ICM. As these lobes evolve, they may reach a pressure-balance stage with their surrounding ICM. Within this framework, we derive two key dynamical timescales.

The first is the lobe expansion time. We assume that the expanded velocity of lobes is on the order of the local sound speed. For simplicity, we assume that the dynamics of the lobes are determined by the expansion of thermal gas. The sound speed is given by $c_s = (\gamma kT/\mu m_{\rm H})^{1/2}$ (here $\gamma=5/3$ and $\mu=0.588$, \citealt{Bir04,Pfr13}). The temperature, $T$ is taken from the temperature profile in \cite{Mat02} at a specific distance $\sim9^\prime$ (i.e., $L_{\rm th}\sim46\rm\,kpc$) from the SMBH to the de-projected lobes' center. This yields a sound speed of $c_{s}\sim 830\,\rm km\,s^{-1}$. The corresponding sound crossing time for the lobes is $t_{s}=L_{\rm th}/c_{s}\sim 54\,\rm Myr$ after accounting for projection effects. Notably, this timescale is consistent with the longest lifetime derived from our JP modeling with a fixed  $\alpha_{\rm inj}$ (from our CI model). This confirms that continuous outflow injection over a timescale of $t_{s}$ is likely responsible for forming the large-scale lobes and supports the adopted viewing angle.


The second is the excavation time, which represents the duration required to carve out the lobes’ diffuse components and the whole source, given a time-averaged outflow power. It is calculated using the equation
\begin{equation}
         t_{\rm exc}=\frac{E_{\rm min;th}}{P_{\rm out}}
\end{equation}
Using our earlier energy estimates, we estimate $t_{\rm exc}\simeq 66/P_{44}\,\rm Myr$ with $\gamma=4/3$ for the lobes' diffuse components and $t_{\rm exc}\simeq (207-332)/P_{44}\,\rm Myr$ with $\gamma \in (4/3,5/3)$ for the whole source. The excavation time would be longer if the lobes contain significant transonic flows or dynamically strong magnetic fields.

By equating the estimated lifetime ($\sim30-50$\,Myr) of M\,87 to the excavation time we obtain outflow power estimates. For the lobes' diffuse components (take $E_{\rm min}$ as the total energy), we obtain outflow power of $P_{\rm out}\sim(1.3-2.2)\times10^{44}\,\ergs$. For the whole source (take $E_{\rm th}$ as the total energy), we obtain outflow power of $P_{\rm out}\sim(4-11)\times10^{44}\,\ergs$. We note that, if a viewing angle of $20\degr$ (or $45\degr$) is adopted, both the estimated $t_{\rm s}$ and $E_{\rm min;th}$ would increase by $\sim50\%$ (or decrease by $\sim30\%$). Correspondingly, the estimated outflow power based on the synchrotron ageing timescale would increase by $\sim50\%$ (or decreases by $\sim30\%$). However, these variations do not affect our main conclusions.

Combining these results with our findings in Section \ref{sec:power}, we conclude that, the lobes' diffuse components of M\,87 are formed by a continuously injected outflow of power of $P_{\rm out}\sim(0.2-2)\times10^{44}\,\ergs$, while the whole source is formed by continuously injected outflows of power of $P_{\rm out}\sim(1-11)\times10^{44}\,\ergs$. In the following section, we will discuss the possible mechanisms responsible for driving outflows, including galactic winds, AGN jets, and/or AGN winds.

\subsection{Origin of the giant diffuse radio lobes in M\,87} \label{sec:mechanisms}

Radio galaxies usually present two large-scale lobes extending from kpc to Mpc (e.g., \citealt{Dre84,Car91,Colbert96,Ro96,Fabian00,McNamara00,Ow00,Kha06,Seb20,Oei24}). The dominant driver of these outflows is believed to be the AGN jet \citep{Lon73,Sch74,Bl74,Bl19}, while material entrainment from AGN winds and the interstellar medium (ISM) can contribute to decelerating the outflow \citep{Bow96,Wyk15,Har20,Bl22}. For instance, \cite{Gas12} traced the bright structures within the lobes of M\,87 and concluded that a continuous injection from the jet, undergoing adiabatic expansion, could explain the large-scale radio structures observed in the lobes. Recently, detailed analysis of EHT and VLBA observations suggests that, there requires an AGN wind associated with the accretion flow near the SMBH of M\,87 \citep{Park19,Bl22,Yuan22,Lu23}. The AGN wind in low-luminosity AGN is also favored by numerical simulations of accretion flows \citep{Yuan12a, Yuan15,Yang21}, whose impact on galaxy scales is now better understood (e.g., \citealt{Weinberger17,Yuan18}). These winds can propogate to up to $\sim 10-100$ kpc from SMBH (\citealt{Zhu23,Zhang25}). Galactic wind from stellar activities must play a role (though maybe negligible) in influencing the large-scale surroundings in M\,87. Here, through the analysis of energy, power, and timescales, we explore the possible mechanisms responsible for the continuous outflows that form the lobes, including galactic winds, AGN jets, and AGN winds.

\subsubsection{Scenario 1: Galactic wind-driven lobes}

Assuming that the lobes of M\,87 were formed through galactic winds energized by supernovae explosions, we can estimate the required energy budget. If we assume the minimum total energy $E_{\rm min}$ of the lobes' diffuse components comes from supernovae, and considering that a typical supernova releases $10^{51} \rm\, \ergs$, we estimate the necessary supernova explosion rate to be approximately $4-7\,\rm yr^{-1}$ over a timescale of $\sim30-50\rm\,Myr$. By adopting a Salpeter initial mass function (IMF) over a mass range $\rm 0.1-100\,\msun$, we derive a corresponding star formation rate (SFR) of $\rm SFR(M\ga0.1\,\msun)\sim(580-960)\,\msun\,{\rm yr}^{-1}$, based on eq. 28 in \cite{Con02} and eq. 20 in \cite{Con92}. Similarly, if we take the total thermal energy $E_{\rm th}$ of the lobes as the energy constraint, we derive a SFR of $\rm SFR(M\ga0.1\,\msun)\sim 1800-4700\,\msun\,{\rm yr}^{-1}$. These values represent the average star formation rate over the past $\sim30-50\rm\,Myr$. 

Observations indicate that the SFR in elliptical galaxies is typically low \citep{Kav14,Sed21}. In contrast, starburst galaxies generally exhibit SFR of around $10\,\rm\msun\,yr^{-1}$ \citep{Ren15}, with extreme cases of massive starburst galaxies reaching $500-1000\,\rm \msun\,y^{-1}$ \citep{McD12,McD13}. However, for M87, far-infrared observations place an upper limit on the SFR at $< 0.081\,\rm \msun\,yr^{-1}$ \citep{Raf06}. Additionally, $H_{\alpha}$ emission analysis by \citet{Tan05} suggests an even lower SFR of approximately $9.5\times10^{-3}\,\rm \msun\,yr^{-1}$.

These observational values are orders of magnitude lower than the estimated SFR required to power the lobes via starburst winds. If the average SFR over the past $\sim30-50\rm\,Myr$ was indeed as high as $\sim 600-5000\,\rm \msun\,yr^{-1}$, it would necessitate an extreme quenching process---reducing star formation by more than five orders of magnitude within a short period. Such a dramatic shift appears highly improbable.

Thus, the starburst-driven lobe scenario is disfavored.

\subsubsection{Scenario 2: AGN-driven lobes}
\label{sec:AGN-driven}
We now consider the AGN-driven lobes scenario. Through radiation, mildly-relativistic wind/outflow, and well-collimated jet, an AGN can significantly influence the dynamical properties of its host galaxy in terms of both energy and momentum transfer \citep{Fabian2012}. However, since the bolometric luminosity of M\,87 is relatively low---approximately $8.5\times 10^{41}\,\ergs$ \citep{EHTM21,Xie23}---we can reasonably omit the impact of radiation.

As discussed in Sec. \ref{sec:power}, the power required to produce the lobes' diffuse components of M\,87 is in the range of $\sim (0.2 - 2)\times 10^{44}\,\ergs$, while the power required to produce the whole radio structures is in the range of $\sim (1 - 11)\times 10^{44}\,\ergs$.

{\bf Energy supply comes from the AGN jet}. M\,87 is most prominent in its large-scale jet. Table \ref{Tab:Contraints} outlines various estimates of the jet power from the literature, covering a broad range of distances from the central black hole and, consequently, different timescales. For a detailed discussion of the limitation process, we refer to the corresponding references in Table \ref{Tab:Contraints}. These estimates, which include constraints from regions near SMBH out to scales of tens-of-kpc (e.g., Bondi radius, jet base, HST-1, the 5 kpc inner lobe, and the whole large-scale lobes), suggest that the jet power falls within the range of $0.1\times10^{44}$ to $10\times10^{44}\,\rm erg\,s^{-1}$, with a median value of approximately $10^{44}\,\rm erg\,s^{-1}$. Therefore, the jet power is sufficient to drive the outflow and generate large-scale structures across the whole lobes, regardless of the lobes' diffuse components.

Additionally, the mass accretion rate near the SMBH ($\dot{M}_{\rm BH}$) of M\,87 
has been estimated using multiple methods, including modeling EHT observations and inner jet morphology within the MAD framework, spectral energy density (SED) modeling, and rotation measure (RM) measurements. These approaches yield an estimated mass accretion rate of $\sim(0.2-2) \times 10^{-3} \,\rm \msun\,yr^{-1}$ (see \citealt{Kuo14,Feng16,Yang21,EHT21,Yang24}). This corresponds to an upper-limit jet power of approximately $\sim10^{43-44} \rm \,erg\,s^{-1}$, assuming $P_{\rm j}=\dot{M}_{\rm BH}c^{2}$ (\citealt{Cruz22}). This is slightly lower than the required jet power, suggesting a varying accretion rate (and extraction energy from SMBH) or/and energy contribution from AGN wind \citep{Bl22}.

{\bf Energy supply comes from the AGN wind}. As pointed out by \citet[][see also \citealt{EHT19,EHT21}]{Yuan22a}, the accretion flow in M\,87 should be a hot magnetically arrested disk (MAD) with saturated magnetic field, the SMBH is expected to have a high spin. Both observations and numerical simulations show that the hot accretion flow can launch strong AGN winds (e.g, \citealt{Yuan12a,Wang13,Che16,Shi21,Shi25}), with most ($\geq98\%$) of the accretion mass being ejected (e.g., \citealt{Wu13,Feng16,Wu20,Bl22}). The AGN wind velocity $v_{\rm w}$ can reach up to $\sim0.2\,c$ (e.g., Figure 8 in \citealt{Yang21}), and they can have a impact on ICM at $\sim 10-100$ kpc away from SMBH (\citealt{Zhu23,Zhang25}). The AGN wind power can be estimated as $P_{\rm w}=0.5\dot{M}_{\rm w}v_{\rm w}^{2}\le2.3\times10^{44}\rm \,erg\, s^{-1}$, where $\dot{M}_{\rm w}$ is the mass loss rate in the form of AGN wind (crudely $\dot{M}_{\rm w}\sim \dot{M}_{\rm B}$, where $\dot{M}_{\rm B}$ is the Bondi accretion rate). $\dot{M}_{\rm B}$ is estimate to be in the range of $0.1-0.2 \,\msun\,{\rm yr}^{-1}$ \citep{Di03,Rus15}. Here, we assume the AGN wind power is dominated by the kinetic power, as the radiative output is much lower at low accretion rate of $\dot{M}< 10^{-5}\,\dot{M}_{\rm Edd}$ (\citealt{Xie12, Xie23}, here $\dot{M}_{\rm Edd}\equiv 10 L_{\rm Edd}/c^2 \approx 22.8\,(M_{\rm BH}/10^{9}\,\msun)\, \msun\,{\rm yr}^{-1}$). Based on this estimation, we find that the AGN wind could provide sufficient power to produce the lobes' diffuse components.

There are two key uncertainties in the above crude estimation, i.e., the AGN-wind mass loss rate and the AGN-wind velocity. Detailed simulations of the hot accretion flow \citep{Yuan12a,Yuan15}, or specifically the MAD model \citep{Yang21} suggest that, unless further accelerated, the terminal velocity of the AGN wind is about $50-70\%$ the Keplerian velocity where it is launched, i.e. $v_{\rm w} \approx 0.5-0.7c\, (R_{\rm launch}/R_{\rm s})^{-1/2}$. Here $R_{\rm s} = 2GM_{\rm BH}/c^2$ is the Schwarzschild radius of SMBH. AGN wind with $v_{\rm w}\sim 0.1-0.2c$ can only be launched within hundreds of $ R_{\rm s}$. 
If we consider a MAD from the Bondi radius down to the SMBH, then most of the AGN wind will be launched at a distance far away from SMBH with a much lower velocity. Assuming $\dot{M}_{\rm w}(R_{\rm launch}) = \dot{M}_{\rm BH}\,(R_{\rm launch}/R_{\rm s})^s$ with the wind strength $s\sim 0.5$ (cf. \citealt{Yuan12a,Yang21} and references therein) and the mass accretion rate near SMBH $\dot{M}_{\rm BH}\approx 10^{-5}\,\dot{M}_{\rm Edd}$ \citep{EHT21,Xie23}, the AGN-wind power can then be estimated as,
\begin{eqnarray}
  P_{\rm w} & \approx & 2\times10^{-5} L_{\rm Edd} \,(R_{\rm launch}/R_{\rm s})^{s-1}\nonumber\\
  &\approx & 1.5\times10^{43} {\rm \,erg\, s^{-1}}\,(R_{\rm launch}/R_{\rm s})^{\sim -0.5}.
\end{eqnarray}
If we take $R_{\rm launch}\approx 10^{2-4} R_{\rm s}$ as a representative value for the launching radius of wind \citep[e.g.,][]{Tom13,Yang21}, then a AGN-wind power of $P_{\rm w}\approx 10^{41-42}\,{\rm \,erg\, s^{-1}}$ can be safely derived.

The above estimation implies that the wind power from current AGN activity can supply only a few percent of the energy required to produce the diffuse components of the two giant lobes. However, when compared with the kinetic power of HST-1, \cite{Pri16} argued that the core power of M\,87 has been two to three orders of magnitude higher in the last 200 years. Additionally, there is compelling evidence that our Galactic center (Sgr A*) was much brighter (by a factor of $\sim 10^{6-7}$) over the past Myr (\citealt{Ponti13} and references therein). Numerical simulations of massive elliptical galaxies also reveal dramatic variations in SMBH activity over a period of $\sim 10^{4-5}$ yr (e.g., \citealt{Yuan18, Zhu23}), and AGN winds can impact ICM at $10-100$ kpc \citep{Zhu23}. An average enhancement (by a factor of $\sim 10^2$) of AGN activity in M\,87 over the past $\sim 30-50$ Myr remains plausible.

\begin{figure}
    \centering
    \includegraphics[width=0.45\textwidth]{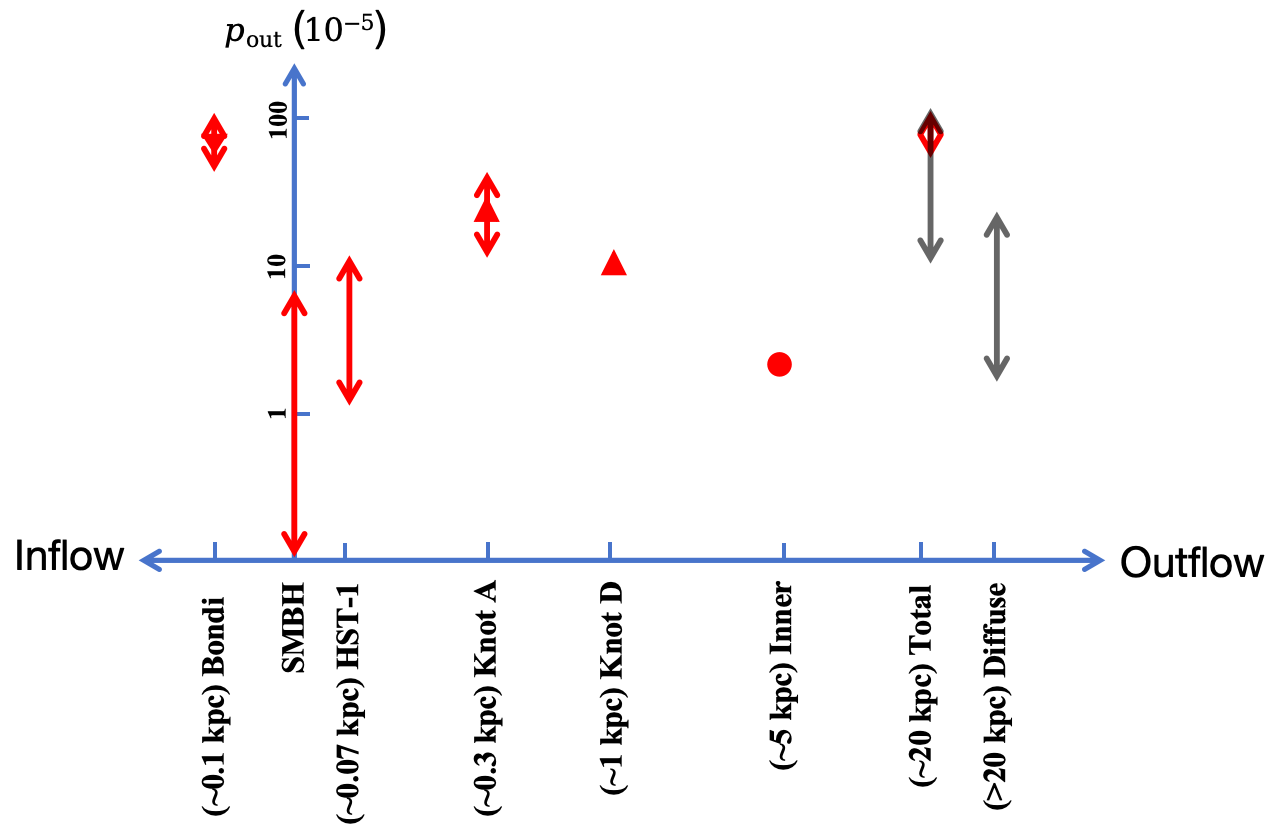}
    \caption{dimensionless outflow power ($p_{\rm out}$) versus (projected) distance from SMBH for some structures of M\,87. Structures presented by red and black symbols are from literature and our work respectively (see Table \ref{Tab:Contraints}). Symbols with upper and lower triangulars present the lower and upper limits respectively.}
    \label{fig:P_distance} 
\end{figure}

\subsection{Variation of AGN activities?}
We calculate the dimensionless outflow power $p_{\rm out}=P_{\rm out}/L_{\rm Edd}$, across various structures of M\,87 at different projected distance from the SMBH, as outlined in Table \ref{Tab:Contraints}. Figure \ref{fig:P_distance} presents the distance dependence of $p_{\rm out}$. The extended outflow structures in M\,87 preserve a imprint of historical AGN activity, whereas circumnuclear accretion flows (inflows) within the Bondi radius provide predictive diagnostics for imminent outflow variability. 

Adopting a fiducial energy coupling efficiency ($\eta$) for outflow production mechanisms---parameterized as $p_{\rm out}\simeq \eta \dot{M}_{\rm BH}/\dot{M}_{\rm Edd}$---the derived accretion rate variability implies two to three orders of magnitude amplitude modulations in AGN activity over $\sim30-50$\,Myr timescales, with at least one cycle of activity escalation and decline. The misalignment between the two lobes and the pc-scale relativistic jets also supports the temporal evolution between the relativistic jet and the large-scale lobes. Such large variability of AGN activity is confirmed by observations of restart or dying radio sources (e.g., \citealt{Cordey87,Ku17,Wu20,Mor21}), which is a natural outcome of AGN feedback as predicted by simulations (e.g., \citealt{Hopkins24,Guo24,Su25}, and reference therein).

\subsection{Future constraints}
The radio lobe structures of M\,87 spanning tens of $\rm kpc$. Combining both micro and macro observations, we can better understand the evolution, energy contributions, and connections between these different scales in M\,87.  

The EHT has probed the core of M\,87 down to the event horizon,, capturing emissions from the hot accretion flow. This data provides constraints on key properties of the accretion flow, yet significant uncertainties remain regarding the accretion rate and the relative contributions of the jet and wind components (see \citealt{EHT19,EHT21,Lu23}). Large-scale jet structures on pc to kpc scales further refine our understanding of the jet’s energy contribution (see \citealt{Rey96,Rus15,Bl19}). Future high-resolution, multi-wavelength observations, e.g., Square Kilometre Array (SKA), Next Generation Event Horizon Telescope (ngEHT), Thirty Meter Telescope (TMT), Very Large Telescope Interferometer (VLTI), NewAthena, will advance our understanding of jet properties at pc-kpc scales.

To further refine our understanding of the AGN wind's contribution, additional observations across optical, ultraviolet (UV) and X-ray wavelengths are essential. Given the high temperatures associated with the hot accretion flow, directly detecting the AGN wind remains challenging (see \citealt{Shi21}). Using UV absorption lines from the Hubble Space Telescope (HST), \cite{Sab03} suggested a link between M\,87’s nuclear outflow and its accretion processes, but more precise constraints require higher spectral resolution. Notably, no direct X-ray evidence of wind near M\,87’s center has been found. However, the AGN wind could serve as an external Faraday rotation measure (RM) screen for the jet \citep{Park19,Yuan22} or interact with the ambient medium, producing forbidden lines observable in UV and soft/hard X-ray wavelengths (see \citealt{Dev19,Shi21}). Future observations with improved spatial and spectral resolution and sensitivity, such as NewAthena, hold great promise for providing direct or indirect insights into the properties of AGN wind in M\,87.

\section{Conclusion}\label{sec:conclusion}
In this study, we utilize well-sampled wideband ($\rm 60\,MHz-10.55\,GHz$) observations from MWA and VLA, supplemented by data from LOFAR and Effelsberg, along with X-ray observations, to conduct a multi-wavelength analysis of the diffuse radio emission in the large-scale radio lobes of M\,87. Through detailed examinations of morphology, flux density, spectral characteristics, spectral analysis, equipartition conditions, dynamical assessments, and power and energy estimations, we establish connections between the micro and macro scales of M\,87. This comprehensive approach allows us to better understand the evolutionary processes and the role of outflow power in generating the lobes’ diffuse components. Key findings from our analysis include:
\begin{itemize}
\item We observe two diffuse and extended large-scale radio lobes ($\sim46\rm\,kpc$ in diameter) in M\,87, both exhibiting a nearly circular surface shape. The lobes display filamentary structures, sharp edges, and similar edges at different frequencies. These characteristics suggest that the lobes are in a turbulent state and are well confined, either by the pressure exerted by the intergalactic medium or by a strong magnetic field at the lobe periphery.

\item We observe the spectral indices distribution in two giant lobes are nearly uniform, moderately steep, narrowly ranged (mostly within $-1.2\leq\alpha\leq-0.8$), with no clear correlation to intensity. 

\item We construct well-sampled wideband spectra ranging from $\rm60\,MHz$ to $\rm 10.55\,GHz$, and conduct a detailed spectral analysis of the lobes' diffuse region. Using the the CI model, we find an injection spectral index of $\alpha_{\rm inj}\simeq -0.86$ and a spectral break frequency of $\nu_{\rm b}\simeq1.72 \rm\,GHz$. Additionally, We apply the JP model to three selected regions (in the lobes' diffuse region) that exhibit steeper spectral indices. When fixing $\alpha_{\rm inj}$ to the value derived from the CI model, we find that $\nu_{\rm b}$ varies from 6 to 13 GHz.

\item Using a revised equipartition analysis (see \citealt{Bec05}), we estimate a typical equipartition magnetic field strength of $\simeq10 \mu\rm G$ and a minimum pressure of $\simeq9\times10^{-12} \rm \,dyn\,cm^{-2}$ in the lobes' diffuse region.

\item Based on the estimated $\nu_{\rm b}$ and $B_{\rm eq}$, we calculate the typical synchrotron age in lobes' diffuse region. The CI model suggests a typical lifetime of $35^{+15}_{-10}\rm\,Myr$, while the longest lifetime among the three selected regions using the JP model is about 15 Myr. Given that the JP model may underestimate the lifetime by a factor of $\sim 2-3$ \citep[e.g.,][]{Tur18c,Mah19} on the lifetime estimation of the JP model, the actual source age (i.e., the longest lifetime) is likely $30-45$ Myr.

\item Utilizing the estimated minimum pressure and the thermal pressure, we calculate the total energy for the lobes' diffuse components to be $\sim2.1\times10^{59}\rm\,erg$, while the total energy for the whole source is $\sim(6.6-10.5)\times10^{59}\rm\,erg$. Additionally, using radio flux measurements and the $P_{\rm kin}$–$L_{\rm151\,MHz}$ relationship, we initially constraint the outflow power required to generate the lobes' diffuse components and the whole source, as outlined in Table \ref{Tab:Contraints}. 

\item We find that the sound crossing-time agrees with the longest lifetime
derived from the JP modeling with a fixed $\alpha_{\rm inj}$ from our CI model, supporting the picture that the lobes are formed by continuous injection. This also constrains the lifetime of the lobes to about $30-50$ Myr. Assuming the excavation time equals the lobe lifetime, we estimate an outflow power of $\sim(1.3-2.2)\times10^{44}\,\ergs$ for the lobes' diffuse components, and $\sim(4-11)\times10^{44}\,\ergs$ for the whole radio structures.

\item Based on the above estimated power, we explore the possible mechanisms driving the outflow. We find that galactic wind driven by stellar process has a negligible effect. The jet can provide sufficient energy to power the whole source. We find that the wind driven by current AGN activity cannot power the two lobes. But we caution that, an average enhancement of AGN activity by a factor of $\sim 10^{2}$ (relative to the current epoch) over the past $\sim 30-50$ Myr remains plausible.

\end{itemize}

The diffuse radio emission in the lobes of M\,87 originates from outflows primarily driven by AGN feedback, with an uncertain energy composition---dominated by a strong AGN jet but accompanied by a weaker AGN wind---rather than by stellar feedback. To better quantify these contributions, future multi-messenger observations with improved spatial and spectral resolution and sensitivity will be crucial for further constraining AGN jet and wind properties in M\,87.

\section*{Acknowledgements}

We appreciate the referee for helpful suggestions. We thank Dr. Frazer Owen for providing the $325\rm\,MHz$ VLA image and the L band VLA data, Dr. Francesco de Gasperin for the LOFAR images at $60\rm\,MHz$ and $140\rm\,MHz$, and the VLA S band data, Dr. Helge Rottman for the Effelsberg image at $10.55$ GHz. We also appreciate Dr. Lin Lin (SHAO) for informative discussions on star formation rate in galaxies. We acknowledge support from the National SKA Program of China (nos. 2020SKA0110100, and 2020SKA0110200), the National Natural Science Foundation of China (NSFC, nos. 12373017, 12192220, 12192223 and 12203085), the State Key Laboratory of Radio Astronomy and Technology (Chinese Acamedy of Sciences), and the China Manned Space Project with NO. CMS-CSST-2021-A01, CMS-CSST-2021-A04. HYS acknowledges the support from NSFC (no. 11973070), Key Research Program of Frontier Sciences, CAS (No. ZDBS-LY-7013) and Program of Shanghai Academic/Technology Research Leader. DH acknowledges the financial support of the GA\v{C}R EXPRO grant No. 21-13491X. CJR acknowledges financial support from the German Science Foundation DFG, via the Collaborative Research Center SFB1491 `Cosmic Interacting Matters – From Source to Signal'.

This work uses data obtained from Inyarrimanha Ilgari Bundara / the Murchison Radio-astronomy Observatory. We acknowledge the Wajarri Yamaji People as the Traditional Owners and native title holders of the Observatory site. Establishment of CSIRO's Murchison Radio-astronomy Observatory and the Pawsey Supercomputing Centre are initiatives of the Australian Government, with support from the Government of Western Australia and the Science and Industry Endowment Fund. Support for the operation of the MWA is provided by the Australian Government (NCRIS), under a contract to Curtin University administered by Astronomy Australia Limited.

\section*{Data Availability}

The data underlying this article will be shared at a reasonable request to LHW. Raw MWA observations can be accessed via the MWA All-Sky Virtual Observatory (ASVO) at \url{https://asvo.mwatelescope.org/}. Raw VLA data can be accessed via the NRAO Data Archive at \url{https://data.nrao.edu/portal/}.

\facilities{MWA, VLA, LOFAR, Effelsberg}
\software{\texttt{piip}, \texttt{WSClean}, \texttt{CASA}, \texttt{synchrofit}} 

\bibliographystyle{aasjournal}
\bibliography{references_draft}

\begin{thebibliography}{}
\expandafter\ifx\csname natexlab\endcsname\relax\def\natexlab#1{#1}\fi
\providecommand{\url}[1]{\href{#1}{#1}}
\providecommand{\dodoi}[1]{doi:~\href{http://doi.org/#1}{\nolinkurl{#1}}}
\providecommand{\doeprint}[1]{\href{http://ascl.net/#1}{\nolinkurl{http://ascl.net/#1}}}
\providecommand{\doarXiv}[1]{\href{https://arxiv.org/abs/#1}{\nolinkurl{https://arxiv.org/abs/#1}}}

\bibitem[{{Allen} {et~al.}(2006){Allen}, {Dunn}, {Fabian}, {Taylor}, \&
  {Reynolds}}]{All06}
{Allen}, S.~W., {Dunn}, R.~J.~H., {Fabian}, A.~C., {Taylor}, G.~B., \&
  {Reynolds}, C.~S. 2006, \mnras, 372, 21,
  \dodoi{10.1111/j.1365-2966.2006.10778.x}

\bibitem[{{Avachat} {et~al.}(2016){Avachat}, {Perlman}, {Adams}, {Cara},
  {Owen}, {Sparks}, \& {Georganopoulos}}]{Ava16}
{Avachat}, S.~S., {Perlman}, E.~S., {Adams}, S.~C., {et~al.} 2016, \apj, 832,
  3, \dodoi{10.3847/0004-637X/832/1/3}

\bibitem[{{Baars} {et~al.}(1977){Baars}, {Genzel}, {Pauliny-Toth}, \&
  {Witzel}}]{Baars77}
{Baars}, J.~W.~M., {Genzel}, R., {Pauliny-Toth}, I.~I.~K., \& {Witzel}, A.
  1977, \aap, 61, 99

\bibitem[{{Beardsley} {et~al.}(2019){Beardsley}, {Johnston-Hollitt}, {Trott},
  {Pober}, {Morgan}, {Oberoi}, {Kaplan}, {Lynch}, {Anderson}, {McCauley},
  {Croft}, {James}, {Wong}, {Tremblay}, {Norris}, {Cairns}, {Lonsdale},
  {Hancock}, {Gaensler}, {Bhat}, {Li}, {Hurley-Walker}, {Callingham},
  {Seymour}, {Yoshiura}, {Joseph}, {Takahashi}, {Sokolowski}, {Miller-Jones},
  {Chauhan}, {Boji{\v{c}}i{\'c}}, {Filipovi{\'c}}, {Leahy}, {Su}, {Tian},
  {McSweeney}, {Meyers}, {Kitaeff}, {Vernstrom}, {G{\"u}rkan}, {Heald}, {Xue},
  {Riseley}, {Duchesne}, {Bowman}, {Jacobs}, {Crosse}, {Emrich}, {Franzen},
  {Horsley}, {Kenney}, {Morales}, {Pallot}, {Steele}, {Tingay}, {Walker},
  {Wayth}, {Williams}, \& {Wu}}]{Bear19}
{Beardsley}, A.~P., {Johnston-Hollitt}, M., {Trott}, C.~M., {et~al.} 2019,
  \pasa, 36, e050, \dodoi{10.1017/pasa.2019.41}

\bibitem[{{Beck} \& {Krause}(2005)}]{Bec05}
{Beck}, R., \& {Krause}, M. 2005, Astronomische Nachrichten, 326, 414,
  \dodoi{10.1002/asna.200510366}

\bibitem[{{Bicknell} \& {Begelman}(1996)}]{Bic96}
{Bicknell}, G.~V., \& {Begelman}, M.~C. 1996, \apj, 467, 597,
  \dodoi{10.1086/177636}

\bibitem[{{B{\^\i}rzan} {et~al.}(2004){B{\^\i}rzan}, {Rafferty}, {McNamara},
  {Wise}, \& {Nulsen}}]{Bir04}
{B{\^\i}rzan}, L., {Rafferty}, D.~A., {McNamara}, B.~R., {Wise}, M.~W., \&
  {Nulsen}, P.~E.~J. 2004, \apj, 607, 800, \dodoi{10.1086/383519}

\bibitem[{{Blandford} \& {Globus}(2022)}]{Bl22}
{Blandford}, R., \& {Globus}, N. 2022, \mnras, 514, 5141,
  \dodoi{10.1093/mnras/stac1682}

\bibitem[{{Blandford} {et~al.}(2019){Blandford}, {Meier}, \& {Readhead}}]{Bl19}
{Blandford}, R., {Meier}, D., \& {Readhead}, A. 2019, \araa, 57, 467,
  \dodoi{10.1146/annurev-astro-081817-051948}

\bibitem[{{Blandford} \& {Rees}(1974)}]{Bl74}
{Blandford}, R.~D., \& {Rees}, M.~J. 1974, \mnras, 169, 395,
  \dodoi{10.1093/mnras/169.3.395}

\bibitem[{{Bowman} {et~al.}(2013){Bowman}, {Cairns}, {Kaplan}, {Murphy},
  {Oberoi}, {Staveley-Smith}, {Arcus}, {Barnes}, {Bernardi}, {Briggs}, {Brown},
  {Bunton}, {Burgasser}, {Cappallo}, {Chatterjee}, {Corey}, {Coster},
  {Deshpande}, {deSouza}, {Emrich}, {Erickson}, {Goeke}, {Gaensler},
  {Greenhill}, {Harvey-Smith}, {Hazelton}, {Herne}, {Hewitt},
  {Johnston-Hollitt}, {Kasper}, {Kincaid}, {Koenig}, {Kratzenberg}, {Lonsdale},
  {Lynch}, {Matthews}, {McWhirter}, {Mitchell}, {Morales}, {Morgan}, {Ord},
  {Pathikulangara}, {Prabu}, {Remillard}, {Robishaw}, {Rogers}, {Roshi},
  {Salah}, {Sault}, {Shankar}, {Srivani}, {Stevens}, {Subrahmanyan}, {Tingay},
  {Wayth}, {Waterson}, {Webster}, {Whitney}, {Williams}, {Williams}, \&
  {Wyithe}}]{Bow13}
{Bowman}, J.~D., {Cairns}, I., {Kaplan}, D.~L., {et~al.} 2013, \pasa, 30, e031,
  \dodoi{10.1017/pas.2013.009}

\bibitem[{{Bowman} {et~al.}(1996){Bowman}, {Leahy}, \& {Komissarov}}]{Bow96}
{Bowman}, M., {Leahy}, J.~P., \& {Komissarov}, S.~S. 1996, \mnras, 279, 899,
  \dodoi{10.1093/mnras/279.3.899}

\bibitem[{{Brienza} {et~al.}(2018){Brienza}, {Morganti}, {Murgia}, {Vilchez},
  {Adebahr}, {Carretti}, {Concu}, {Govoni}, {Harwood}, {Intema}, {Loi},
  {Melis}, {Paladino}, {Poppi}, {Shulevski}, {Vacca}, \& {Valente}}]{Bri18}
{Brienza}, M., {Morganti}, R., {Murgia}, M., {et~al.} 2018, \aap, 618, A45,
  \dodoi{10.1051/0004-6361/201832846}

\bibitem[{{Briggs}(1995)}]{Bri95}
{Briggs}, D.~S. 1995, in American Astronomical Society Meeting Abstracts, Vol.
  187, American Astronomical Society Meeting Abstracts, 112.02

\bibitem[{{Carilli} {et~al.}(1991){Carilli}, {Perley}, {Dreher}, \&
  {Leahy}}]{Car91}
{Carilli}, C.~L., {Perley}, R.~A., {Dreher}, J.~W., \& {Leahy}, J.~P. 1991,
  \apj, 383, 554, \dodoi{10.1086/170813}

\bibitem[{{Cheung} {et~al.}(2016){Cheung}, {Bundy}, {Cappellari}, {Peirani},
  {Rujopakarn}, {Westfall}, {Yan}, {Bershady}, {Greene}, {Heckman}, {Drory},
  {Law}, {Masters}, {Thomas}, {Wake}, {Weijmans}, {Rubin}, {Belfiore},
  {Vulcani}, {Chen}, {Zhang}, {Gelfand}, {Bizyaev}, {Roman-Lopes}, \&
  {Schneider}}]{Che16}
{Cheung}, E., {Bundy}, K., {Cappellari}, M., {et~al.} 2016, \nat, 533, 504,
  \dodoi{10.1038/nature18006}

\bibitem[{{Colbert} {et~al.}(1996){Colbert}, {Baum}, {Gallimore}, {O'Dea}, \&
  {Christensen}}]{Colbert96}
{Colbert}, E. J.~M., {Baum}, S.~A., {Gallimore}, J.~F., {O'Dea}, C.~P., \&
  {Christensen}, J.~A. 1996, \apj, 467, 551, \dodoi{10.1086/177633}

\bibitem[{{Condon}(1992)}]{Con92}
{Condon}, J.~J. 1992, \araa, 30, 575,
  \dodoi{10.1146/annurev.aa.30.090192.003043}

\bibitem[{{Condon} {et~al.}(2002){Condon}, {Cotton}, \& {Broderick}}]{Con02}
{Condon}, J.~J., {Cotton}, W.~D., \& {Broderick}, J.~J. 2002, \aj, 124, 675,
  \dodoi{10.1086/341650}

\bibitem[{{Cordey}(1987)}]{Cordey87}
{Cordey}, R.~A. 1987, \mnras, 227, 695, \dodoi{10.1093/mnras/227.3.695}

\bibitem[{{Cornwell}(2008)}]{Cor08}
{Cornwell}, T.~J. 2008, IEEE Journal of Selected Topics in Signal Processing,
  2, 793, \dodoi{10.1109/JSTSP.2008.2006388}

\bibitem[{{Crocker} \& {Aharonian}(2011)}]{Cro11}
{Crocker}, R.~M., \& {Aharonian}, F. 2011, \prl, 106, 101102,
  \dodoi{10.1103/PhysRevLett.106.101102}

\bibitem[{{Cruz-Osorio} {et~al.}(2022){Cruz-Osorio}, {Fromm}, {Mizuno},
  {Nathanail}, {Younsi}, {Porth}, {Davelaar}, {Falcke}, {Kramer}, \&
  {Rezzolla}}]{Cruz22}
{Cruz-Osorio}, A., {Fromm}, C.~M., {Mizuno}, Y., {et~al.} 2022, Nature
  Astronomy, 6, 103, \dodoi{10.1038/s41550-021-01506-w}

\bibitem[{{de Gasperin} {et~al.}(2012){de Gasperin}, {Orr{\'u}}, {Murgia},
  {Merloni}, {Falcke}, {Beck}, {Beswick}, {B{\^\i}rzan}, {Bonafede},
  {Br{\"u}ggen}, {Brunetti}, {Chy{\.z}y}, {Conway}, {Croston}, {En{\ss}lin},
  {Ferrari}, {Heald}, {Heidenreich}, {Jackson}, {Macario}, {McKean}, {Miley},
  {Morganti}, {Offringa}, {Pizzo}, {Rafferty}, {R{\"o}ttgering}, {Shulevski},
  {Steinmetz}, {Tasse}, {van der Tol}, {van Driel}, {van Weeren}, {van
  Zwieten}, {Alexov}, {Anderson}, {Asgekar}, {Avruch}, {Bell}, {Bell},
  {Bentum}, {Bernardi}, {Best}, {Breitling}, {Broderick}, {Butcher}, {Ciardi},
  {Dettmar}, {Eisloeffel}, {Frieswijk}, {Gankema}, {Garrett}, {Gerbers},
  {Griessmeier}, {Gunst}, {Hassall}, {Hessels}, {Hoeft}, {Horneffer},
  {Karastergiou}, {K{\"o}hler}, {Koopman}, {Kuniyoshi}, {Kuper}, {Maat},
  {Mann}, {Mevius}, {Mulcahy}, {Munk}, {Nijboer}, {Noordam}, {Paas}, {Pandey},
  {Pandey}, {Polatidis}, {Reich}, {Schoenmakers}, {Sluman}, {Smirnov}, {Sobey},
  {Stappers}, {Swinbank}, {Tagger}, {Tang}, {van Bemmel}, {van Cappellen}, {van
  Duin}, {van Haarlem}, {van Leeuwen}, {Vermeulen}, {Vocks}, {White}, {Wise},
  {Wucknitz}, \& {Zarka}}]{Gas12}
{de Gasperin}, F., {Orr{\'u}}, E., {Murgia}, M., {et~al.} 2012, \aap, 547, A56,
  \dodoi{10.1051/0004-6361/201220209}

\bibitem[{{de Gasperin} {et~al.}(2020){de Gasperin}, {Vink}, {McKean},
  {Asgekar}, {Avruch}, {Bentum}, {Blaauw}, {Bonafede}, {Broderick},
  {Br{\"u}ggen}, {Breitling}, {Brouw}, {Butcher}, {Ciardi}, {Cuciti}, {de Vos},
  {Duscha}, {Eisl{\"o}ffel}, {Engels}, {Fallows}, {Franzen}, {Garrett},
  {Gunst}, {H{\"o}randel}, {Heald}, {Hoeft}, {Iacobelli}, {Koopmans},
  {Krankowski}, {Maat}, {Mann}, {Mevius}, {Miley}, {Morganti}, {Nelles},
  {Norden}, {Offringa}, {Orr{\'u}}, {Paas}, {Pandey}, {Pandey-Pommier},
  {Pekal}, {Pizzo}, {Reich}, {Rowlinson}, {Rottgering}, {Schwarz}, {Shulevski},
  {Smirnov}, {Sobey}, {Soida}, {Steinmetz}, {Tagger}, {Toribio}, {van Ardenne},
  {van der Horst}, {van Haarlem}, {van Weeren}, {Vocks}, {Wucknitz}, {Zarka},
  \& {Zucca}}]{Gas20}
{de Gasperin}, F., {Vink}, J., {McKean}, J.~P., {et~al.} 2020, \aap, 635, A150,
  \dodoi{10.1051/0004-6361/201936844}

\bibitem[{{de Gasperin} {et~al.}(2025){de Gasperin}, {Edler}, {Boselli},
  {Serra}, {Fossati}, {Heesen}, {Merloni}, {Murgia}, {Reiprich}, {Spasic}, \&
  {Zabel}}]{Gas25}
{de Gasperin}, F., {Edler}, H.~W., {Boselli}, A., {et~al.} 2025, \aap, 693,
  A189, \dodoi{10.1051/0004-6361/202452060}

\bibitem[{{Devereux}(2019)}]{Dev19}
{Devereux}, N. 2019, \mnras, 488, 1199, \dodoi{10.1093/mnras/stz1761}

\bibitem[{{Dey} {et~al.}(2019){Dey}, {Schlegel}, {Lang}, {Blum}, {Burleigh},
  {Fan}, {Findlay}, {Finkbeiner}, {Herrera}, {Juneau}, {Landriau}, {Levi},
  {McGreer}, {Meisner}, {Myers}, {Moustakas}, {Nugent}, {Patej}, {Schlafly},
  {Walker}, {Valdes}, {Weaver}, {Y{\`e}che}, {Zou}, {Zhou}, {Abareshi},
  {Abbott}, {Abolfathi}, {Aguilera}, {Alam}, {Allen}, {Alvarez}, {Annis},
  {Ansarinejad}, {Aubert}, {Beechert}, {Bell}, {BenZvi}, {Beutler}, {Bielby},
  {Bolton}, {Brice{\~n}o}, {Buckley-Geer}, {Butler}, {Calamida}, {Carlberg},
  {Carter}, {Casas}, {Castander}, {Choi}, {Comparat}, {Cukanovaite}, {Delubac},
  {DeVries}, {Dey}, {Dhungana}, {Dickinson}, {Ding}, {Donaldson}, {Duan},
  {Duckworth}, {Eftekharzadeh}, {Eisenstein}, {Etourneau}, {Fagrelius},
  {Farihi}, {Fitzpatrick}, {Font-Ribera}, {Fulmer}, {G{\"a}nsicke},
  {Gaztanaga}, {George}, {Gerdes}, {Gontcho}, {Gorgoni}, {Green}, {Guy},
  {Harmer}, {Hernandez}, {Honscheid}, {Huang}, {James}, {Jannuzi}, {Jiang},
  {Joyce}, {Karcher}, {Karkar}, {Kehoe}, {Kneib}, {Kueter-Young}, {Lan},
  {Lauer}, {Le Guillou}, {Le Van Suu}, {Lee}, {Lesser}, {Perreault Levasseur},
  {Li}, {Mann}, {Marshall}, {Mart{\'\i}nez-V{\'a}zquez}, {Martini}, {du Mas des
  Bourboux}, {McManus}, {Meier}, {M{\'e}nard}, {Metcalfe},
  {Mu{\~n}oz-Guti{\'e}rrez}, {Najita}, {Napier}, {Narayan}, {Newman}, {Nie},
  {Nord}, {Norman}, {Olsen}, {Paat}, {Palanque-Delabrouille}, {Peng},
  {Poppett}, {Poremba}, {Prakash}, {Rabinowitz}, {Raichoor}, {Rezaie},
  {Robertson}, {Roe}, {Ross}, {Ross}, {Rudnick}, {Safonova}, {Saha},
  {S{\'a}nchez}, {Savary}, {Schweiker}, {Scott}, {Seo}, {Shan}, {Silva},
  {Slepian}, {Soto}, {Sprayberry}, {Staten}, {Stillman}, {Stupak}, {Summers},
  {Sien Tie}, {Tirado}, {Vargas-Maga{\~n}a}, {Vivas}, {Wechsler}, {Williams},
  {Yang}, {Yang}, {Yapici}, {Zaritsky}, {Zenteno}, {Zhang}, {Zhang}, {Zhou}, \&
  {Zhou}}]{Dey19}
{Dey}, A., {Schlegel}, D.~J., {Lang}, D., {et~al.} 2019, \aj, 157, 168,
  \dodoi{10.3847/1538-3881/ab089d}

\bibitem[{{Di Matteo} {et~al.}(2003){Di Matteo}, {Allen}, {Fabian}, {Wilson},
  \& {Young}}]{Di03}
{Di Matteo}, T., {Allen}, S.~W., {Fabian}, A.~C., {Wilson}, A.~S., \& {Young},
  A.~J. 2003, \apj, 582, 133, \dodoi{10.1086/344504}

\bibitem[{{Dreher} \& {Feigelson}(1984)}]{Dre84}
{Dreher}, J.~W., \& {Feigelson}, E.~D. 1984, \nat, 308, 43,
  \dodoi{10.1038/308043a0}

\bibitem[{{Duan} \& {Guo}(2020)}]{Duan20}
{Duan}, X., \& {Guo}, F. 2020, \apj, 896, 114, \dodoi{10.3847/1538-4357/ab93b3}

\bibitem[{{Duan} {et~al.}(2024){Duan}, {Wu}, {Zhang}, \& {Li}}]{Duan24}
{Duan}, X., {Wu}, L., {Zhang}, R., \& {Li}, J. 2024, arXiv e-prints,
  arXiv:2410.04467, \dodoi{10.48550/arXiv.2410.04467}

\bibitem[{{Duchesne} {et~al.}(2020){Duchesne}, {Johnston-Hollitt}, {Zhu},
  {Wayth}, \& {Line}}]{Du20}
{Duchesne}, S.~W., {Johnston-Hollitt}, M., {Zhu}, Z., {Wayth}, R.~B., \&
  {Line}, J.~L.~B. 2020, \pasa, 37, e037, \dodoi{10.1017/pasa.2020.29}

\bibitem[{{Ehlert} {et~al.}(2018){Ehlert}, {Weinberger}, {Pfrommer}, {Pakmor},
  \& {Springel}}]{Ehl18}
{Ehlert}, K., {Weinberger}, R., {Pfrommer}, C., {Pakmor}, R., \& {Springel}, V.
  2018, \mnras, 481, 2878, \dodoi{10.1093/mnras/sty2397}

\bibitem[{{Ehlert} {et~al.}(2021){Ehlert}, {Weinberger}, {Pfrommer}, \&
  {Springel}}]{Ehl21}
{Ehlert}, K., {Weinberger}, R., {Pfrommer}, C., \& {Springel}, V. 2021, \mnras,
  503, 1327, \dodoi{10.1093/mnras/stab551}

\bibitem[{{EHT MWL Science Working Group} {et~al.}(2021){EHT MWL Science
  Working Group}, {Algaba}, {Anczarski}, {Asada}, {Balokovi{\'c}}, {Chandra},
  {Cui}, {Falcone}, {Giroletti}, {Goddi}, {Hada}, {Haggard}, {Jorstad}, {Kaur},
  {Kawashima}, {Keating}, {Kim}, {Kino}, {Komossa}, {Kravchenko}, {Krichbaum},
  {Lee}, {Lu}, {Lucchini}, {Markoff}, {Neilsen}, {Nowak}, {Park}, {Principe},
  {Ramakrishnan}, {Reynolds}, {Sasada}, {Savchenko}, {Williamson}, {Event
  Horizon Telescope Collaboration}, {Akiyama}, {Alberdi}, {Alef}, {Anantua},
  {Azulay}, {Baczko}, {Ball}, {Barrett}, {Bintley}, {Benson}, {Blackburn},
  {Blundell}, {Boland}, {Bouman}, {Bower}, {Boyce}, {Bremer}, {Brinkerink},
  {Brissenden}, {Britzen}, {Broderick}, {Broguiere}, {Bronzwaer}, {Byun},
  {Carlstrom}, {Chael}, {Chan}, {Chatterjee}, {Chatterjee}, {Chen}, {Chen},
  {Chesler}, {Cho}, {Christian}, {Conway}, {Cordes}, {Crawford}, {Crew},
  {Cruz-Osorio}, {Davelaar}, {de Laurentis}, {Deane}, {Dempsey}, {Desvignes},
  {Dexter}, {Doeleman}, {Eatough}, {Falcke}, {Farah}, {Fish}, {Fomalont},
  {Ford}, {Fraga-Encinas}, {Friberg}, {Fromm}, {Fuentes}, {Galison}, {Gammie},
  {Garc{\'\i}a}, {Gentaz}, {Georgiev}, {Gold}, {G{\'o}mez}, {G{\'o}mez-Ruiz},
  {Gu}, {Gurwell}, {Hecht}, {Hesper}, {Ho}, {Ho}, {Honma}, {Huang}, {Huang},
  {Hughes}, {Ikeda}, {Inoue}, {Issaoun}, {James}, {Jannuzi}, {Janssen},
  {Jeter}, {Jiang}, {Jim{\'e}nez-Rosales}, {Johnson}, {Jung}, {Karami},
  {Karuppusamy}, {Kettenis}, {Kim}, {Kim}, {Kim}, {Koay}, {Kofuji}, {Koch},
  {Koyama}, {Kramer}, {Kramer}, {Kuo}, {Lauer}, {Levis}, {Li}, {Li},
  {Lindqvist}, {Lico}, {Lindahl}, {Liu}, {Liu}, {Liuzzo}, {Lo}, {Lobanov},
  {Loinard}, {Lonsdale}, {MacDonald}, {Mao}, {Marchili}, {Marrone}, {Marscher},
  {Mart{\'\i}-Vidal}, {Matsushita}, {Matthews}, {Medeiros}, {Menten}, {Mizuno},
  {Mizuno}, {Moran}, {Moriyama}, {Moscibrodzka}, {M{\"u}ller}, {Musoke},
  {Mej{\'\i}as}, {Nagai}, {Nagar}, {Nakamura}, {Narayan}, {Narayanan},
  {Natarajan}, {Nathanail}, {Neri}, {Ni}, {Noutsos}, {Okino}, {Olivares},
  {Ortiz-Le{\'o}n}, {Oyama}, {{\"O}zel}, {Palumbo}, {Patel}, {Pen}, {Pesce},
  {Pi{\'e}tu}, {Plambeck}, {Popstefanija}, {Porth}, {P{\"o}tzl}, {Prather},
  {Preciado-L{\'o}pez}, {Psaltis}, {Pu}, {Rao}, {Rawlings}, {Raymond},
  {Rezzolla}, {Ricarte}, {Ripperda}, {Roelofs}, {Rogers}, {Ros}, {Rose},
  {Roshanineshat}, {Rottmann}, {Roy}, {Ruszczyk}, {Rygl}, {S{\'a}nchez},
  {S{\'a}nchez-Arguelles}, {Savolainen}, {Schloerb}, {Schuster}, {Shao},
  {Shen}, {Small}, {Sohn}, {Soohoo}, {Sun}, {Tazaki}, {Tetarenko}, {Tiede},
  {Tilanus}, {Titus}, {Toma}, {Torne}, {Trent}, {Traianou}, {Trippe}, {van
  Bemmel}, {van Langevelde}, {van Rossum}, {Wagner}, {Ward-Thompson}, {Wardle},
  {Weintroub}, {Wex}, {Wharton}, {Wielgus}, {Wong}, {Wu}, {Yoon}, {Young},
  {Young}, {Younsi}, {Yuan}, {Yuan}, {Zensus}, {Zhao}, {Zhao}, {Fermi Large
  Area Telescope Collaboration}, {Principe}, {Giroletti}, {D'Ammando},
  {Orienti}, {H.~E.~S.~S. Collaboration}, {Abdalla}, {Adam}, {Aharonian},
  {Benkhali}, {Ang{\"u}ner}, {Arcaro}, {Armand}, {Armstrong}, {Ashkar},
  {Backes}, {Baghmanyan}, {Barbosa Martins}, {Barnacka}, {Barnard},
  {Becherini}, {Berge}, {Bernl{\"o}hr}, {Bi}, {B{\"o}ttcher}, {Boisson},
  {Bolmont}, {de Lavergne}, {Breuhaus}, {Brun}, {Brun}, {Bryan}, {B{\"u}chele},
  {Bulik}, {Bylund}, {Caroff}, {Carosi}, {Casanova}, {Chand}, {Chen}, {Cotter},
  {Cury{\l}o}, {Damascene Mbarubucyeye}, {Davids}, {Davies}, {Deil}, {Devin},
  {Dewilt}, {Dirson}, {Djannati-Ata{\"\i}}, {Dmytriiev}, {Donath},
  {Doroshenko}, {Duffy}, {Dyks}, {Egberts}, {Eichhorn}, {Einecke}, {Emery},
  {Ernenwein}, {Feijen}, {Fegan}, {Fiasson}, {de Clairfontaine}, {Fontaine},
  {Funk}, {F{\"u}{\ss}ling}, {Gabici}, {Gallant}, {Giavitto}, {Giunti},
  {Glawion}, {Glicenstein}, {Gottschall}, {Grondin}, {Hahn}, {Haupt},
  {Hermann}, {Hinton}, {Hofmann}, {Hoischen}, {Holch}, {Holler}, {H{\"o}rbe},
  {Horns}, {Huber}, {Jamrozy}, {Jankowsky}, {Jankowsky}, {Jardin-Blicq},
  {Joshi}, {Jung-Richardt}, {Kasai}, {Kastendieck}, {Katarzy{\'n}ski}, {Katz},
  {Khangulyan}, {Kh{\'e}lifi}, {Klepser}, {Klu{\'z}niak}, {Komin}, {Konno},
  {Kosack}, {Kostunin}, {Kreter}, {Lamanna}, {Lemi{\`e}re}, {Lemoine-Goumard},
  {Lenain}, {Levy}, {Lohse}, {Lypova}, {Mackey}, {Majumdar}, {Malyshev},
  {Malyshev}, {Marandon}, {Marchegiani}, {Marcowith}, {Mares},
  {Mart{\'\i}-Devesa}, {Marx}, {Maurin}, {Meintjes}, {Meyer}, {Moderski},
  {Mohamed}, {Mohrmann}, {Montanari}, {Moore}, {Morris}, {Moulin}, {Muller},
  {Murach}, {Nakashima}, {Nayerhoda}, {de Naurois}, {Ndiyavala},
  {Niederwanger}, {Niemiec}, {Oakes}, {O'Brien}, {Odaka}, {Ohm},
  {Olivera-Nieto}, {de Ona Wilhelmi}, {Ostrowski}, {Panter}, {Panny},
  {Parsons}, {Peron}, {Peyaud}, {Piel}, {Pita}, {Poireau}, {Noel}, {Prokhorov},
  {Prokoph}, {P{\"u}hlhofer}, {Punch}, {Quirrenbach}, {Rauth}, {Reichherzer},
  {Reimer}, {Reimer}, {Remy}, {Renaud}, {Rieger}, {Rinchiuso}, {Romoli},
  {Rowell}, {Rudak}, {Ruiz-Velasco}, {Sahakian}, {Sailer}, {Sanchez},
  {Santangelo}, {Sasaki}, {Scalici}, {Schutte}, {Schwanke}, {Schwemmer},
  {Seglar-Arroyo}, {Senniappan}, {Seyffert}, {Shafi}, {Shiningayamwe},
  {Simoni}, {Sinha}, {Sol}, {Specovius}, {Spencer}, {Spir-Jacob}, {Stawarz},
  {Sun}, {Steenkamp}, {Stegmann}, {Steinmassl}, {Steppa}, {Takahashi},
  {Tavernier}, {Taylor}, {Terrier}, {Tiziani}, {Tluczykont}, {Tomankova},
  {Trichard}, {Tsirou}, {Tuffs}, {Uchiyama}, {van der Walt}, {van Eldik}, {van
  Rensburg}, {van Soelen}, {Vasileiadis}, {Veh}, {Venter}, {Vincent}, {Vink},
  {V{\"o}lk}, {Vuillaume}, {Wadiasingh}, {Wagner}, {Watson}, {Werner}, {White},
  {Wierzcholska}, {Wong}, {Yusafzai}, {Zacharias}, {Zanin}, {Zargaryan},
  {Zdziarski}, {Zech}, {Zhu}, {Zorn}, {Zouari}, {{\.Z}ywucka}, {MAGIC
  Collaboration}, {Acciari}, {Ansoldi}, {Antonelli}, {Engels}, {Artero},
  {Asano}, {Baack}, {Babi{\'c}}, {Baquero}, {de Almeida}, {Barrio}, {Becerra
  Gonz{\'a}lez}, {Bednarek}, {Bellizzi}, {Bernardini}, {Bernardos}, {Berti},
  {Besenrieder}, {Bhattacharyya}, {Bigongiari}, {Biland}, {Blanch}, {Bonnoli},
  {Bo{\v{s}}njak}, {Busetto}, {Carosi}, {Ceribella}, {Cerruti}, {Chai},
  {Chilingarian}, {Cikota}, {Colak}, {Colombo}, {Contreras}, {Cortina},
  {Covino}, {D'Amico}, {D'Elia}, {da Vela}, {Dazzi}, {de Angelis}, {de Lotto},
  {Delfino}, {Delgado}, {Delgado Mendez}, {Depaoli}, {di Pierro}, {di Venere},
  {Do Souto Espi{\~n}eira}, {Dominis Prester}, {Donini}, {Dorner}, {Doro},
  {Elsaesser}, {Ramazani}, {Fattorini}, {Ferrara}, {Fonseca}, {Font}, {Fruck},
  {Fukami}, {Garc{\'\i}a L{\'o}pez}, {Garczarczyk}, {Gasparyan}, {Gaug},
  {Giglietto}, {Giordano}, {Gliwny}, {Godinovi{\'c}}, {Green}, {Green},
  {Hadasch}, {Hahn}, {Heckmann}, {Herrera}, {Hoang}, {Hrupec}, {H{\"u}tten},
  {Inada}, {Inoue}, {Ishio}, {Iwamura}, {Jim{\'e}nez}, {Jormanainen}, {Jouvin},
  {Kajiwara}, {Karjalainen}, {Kerszberg}, {Kobayashi}, {Kubo}, {Kushida},
  {Lamastra}, {Lelas}, {Leone}, {Lindfors}, {Lombardi}, {Longo},
  {L{\'o}pez-Coto}, {L{\'o}pez-Moya}, {L{\'o}pez-Oramas}, {Loporchio}, {Machado
  de Oliveira Fraga}, {Maggio}, {Majumdar}, {Makariev}, {Mallamaci}, {Maneva},
  {Manganaro}, {Mannheim}, {Maraschi}, {Mariotti}, {Mart{\'\i}nez}, {Mazin},
  {Menchiari}, {Mender}, {Mi{\'c}anovi{\'c}}, {Miceli}, {Miener}, {Minev},
  {Miranda}, {Mirzoyan}, {Molina}, {Moralejo}, {Morcuende}, {Moreno},
  {Moretti}, {Neustroev}, {Nigro}, {Nilsson}, {Nishijima}, {Noda}, {Nozaki},
  {Ohtani}, {Oka}, {Otero-Santos}, {Paiano}, {Palatiello}, {Paneque},
  {Paoletti}, {Paredes}, {Pavleti{\'c}}, {Pe{\~n}il}, {Perennes}, {Persic},
  {Moroni}, {Prandini}, {Priyadarshi}, {Puljak}, {Rhode}, {Rib{\'o}}, {Rico},
  {Righi}, {Rugliancich}, {Saha}, {Sahakyan}, {Saito}, {Sakurai}, {Satalecka},
  {Saturni}, {Schleicher}, {Schmidt}, {Schweizer}, {Sitarek},
  {{\v{S}}nidari{\'c}}, {Sobczynska}, {Spolon}, {Stamerra}, {Strom}, {Strzys},
  {Suda}, {Suri{\'c}}, {Takahashi}, {Tavecchio}, {Temnikov}, {Terzi{\'c}},
  {Teshima}, {Tosti}, {Truzzi}, {Tutone}, {Ubach}, {van Scherpenberg}, {Vanzo},
  {Vazquez Acosta}, {Ventura}, {Verguilov}, {Vigorito}, {Vitale}, {Vovk},
  {Will}, {Wunderlich}, {Zari{\'c}}, {VERITAS Collaboration}, {Adams},
  {Benbow}, {Brill}, {Capasso}, {Christiansen}, {Chromey}, {Daniel}, {Errando},
  {Farrell}, {Feng}, {Finley}, {Fortson}, {Furniss}, {Gent}, {Giuri}, {Hassan},
  {Hervet}, {Holder}, {Hughes}, {Humensky}, {Jin}, {Kaaret}, {Kertzman},
  {Kieda}, {Kumar}, {Lang}, {Lundy}, {Maier}, {Moriarty}, {Mukherjee}, {Nieto},
  {Nievas-Rosillo}, {O'Brien}, {Ong}, {Otte}, {Patel}, {Pfrang}, {Pohl},
  {Prado}, {Pueschel}, {Quinn}, {Ragan}, {Reynolds}, {Ribeiro}, {Richards},
  {Roache}, {Rulten}, {Ryan}, {Santander}, {Sembroski}, {Shang}, {Weinstein},
  {Williams}, {Williamson}, {Eavn Collaboration}, {Hirota}, {Cui}, {Niinuma},
  {Ro}, {Sakai}, {Sawada-Satoh}, {Wajima}, {Wang}, {Liu}, \&
  {Yonekura}}]{EHTM21}
{EHT MWL Science Working Group}, {Algaba}, J.~C., {Anczarski}, J., {et~al.}
  2021, \apjl, 911, L11, \dodoi{10.3847/2041-8213/abef71}

\bibitem[{{Event Horizon Telescope Collaboration} {et~al.}(2019){Event Horizon
  Telescope Collaboration}, {Akiyama}, {Alberdi}, {Alef}, {Asada}, {Azulay},
  {Baczko}, {Ball}, {Balokovi{\'c}}, {Barrett}, {Bintley}, {Blackburn},
  {Boland}, {Bouman}, {Bower}, {Bremer}, {Brinkerink}, {Brissenden}, {Britzen},
  {Broderick}, {Broguiere}, {Bronzwaer}, {Byun}, {Carlstrom}, {Chael}, {Chan},
  {Chatterjee}, {Chatterjee}, {Chen}, {Chen}, {Cho}, {Christian}, {Conway},
  {Cordes}, {Crew}, {Cui}, {Davelaar}, {De Laurentis}, {Deane}, {Dempsey},
  {Desvignes}, {Dexter}, {Doeleman}, {Eatough}, {Falcke}, {Fish}, {Fomalont},
  {Fraga-Encinas}, {Friberg}, {Fromm}, {G{\'o}mez}, {Galison}, {Gammie},
  {Garc{\'\i}a}, {Gentaz}, {Georgiev}, {Goddi}, {Gold}, {Gu}, {Gurwell},
  {Hada}, {Hecht}, {Hesper}, {Ho}, {Ho}, {Honma}, {Huang}, {Huang}, {Hughes},
  {Ikeda}, {Inoue}, {Issaoun}, {James}, {Jannuzi}, {Janssen}, {Jeter}, {Jiang},
  {Johnson}, {Jorstad}, {Jung}, {Karami}, {Karuppusamy}, {Kawashima},
  {Keating}, {Kettenis}, {Kim}, {Kim}, {Kim}, {Kino}, {Koay}, {Koch}, {Koyama},
  {Kramer}, {Kramer}, {Krichbaum}, {Kuo}, {Lauer}, {Lee}, {Li}, {Li},
  {Lindqvist}, {Liu}, {Liuzzo}, {Lo}, {Lobanov}, {Loinard}, {Lonsdale}, {Lu},
  {MacDonald}, {Mao}, {Markoff}, {Marrone}, {Marscher}, {Mart{\'\i}-Vidal},
  {Matsushita}, {Matthews}, {Medeiros}, {Menten}, {Mizuno}, {Mizuno}, {Moran},
  {Moriyama}, {Moscibrodzka}, {Mul{\ensuremath{\ddot{}}}ler}, {Nagai}, {Nagar},
  {Nakamura}, {Narayan}, {Narayanan}, {Natarajan}, {Neri}, {Ni}, {Noutsos},
  {Okino}, {Olivares}, {Oyama}, {{\"O}zel}, {Palumbo}, {Patel}, {Pen}, {Pesce},
  {Pi{\'e}tu}, {Plambeck}, {PopStefanija}, {Porth}, {Prather},
  {Preciado-L{\'o}pez}, {Psaltis}, {Pu}, {Ramakrishnan}, {Rao}, {Rawlings},
  {Raymond}, {Rezzolla}, {Ripperda}, {Roelofs}, {Rogers}, {Ros}, {Rose},
  {Roshanineshat}, {Rottmann}, {Roy}, {Ruszczyk}, {Ryan}, {Rygl},
  {S{\'a}nchez}, {S{\'a}nchez-Arguelles}, {Sasada}, {Savolainen}, {Schloerb},
  {Schuster}, {Shao}, {Shen}, {Small}, {Sohn}, {SooHoo}, {Tazaki}, {Tiede},
  {Tilanus}, {Titus}, {Toma}, {Torne}, {Trent}, {Trippe}, {Tsuda}, {van
  Bemmel}, {van Langevelde}, {van Rossum}, {Wagner}, {Wardle}, {Weintroub},
  {Wex}, {Wharton}, {Wielgus}, {Wong}, {Wu}, {Young}, {Young}, {Younsi},
  {Yuan}, {Yuan}, {Zensus}, {Zhao}, {Zhao}, {Zhu}, {Anczarski}, {Baganoff},
  {Eckart}, {Farah}, {Haggard}, {Meyer-Zhao}, {Michalik}, {Nadolski},
  {Neilsen}, {Nishioka}, {Nowak}, {Pradel}, {Primiani}, {Souccar},
  {Vertatschitsch}, {Yamaguchi}, \& {Zhang}}]{EHT19}
{Event Horizon Telescope Collaboration}, {Akiyama}, K., {Alberdi}, A., {et~al.}
  2019, \apjl, 875, L5, \dodoi{10.3847/2041-8213/ab0f43}

\bibitem[{{Event Horizon Telescope Collaboration} {et~al.}(2021){Event Horizon
  Telescope Collaboration}, {Akiyama}, {Algaba}, {Alberdi}, {Alef}, {Anantua},
  {Asada}, {Azulay}, {Baczko}, {Ball}, {Balokovi{\'c}}, {Barrett}, {Benson},
  {Bintley}, {Blackburn}, {Blundell}, {Boland}, {Bouman}, {Bower}, {Boyce},
  {Bremer}, {Brinkerink}, {Brissenden}, {Britzen}, {Broderick}, {Broguiere},
  {Bronzwaer}, {Byun}, {Carlstrom}, {Chael}, {Chan}, {Chatterjee},
  {Chatterjee}, {Chen}, {Chen}, {Chesler}, {Cho}, {Christian}, {Conway},
  {Cordes}, {Crawford}, {Crew}, {Cruz-Osorio}, {Cui}, {Davelaar}, {De
  Laurentis}, {Deane}, {Dempsey}, {Desvignes}, {Dexter}, {Doeleman}, {Eatough},
  {Falcke}, {Farah}, {Fish}, {Fomalont}, {Ford}, {Fraga-Encinas}, {Friberg},
  {Fromm}, {Fuentes}, {Galison}, {Gammie}, {Garc{\'\i}a}, {Gelles}, {Gentaz},
  {Georgiev}, {Goddi}, {Gold}, {G{\'o}mez}, {G{\'o}mez-Ruiz}, {Gu}, {Gurwell},
  {Hada}, {Haggard}, {Hecht}, {Hesper}, {Himwich}, {Ho}, {Ho}, {Honma},
  {Huang}, {Huang}, {Hughes}, {Ikeda}, {Inoue}, {Issaoun}, {James}, {Jannuzi},
  {Janssen}, {Jeter}, {Jiang}, {Jimenez-Rosales}, {Johnson}, {Jorstad}, {Jung},
  {Karami}, {Karuppusamy}, {Kawashima}, {Keating}, {Kettenis}, {Kim}, {Kim},
  {Kim}, {Kim}, {Kino}, {Koay}, {Kofuji}, {Koch}, {Koyama}, {Kramer}, {Kramer},
  {Krichbaum}, {Kuo}, {Lauer}, {Lee}, {Levis}, {Li}, {Li}, {Lindqvist}, {Lico},
  {Lindahl}, {Liu}, {Liu}, {Liuzzo}, {Lo}, {Lobanov}, {Loinard}, {Lonsdale},
  {Lu}, {MacDonald}, {Mao}, {Marchili}, {Markoff}, {Marrone}, {Marscher},
  {Mart{\'\i}-Vidal}, {Matsushita}, {Matthews}, {Medeiros}, {Menten}, {Mizuno},
  {Mizuno}, {Moran}, {Moriyama}, {Moscibrodzka}, {M{\"u}ller}, {Musoke}, {Mus
  Mej{\'\i}as}, {Michalik}, {Nadolski}, {Nagai}, {Nagar}, {Nakamura},
  {Narayan}, {Narayanan}, {Natarajan}, {Nathanail}, {Neilsen}, {Neri}, {Ni},
  {Noutsos}, {Nowak}, {Okino}, {Olivares}, {Ortiz-Le{\'o}n}, {Oyama},
  {{\"O}zel}, {Palumbo}, {Park}, {Patel}, {Pen}, {Pesce}, {Pi{\'e}tu},
  {Plambeck}, {PopStefanija}, {Porth}, {P{\"o}tzl}, {Prather},
  {Preciado-L{\'o}pez}, {Psaltis}, {Pu}, {Ramakrishnan}, {Rao}, {Rawlings},
  {Raymond}, {Rezzolla}, {Ricarte}, {Ripperda}, {Roelofs}, {Rogers}, {Ros},
  {Rose}, {Roshanineshat}, {Rottmann}, {Roy}, {Ruszczyk}, {Rygl},
  {S{\'a}nchez}, {S{\'a}nchez-Arguelles}, {Sasada}, {Savolainen}, {Schloerb},
  {Schuster}, {Shao}, {Shen}, {Small}, {Sohn}, {SooHoo}, {Sun}, {Tazaki},
  {Tetarenko}, {Tiede}, {Tilanus}, {Titus}, {Toma}, {Torne}, {Trent},
  {Traianou}, {Trippe}, {van Bemmel}, {van Langevelde}, {van Rossum}, {Wagner},
  {Ward-Thompson}, {Wardle}, {Weintroub}, {Wex}, {Wharton}, {Wielgus}, {Wong},
  {Wu}, {Yoon}, {Young}, {Young}, {Younsi}, {Yuan}, {Yuan}, {Zensus}, {Zhao},
  \& {Zhao}}]{EHT21}
{Event Horizon Telescope Collaboration}, {Akiyama}, K., {Algaba}, J.~C.,
  {et~al.} 2021, \apjl, 910, L13, \dodoi{10.3847/2041-8213/abe4de}

\bibitem[{{Fabian}(2012)}]{Fabian2012}
{Fabian}, A.~C. 2012, \araa, 50, 455,
  \dodoi{10.1146/annurev-astro-081811-125521}

\bibitem[{{Fabian} {et~al.}(2000){Fabian}, {Sanders}, {Ettori}, {Taylor},
  {Allen}, {Crawford}, {Iwasawa}, {Johnstone}, \& {Ogle}}]{Fabian00}
{Fabian}, A.~C., {Sanders}, J.~S., {Ettori}, S., {et~al.} 2000, \mnras, 318,
  L65, \dodoi{10.1046/j.1365-8711.2000.03904.x}

\bibitem[{{Fan} {et~al.}(2018){Fan}, {Wu}, \& {Liao}}]{Fan18}
{Fan}, X.-L., {Wu}, Q., \& {Liao}, N.-H. 2018, \apj, 861, 97,
  \dodoi{10.3847/1538-4357/aac959}

\bibitem[{{Feng} {et~al.}(2016){Feng}, {Wu}, \& {Lu}}]{Feng16}
{Feng}, J., {Wu}, Q., \& {Lu}, R.-S. 2016, \apj, 830, 6,
  \dodoi{10.3847/0004-637X/830/1/6}

\bibitem[{{Forman} {et~al.}(2007){Forman}, {Jones}, {Churazov}, {Markevitch},
  {Nulsen}, {Vikhlinin}, {Begelman}, {B{\"o}hringer}, {Eilek}, {Heinz},
  {Kraft}, {Owen}, \& {Pahre}}]{For07}
{Forman}, W., {Jones}, C., {Churazov}, E., {et~al.} 2007, \apj, 665, 1057,
  \dodoi{10.1086/519480}

\bibitem[{{Gebhardt} {et~al.}(2011){Gebhardt}, {Adams}, {Richstone}, {Lauer},
  {Faber}, {G{\"u}ltekin}, {Murphy}, \& {Tremaine}}]{Geb11}
{Gebhardt}, K., {Adams}, J., {Richstone}, D., {et~al.} 2011, \apj, 729, 119,
  \dodoi{10.1088/0004-637X/729/2/119}

\bibitem[{{Genzel} {et~al.}(1995){Genzel}, {Weitzel}, {Tacconi-Garman},
  {Blietz}, {Cameron}, {Krabbe}, {Lutz}, \& {Sternberg}}]{Gen95}
{Genzel}, R., {Weitzel}, L., {Tacconi-Garman}, L.~E., {et~al.} 1995, \apj, 444,
  129, \dodoi{10.1086/175588}

\bibitem[{{Gilli} {et~al.}(2019){Gilli}, {Mignoli}, {Peca}, {Nanni},
  {Prandoni}, {Liuzzo}, {D'Amato}, {Brusa}, {Calura}, {Caminha}, {Chiaberge},
  {Comastri}, {Cucciati}, {Cusano}, {Grandi}, {Decarli}, {Lanzuisi},
  {Mannucci}, {Pinna}, {Tozzi}, {Vanzella}, {Vignali}, {Vito}, {Balmaverde},
  {Citro}, {Cappelluti}, {Zamorani}, \& {Norman}}]{Gilli19}
{Gilli}, R., {Mignoli}, M., {Peca}, A., {et~al.} 2019, \aap, 632, A26,
  \dodoi{10.1051/0004-6361/201936121}

\bibitem[{{Godfrey} \& {Shabala}(2013)}]{God13}
{Godfrey}, L.~E.~H., \& {Shabala}, S.~S. 2013, \apj, 767, 12,
  \dodoi{10.1088/0004-637X/767/1/12}

\bibitem[{{Guo} \& {Mathews}(2010)}]{Guo10}
{Guo}, F., \& {Mathews}, W.~G. 2010, \apj, 717, 937,
  \dodoi{10.1088/0004-637X/717/2/937}

\bibitem[{{Guo} \& {Mathews}(2012)}]{Guo12a}
---. 2012, \apj, 756, 181, \dodoi{10.1088/0004-637X/756/2/181}

\bibitem[{{Guo} {et~al.}(2012){Guo}, {Mathews}, {Dobler}, \& {Oh}}]{Guo12b}
{Guo}, F., {Mathews}, W.~G., {Dobler}, G., \& {Oh}, S.~P. 2012, \apj, 756, 182,
  \dodoi{10.1088/0004-637X/756/2/182}

\bibitem[{{Guo} \& {Oh}(2008)}]{Guo08a}
{Guo}, F., \& {Oh}, S.~P. 2008, \mnras, 384, 251,
  \dodoi{10.1111/j.1365-2966.2007.12692.x}

\bibitem[{{Guo} {et~al.}(2024){Guo}, {Stone}, {Quataert}, \& {Kim}}]{Guo24}
{Guo}, M., {Stone}, J.~M., {Quataert}, E., \& {Kim}, C.-G. 2024, \apj, 973,
  141, \dodoi{10.3847/1538-4357/ad5fe7}

\bibitem[{{Hardcastle} \& {Croston}(2020)}]{Har20}
{Hardcastle}, M.~J., \& {Croston}, J.~H. 2020, \nar, 88, 101539,
  \dodoi{10.1016/j.newar.2020.101539}

\bibitem[{{Harwood} {et~al.}(2020){Harwood}, {Vernstrom}, \&
  {Stroe}}]{Harwood20}
{Harwood}, J.~J., {Vernstrom}, T., \& {Stroe}, A. 2020, \mnras, 491, 803,
  \dodoi{10.1093/mnras/stz3069}

\bibitem[{{Heckman} {et~al.}(1993){Heckman}, {Lehnert}, \& {Armus}}]{Hec93}
{Heckman}, T.~M., {Lehnert}, M.~D., \& {Armus}, L. 1993, in Astrophysics and
  Space Science Library, Vol. 188, The Environment and Evolution of Galaxies,
  ed. J.~M. {Shull} \& H.~A. {Thronson}, 455,
  \dodoi{10.1007/978-94-011-1882-8_25}

\bibitem[{{Hill} \& {Zakamska}(2014)}]{Hil14}
{Hill}, M.~J., \& {Zakamska}, N.~L. 2014, \mnras, 439, 2701,
  \dodoi{10.1093/mnras/stu123}

\bibitem[{{Hindson} {et~al.}(2016){Hindson}, {Johnston-Hollitt},
  {Hurley-Walker}, {Callingham}, {Su}, {Morgan}, {Bell}, {Bernardi}, {Bowman},
  {Briggs}, {Cappallo}, {Deshpande}, {Dwarakanath}, {For}, {Gaensler},
  {Greenhill}, {Hancock}, {Hazelton}, {Kapi{\'n}ska}, {Kaplan}, {Lenc},
  {Lonsdale}, {Mckinley}, {McWhirter}, {Mitchell}, {Morales}, {Morgan},
  {Oberoi}, {Offringa}, {Ord}, {Procopio}, {Prabu}, {Shankar}, {Srivani},
  {Staveley-Smith}, {Subrahmanyan}, {Tingay}, {Wayth}, {Webster}, {Williams},
  {Williams}, {Wu}, \& {Zheng}}]{Hin16}
{Hindson}, L., {Johnston-Hollitt}, M., {Hurley-Walker}, N., {et~al.} 2016,
  \pasa, 33, e020, \dodoi{10.1017/pasa.2016.19}

\bibitem[{{Hines} {et~al.}(1989){Hines}, {Owen}, \& {Eilek}}]{Hines89}
{Hines}, D.~C., {Owen}, F.~N., \& {Eilek}, J.~A. 1989, \apj, 347, 713,
  \dodoi{10.1086/168163}

\bibitem[{{Hopkins} {et~al.}(2024){Hopkins}, {Squire}, {Su}, {Steinwandel},
  {Kremer}, {Shi}, {Grudic}, {Wellons}, {Faucher-Giguere}, {Angles-Alcazar},
  {Murray}, \& {Quataert}}]{Hopkins24}
{Hopkins}, P.~F., {Squire}, J., {Su}, K.-Y., {et~al.} 2024, The Open Journal of
  Astrophysics, 7, 19, \dodoi{10.21105/astro.2310.04506}

\bibitem[{{Hurley-Walker} {et~al.}(2017){Hurley-Walker}, {Callingham},
  {Hancock}, {Franzen}, {Hindson}, {Kapi{\'n}ska}, {Morgan}, {Offringa},
  {Wayth}, {Wu}, {Zheng}, {Murphy}, {Bell}, {Dwarakanath}, {For}, {Gaensler},
  {Johnston-Hollitt}, {Lenc}, {Procopio}, {Staveley-Smith}, {Ekers}, {Bowman},
  {Briggs}, {Cappallo}, {Deshpande}, {Greenhill}, {Hazelton}, {Kaplan},
  {Lonsdale}, {McWhirter}, {Mitchell}, {Morales}, {Morgan}, {Oberoi}, {Ord},
  {Prabu}, {Shankar}, {Srivani}, {Subrahmanyan}, {Tingay}, {Webster},
  {Williams}, \& {Williams}}]{Hur17}
{Hurley-Walker}, N., {Callingham}, J.~R., {Hancock}, P.~J., {et~al.} 2017,
  \mnras, 464, 1146, \dodoi{10.1093/mnras/stw2337}

\bibitem[{{Jaffe} \& {Perola}(1973)}]{Jaf73}
{Jaffe}, W.~J., \& {Perola}, G.~C. 1973, \aap, 26, 423

\bibitem[{{Kassim} {et~al.}(1993){Kassim}, {Perley}, {Erickson}, \&
  {Dwarakanath}}]{Ka93}
{Kassim}, N.~E., {Perley}, R.~A., {Erickson}, W.~C., \& {Dwarakanath}, K.~S.
  1993, \aj, 106, 2218, \dodoi{10.1086/116795}

\bibitem[{{Kaviraj}(2014)}]{Kav14}
{Kaviraj}, S. 2014, \mnras, 437, L41, \dodoi{10.1093/mnrasl/slt136}

\bibitem[{{Kellermann} {et~al.}(1969){Kellermann}, {Pauliny-Toth}, \&
  {Williams}}]{Ke96}
{Kellermann}, K.~I., {Pauliny-Toth}, I.~I.~K., \& {Williams}, P.~J.~S. 1969,
  \apj, 157, 1, \dodoi{10.1086/150046}

\bibitem[{{Kharb} {et~al.}(2006){Kharb}, {O'Dea}, {Baum}, {Colbert}, \&
  {Xu}}]{Kha06}
{Kharb}, P., {O'Dea}, C.~P., {Baum}, S.~A., {Colbert}, E.~J.~M., \& {Xu}, C.
  2006, \apj, 652, 177, \dodoi{10.1086/507945}

\bibitem[{{Kim} {et~al.}(2023){Kim}, {Savolainen}, {Voitsik}, {Kravchenko},
  {Lisakov}, {Kovalev}, {M{\"u}ller}, {Lobanov}, {Sokolovsky}, {Bruni},
  {Edwards}, {Reynolds}, {Bach}, {Gurvits}, {Krichbaum}, {Hada}, {Giroletti},
  {Orienti}, {Anderson}, {Lee}, {Sohn}, \& {Zensus}}]{Kim23}
{Kim}, J.-Y., {Savolainen}, T., {Voitsik}, P., {et~al.} 2023, \apj, 952, 34,
  \dodoi{10.3847/1538-4357/accf17}

\bibitem[{{Komissarov} \& {Gubanov}(1994)}]{Kom94}
{Komissarov}, S.~S., \& {Gubanov}, A.~G. 1994, \aap, 285, 27

\bibitem[{{Kuo} {et~al.}(2014){Kuo}, {Asada}, {Rao}, {Nakamura}, {Algaba},
  {Liu}, {Inoue}, {Koch}, {Ho}, {Matsushita}, {Pu}, {Akiyama}, {Nishioka}, \&
  {Pradel}}]{Kuo14}
{Kuo}, C.~Y., {Asada}, K., {Rao}, R., {et~al.} 2014, \apjl, 783, L33,
  \dodoi{10.1088/2041-8205/783/2/L33}

\bibitem[{{Ku{\'z}micz} {et~al.}(2017){Ku{\'z}micz}, {Jamrozy},
  {Kozie{\l}-Wierzbowska}, \& {We{\.z}gowiec}}]{Ku17}
{Ku{\'z}micz}, A., {Jamrozy}, M., {Kozie{\l}-Wierzbowska}, D., \&
  {We{\.z}gowiec}, M. 2017, \mnras, 471, 3806, \dodoi{10.1093/mnras/stx1830}

\bibitem[{{Liu} {et~al.}(2013){Liu}, {Zakamska}, {Greene}, {Nesvadba}, \&
  {Liu}}]{Liu13}
{Liu}, G., {Zakamska}, N.~L., {Greene}, J.~E., {Nesvadba}, N. P.~H., \& {Liu},
  X. 2013, \mnras, 436, 2576, \dodoi{10.1093/mnras/stt1755}

\bibitem[{{Longair} {et~al.}(1973){Longair}, {Ryle}, \& {Scheuer}}]{Lon73}
{Longair}, M.~S., {Ryle}, M., \& {Scheuer}, P.~A.~G. 1973, \mnras, 164, 243,
  \dodoi{10.1093/mnras/164.3.243}

\bibitem[{{Lu} {et~al.}(2023){Lu}, {Asada}, {Krichbaum}, {Park}, {Tazaki},
  {Pu}, {Nakamura}, {Lobanov}, {Hada}, {Akiyama}, {Kim}, {Marti-Vidal},
  {G{\'o}mez}, {Kawashima}, {Yuan}, {Ros}, {Alef}, {Britzen}, {Bremer},
  {Broderick}, {Doi}, {Giovannini}, {Giroletti}, {Ho}, {Honma}, {Hughes},
  {Inoue}, {Jiang}, {Kino}, {Koyama}, {Lindqvist}, {Liu}, {Marscher},
  {Matsushita}, {Nagai}, {Rottmann}, {Savolainen}, {Schuster}, {Shen}, {de
  Vicente}, {Walker}, {Yang}, {Zensus}, {Algaba}, {Allardi}, {Bach},
  {Berthold}, {Bintley}, {Byun}, {Casadio}, {Chang}, {Chang}, {Chang}, {Chen},
  {Chen}, {Chilson}, {Chuter}, {Conway}, {Crew}, {Dempsey}, {Dornbusch},
  {Faber}, {Friberg}, {Garc{\'\i}a}, {Garrido}, {Han}, {Han}, {Hasegawa},
  {Herrero-Illana}, {Huang}, {Huang}, {Impellizzeri}, {Jiang}, {Jinchi},
  {Jung}, {Kallunki}, {Kirves}, {Kimura}, {Koay}, {Koch}, {Kramer}, {Kraus},
  {Kubo}, {Kuo}, {Li}, {Lin}, {Liu}, {Liu}, {Lo}, {Lu}, {MacDonald},
  {Martin-Cocher}, {Messias}, {Meyer-Zhao}, {Minter}, {Nair}, {Nishioka},
  {Norton}, {Nystrom}, {Ogawa}, {Oshiro}, {Patel}, {Pen}, {Pidopryhora},
  {Pradel}, {Raffin}, {Rao}, {Ruiz}, {Sanchez}, {Shaw}, {Snow}, {Sridharan},
  {Srinivasan}, {Tercero}, {Torne}, {Traianou}, {Wagner}, {Walther}, {Wei},
  {Yang}, \& {Yu}}]{Lu23}
{Lu}, R.-S., {Asada}, K., {Krichbaum}, T.~P., {et~al.} 2023, \nat, 616, 686,
  \dodoi{10.1038/s41586-023-05843-w}

\bibitem[{{Macfarlane} {et~al.}(2021){Macfarlane}, {Best}, {Sabater},
  {G{\"u}rkan}, {Jarvis}, {R{\"o}ttgering}, {Baldi}, {Calistro Rivera},
  {Duncan}, {Morabito}, {Prandoni}, \& {Retana-Montenegro}}]{Macfarlane21}
{Macfarlane}, C., {Best}, P.~N., {Sabater}, J., {et~al.} 2021, \mnras, 506,
  5888, \dodoi{10.1093/mnras/stab1998}

\bibitem[{{Magliocchetti}(2022)}]{Manuela22}
{Magliocchetti}, M. 2022, \aapr, 30, 6, \dodoi{10.1007/s00159-022-00142-1}

\bibitem[{{Mahatma} {et~al.}(2019){Mahatma}, {Hardcastle}, {Williams}, {Best},
  {Croston}, {Duncan}, {Mingo}, {Morganti}, {Brienza}, {Cochrane},
  {G{\"u}rkan}, {Harwood}, {Jarvis}, {Jamrozy}, {Jurlin}, {Morabito},
  {R{\"o}ttgering}, {Sabater}, {Shimwell}, {Smith}, {Shulevski}, \&
  {Tasse}}]{Mah19}
{Mahatma}, V.~H., {Hardcastle}, M.~J., {Williams}, W.~L., {et~al.} 2019, \aap,
  622, A13, \dodoi{10.1051/0004-6361/201833973}

\bibitem[{{Maiolino} {et~al.}(1998){Maiolino}, {Krabbe}, {Thatte}, \&
  {Genzel}}]{Mai98}
{Maiolino}, R., {Krabbe}, A., {Thatte}, N., \& {Genzel}, R. 1998, \apj, 493,
  650, \dodoi{10.1086/305150}

\bibitem[{{Martin}(2005)}]{Mar05}
{Martin}, C.~L. 2005, \apj, 621, 227, \dodoi{10.1086/427277}

\bibitem[{{Matsushita} {et~al.}(2002){Matsushita}, {Belsole}, {Finoguenov}, \&
  {B{\"o}hringer}}]{Mat02}
{Matsushita}, K., {Belsole}, E., {Finoguenov}, A., \& {B{\"o}hringer}, H. 2002,
  \aap, 386, 77, \dodoi{10.1051/0004-6361:20020087}

\bibitem[{{McDonald} {et~al.}(2013){McDonald}, {Benson}, {Veilleux}, {Bautz},
  \& {Reichardt}}]{McD13}
{McDonald}, M., {Benson}, B., {Veilleux}, S., {Bautz}, M.~W., \& {Reichardt},
  C.~L. 2013, \apjl, 765, L37, \dodoi{10.1088/2041-8205/765/2/L37}

\bibitem[{{McDonald} {et~al.}(2012){McDonald}, {Bayliss}, {Benson}, {Foley},
  {Ruel}, {Sullivan}, {Veilleux}, {Aird}, {Ashby}, {Bautz}, {Bazin}, {Bleem},
  {Brodwin}, {Carlstrom}, {Chang}, {Cho}, {Clocchiatti}, {Crawford}, {Crites},
  {de Haan}, {Desai}, {Dobbs}, {Dudley}, {Egami}, {Forman}, {Garmire},
  {George}, {Gladders}, {Gonzalez}, {Halverson}, {Harrington}, {High},
  {Holder}, {Holzapfel}, {Hoover}, {Hrubes}, {Jones}, {Joy}, {Keisler}, {Knox},
  {Lee}, {Leitch}, {Liu}, {Lueker}, {Luong-van}, {Mantz}, {Marrone}, {McMahon},
  {Mehl}, {Meyer}, {Miller}, {Mocanu}, {Mohr}, {Montroy}, {Murray}, {Natoli},
  {Padin}, {Plagge}, {Pryke}, {Rawle}, {Reichardt}, {Rest}, {Rex}, {Ruhl},
  {Saliwanchik}, {Saro}, {Sayre}, {Schaffer}, {Shaw}, {Shirokoff}, {Simcoe},
  {Song}, {Spieler}, {Stalder}, {Staniszewski}, {Stark}, {Story}, {Stubbs},
  {{\v{S}}uhada}, {van Engelen}, {Vanderlinde}, {Vieira}, {Vikhlinin},
  {Williamson}, {Zahn}, \& {Zenteno}}]{McD12}
{McDonald}, M., {Bayliss}, M., {Benson}, B.~A., {et~al.} 2012, \nat, 488, 349,
  \dodoi{10.1038/nature11379}

\bibitem[{{McMullin} {et~al.}(2007){McMullin}, {Waters}, {Schiebel}, {Young},
  \& {Golap}}]{Mc07}
{McMullin}, J.~P., {Waters}, B., {Schiebel}, D., {Young}, W., \& {Golap}, K.
  2007, in Astronomical Society of the Pacific Conference Series, Vol. 376,
  Astronomical Data Analysis Software and Systems XVI, ed. R.~A. {Shaw},
  F.~{Hill}, \& D.~J. {Bell}, 127

\bibitem[{{McNamara} \& {Nulsen}(2012)}]{McNamara12}
{McNamara}, B.~R., \& {Nulsen}, P.~E.~J. 2012, New Journal of Physics, 14,
  055023, \dodoi{10.1088/1367-2630/14/5/055023}

\bibitem[{{McNamara} {et~al.}(2000){McNamara}, {Wise}, {Nulsen}, {David},
  {Sarazin}, {Bautz}, {Markevitch}, {Vikhlinin}, {Forman}, {Jones}, \&
  {Harris}}]{McNamara00}
{McNamara}, B.~R., {Wise}, M., {Nulsen}, P.~E.~J., {et~al.} 2000, \apjl, 534,
  L135, \dodoi{10.1086/312662}

\bibitem[{{Mertens} {et~al.}(2016){Mertens}, {Lobanov}, {Walker}, \&
  {Hardee}}]{Mertens16}
{Mertens}, F., {Lobanov}, A.~P., {Walker}, R.~C., \& {Hardee}, P.~E. 2016,
  \aap, 595, A54, \dodoi{10.1051/0004-6361/201628829}

\bibitem[{{Million} {et~al.}(2010){Million}, {Werner}, {Simionescu}, {Allen},
  {Nulsen}, {Fabian}, {B{\"o}hringer}, \& {Sanders}}]{Mil10}
{Million}, E.~T., {Werner}, N., {Simionescu}, A., {et~al.} 2010, \mnras, 407,
  2046, \dodoi{10.1111/j.1365-2966.2010.17220.x}

\bibitem[{{Morganti} {et~al.}(2021){Morganti}, {Oosterloo}, {Brienza},
  {Jurlin}, {Prandoni}, {Orr{\`u}}, {Shabala}, {Adams}, {Adebahr}, {Best},
  {Coolen}, {Damstra}, {de Blok}, {de Gasperin}, {D{\'e}nes}, {Hardcastle},
  {Hess}, {Hut}, {Kondapally}, {Kutkin}, {Loose}, {Lucero}, {Maan}, {Maccagni},
  {Mingo}, {Moss}, {Mostert}, {Norden}, {Oostrum}, {R{\"o}ttgering}, {Ruiter},
  {Shimwell}, {Schulz}, {Vermaas}, {Vohl}, {van der Hulst}, {van Diepen}, {van
  Leeuwen}, \& {Ziemke}}]{Mor21}
{Morganti}, R., {Oosterloo}, T.~A., {Brienza}, M., {et~al.} 2021, \aap, 648,
  A9, \dodoi{10.1051/0004-6361/202039102}

\bibitem[{{Murgia} {et~al.}(2011){Murgia}, {Parma}, {Mack}, {de Ruiter},
  {Fanti}, {Govoni}, {Tarchi}, {Giacintucci}, \& {Markevitch}}]{Mur11}
{Murgia}, M., {Parma}, P., {Mack}, K.~H., {et~al.} 2011, \aap, 526, A148,
  \dodoi{10.1051/0004-6361/201015302}

\bibitem[{{O'Dea}(1985)}]{Odea85}
{O'Dea}, C.~P. 1985, \apj, 295, 80, \dodoi{10.1086/163351}

\bibitem[{{Oei} {et~al.}(2024){Oei}, {Hardcastle}, {Timmerman}, {Gast},
  {Botteon}, {Rodriguez}, {Stern}, {Calistro Rivera}, {van Weeren},
  {R{\"o}ttgering}, {Intema}, {de Gasperin}, \& {Djorgovski}}]{Oei24}
{Oei}, M. S.~S.~L., {Hardcastle}, M.~J., {Timmerman}, R., {et~al.} 2024, \nat,
  633, 537, \dodoi{10.1038/s41586-024-07879-y}

\bibitem[{{Offringa} \& {Smirnov}(2017)}]{Off17}
{Offringa}, A.~R., \& {Smirnov}, O. 2017, \mnras, 471, 301,
  \dodoi{10.1093/mnras/stx1547}

\bibitem[{{Offringa} {et~al.}(2012){Offringa}, {van de Gronde}, \&
  {Roerdink}}]{Off12}
{Offringa}, A.~R., {van de Gronde}, J.~J., \& {Roerdink}, J.~B.~T.~M. 2012,
  \aap, 539, A95, \dodoi{10.1051/0004-6361/201118497}

\bibitem[{{Offringa} {et~al.}(2014){Offringa}, {McKinley}, {Hurley-Walker},
  {Briggs}, {Wayth}, {Kaplan}, {Bell}, {Feng}, {Neben}, {Hughes}, {Rhee},
  {Murphy}, {Bhat}, {Bernardi}, {Bowman}, {Cappallo}, {Corey}, {Deshpande},
  {Emrich}, {Ewall-Wice}, {Gaensler}, {Goeke}, {Greenhill}, {Hazelton},
  {Hindson}, {Johnston-Hollitt}, {Jacobs}, {Kasper}, {Kratzenberg}, {Lenc},
  {Lonsdale}, {Lynch}, {McWhirter}, {Mitchell}, {Morales}, {Morgan},
  {Kudryavtseva}, {Oberoi}, {Ord}, {Pindor}, {Procopio}, {Prabu}, {Riding},
  {Roshi}, {Shankar}, {Srivani}, {Subrahmanyan}, {Tingay}, {Waterson},
  {Webster}, {Whitney}, {Williams}, \& {Williams}}]{Off14}
{Offringa}, A.~R., {McKinley}, B., {Hurley-Walker}, N., {et~al.} 2014, \mnras,
  444, 606, \dodoi{10.1093/mnras/stu1368}

\bibitem[{{Offringa} {et~al.}(2015){Offringa}, {Wayth}, {Hurley-Walker},
  {Kaplan}, {Barry}, {Beardsley}, {Bell}, {Bernardi}, {Bowman}, {Briggs},
  {Callingham}, {Cappallo}, {Carroll}, {Deshpande}, {Dillon}, {Dwarakanath},
  {Ewall-Wice}, {Feng}, {For}, {Gaensler}, {Greenhill}, {Hancock}, {Hazelton},
  {Hewitt}, {Hindson}, {Jacobs}, {Johnston-Hollitt}, {Kapi{\'n}ska}, {Kim},
  {Kittiwisit}, {Lenc}, {Line}, {Loeb}, {Lonsdale}, {McKinley}, {McWhirter},
  {Mitchell}, {Morales}, {Morgan}, {Morgan}, {Neben}, {Oberoi}, {Ord}, {Paul},
  {Pindor}, {Pober}, {Prabu}, {Procopio}, {Riding}, {Udaya Shankar}, {Sethi},
  {Srivani}, {Staveley-Smith}, {Subrahmanyan}, {Sullivan}, {Tegmark},
  {Thyagarajan}, {Tingay}, {Trott}, {Webster}, {Williams}, {Williams}, {Wu},
  {Wyithe}, \& {Zheng}}]{Off15}
{Offringa}, A.~R., {Wayth}, R.~B., {Hurley-Walker}, N., {et~al.} 2015, \pasa,
  32, e008, \dodoi{10.1017/pasa.2015.7}

\bibitem[{{Offringa} {et~al.}(2016){Offringa}, {Trott}, {Hurley-Walker},
  {Johnston-Hollitt}, {McKinley}, {Barry}, {Beardsley}, {Bowman}, {Briggs},
  {Carroll}, {Dillon}, {Ewall-Wice}, {Feng}, {Gaensler}, {Greenhill},
  {Hazelton}, {Hewitt}, {Jacobs}, {Kim}, {Kittiwisit}, {Lenc}, {Line}, {Loeb},
  {Mitchell}, {Morales}, {Neben}, {Paul}, {Pindor}, {Pober}, {Procopio},
  {Riding}, {Sethi}, {Shankar}, {Subrahmanyan}, {Sullivan}, {Tegmark},
  {Thyagarajan}, {Tingay}, {Wayth}, {Webster}, \& {Wyithe}}]{Off16}
{Offringa}, A.~R., {Trott}, C.~M., {Hurley-Walker}, N., {et~al.} 2016, \mnras,
  458, 1057, \dodoi{10.1093/mnras/stw310}

\bibitem[{{Owen} {et~al.}(2000){Owen}, {Eilek}, \& {Kassim}}]{Ow00}
{Owen}, F.~N., {Eilek}, J.~A., \& {Kassim}, N.~E. 2000, \apj, 543, 611,
  \dodoi{10.1086/317151}

\bibitem[{{Pacholczyk}(1970)}]{Pac70}
{Pacholczyk}, A.~G. 1970, {Radio astrophysics. Nonthermal processes in galactic
  and extragalactic sources}

\bibitem[{{Park} {et~al.}(2019){Park}, {Hada}, {Kino}, {Nakamura}, {Ro}, \&
  {Trippe}}]{Park19}
{Park}, J., {Hada}, K., {Kino}, M., {et~al.} 2019, \apj, 871, 257,
  \dodoi{10.3847/1538-4357/aaf9a9}

\bibitem[{{Parma} {et~al.}(2007){Parma}, {Murgia}, {de Ruiter}, {Fanti},
  {Mack}, \& {Govoni}}]{Pa07}
{Parma}, P., {Murgia}, M., {de Ruiter}, H.~R., {et~al.} 2007, \aap, 470, 875,
  \dodoi{10.1051/0004-6361:20077592}

\bibitem[{{Pasetto} {et~al.}(2021){Pasetto}, {Carrasco-Gonz{\'a}lez},
  {G{\'o}mez}, {Mart{\'\i}}, {Perucho}, {O'Sullivan}, {Anderson},
  {D{\'\i}az-Gonz{\'a}lez}, {Fuentes}, \& {Wardle}}]{Pas21}
{Pasetto}, A., {Carrasco-Gonz{\'a}lez}, C., {G{\'o}mez}, J.~L., {et~al.} 2021,
  \apjl, 923, L5, \dodoi{10.3847/2041-8213/ac3a88}

\bibitem[{{Perley} \& {Butler}(2013)}]{Per13}
{Perley}, R.~A., \& {Butler}, B.~J. 2013, \apjs, 204, 19,
  \dodoi{10.1088/0067-0049/204/2/19}

\bibitem[{{Perley} {et~al.}(2011){Perley}, {Chandler}, {Butler}, \&
  {Wrobel}}]{Perley11}
{Perley}, R.~A., {Chandler}, C.~J., {Butler}, B.~J., \& {Wrobel}, J.~M. 2011,
  \apjl, 739, L1, \dodoi{10.1088/2041-8205/739/1/L1}

\bibitem[{{Pfrommer}(2013)}]{Pfr13}
{Pfrommer}, C. 2013, \apj, 779, 10, \dodoi{10.1088/0004-637X/779/1/10}

\bibitem[{{Ponti} {et~al.}(2013){Ponti}, {Morris}, {Terrier}, \&
  {Goldwurm}}]{Ponti13}
{Ponti}, G., {Morris}, M.~R., {Terrier}, R., \& {Goldwurm}, A. 2013, in
  Astrophysics and Space Science Proceedings, Vol.~34, Cosmic Rays in
  Star-Forming Environments, ed. D.~F. {Torres} \& O.~{Reimer}, 331,
  \dodoi{10.1007/978-3-642-35410-6_26}

\bibitem[{{Prieto} {et~al.}(2016){Prieto}, {Fern{\'a}ndez-Ontiveros},
  {Markoff}, {Espada}, \& {Gonz{\'a}lez-Mart{\'\i}n}}]{Pri16}
{Prieto}, M.~A., {Fern{\'a}ndez-Ontiveros}, J.~A., {Markoff}, S., {Espada}, D.,
  \& {Gonz{\'a}lez-Mart{\'\i}n}, O. 2016, \mnras, 457, 3801,
  \dodoi{10.1093/mnras/stw166}

\bibitem[{{Quici} {et~al.}(2022){Quici}, {Turner}, {Seymour}, {Hurley-Walker},
  {Shabala}, \& {Ishwara-Chandra}}]{Quici2022}
{Quici}, B., {Turner}, R.~J., {Seymour}, N., {et~al.} 2022, \mnras,
  \dodoi{10.1093/mnras/stac1328}

\bibitem[{{Rafferty} {et~al.}(2006){Rafferty}, {McNamara}, {Nulsen}, \&
  {Wise}}]{Raf06}
{Rafferty}, D.~A., {McNamara}, B.~R., {Nulsen}, P.~E.~J., \& {Wise}, M.~W.
  2006, \apj, 652, 216, \dodoi{10.1086/507672}

\bibitem[{{Renzini} \& {Peng}(2015)}]{Ren15}
{Renzini}, A., \& {Peng}, Y.-j. 2015, \apjl, 801, L29,
  \dodoi{10.1088/2041-8205/801/2/L29}

\bibitem[{{Reynolds} {et~al.}(1996){Reynolds}, {Fabian}, {Celotti}, \&
  {Rees}}]{Rey96}
{Reynolds}, C.~S., {Fabian}, A.~C., {Celotti}, A., \& {Rees}, M.~J. 1996,
  \mnras, 283, 873, \dodoi{10.1093/mnras/283.3.873}

\bibitem[{{Roger} {et~al.}(1973){Roger}, {Costain}, \& {Bridle}}]{Ro73}
{Roger}, R.~S., {Costain}, C.~H., \& {Bridle}, A.~H. 1973, \aj, 78, 1030,
  \dodoi{10.1086/111506}

\bibitem[{{Rottmann} {et~al.}(1996){Rottmann}, {Mack}, {Klein}, \&
  {Wielebinski}}]{Ro96}
{Rottmann}, H., {Mack}, K.~H., {Klein}, U., \& {Wielebinski}, R. 1996, \aap,
  309, L19

\bibitem[{{Rupke} {et~al.}(2005){Rupke}, {Veilleux}, \& {Sanders}}]{Rup05}
{Rupke}, D.~S., {Veilleux}, S., \& {Sanders}, D.~B. 2005, \apjs, 160, 115,
  \dodoi{10.1086/432889}

\bibitem[{{Russell} {et~al.}(2015){Russell}, {Fabian}, {McNamara}, \&
  {Broderick}}]{Rus15}
{Russell}, H.~R., {Fabian}, A.~C., {McNamara}, B.~R., \& {Broderick}, A.~E.
  2015, \mnras, 451, 588, \dodoi{10.1093/mnras/stv954}

\bibitem[{{Sabra} {et~al.}(2003){Sabra}, {Shields}, {Ho}, {Barth}, \&
  {Filippenko}}]{Sab03}
{Sabra}, B.~M., {Shields}, J.~C., {Ho}, L.~C., {Barth}, A.~J., \& {Filippenko},
  A.~V. 2003, \apj, 584, 164, \dodoi{10.1086/345664}

\bibitem[{{Scheuer}(1974)}]{Sch74}
{Scheuer}, P.~A.~G. 1974, \mnras, 166, 513, \dodoi{10.1093/mnras/166.3.513}

\bibitem[{{Sebastian} {et~al.}(2019{\natexlab{a}}){Sebastian}, {Kharb},
  {O'Dea}, {Colbert}, \& {Baum}}]{Sebastian19}
{Sebastian}, B., {Kharb}, P., {O'Dea}, C.~P., {Colbert}, E.~J.~M., \& {Baum},
  S.~A. 2019{\natexlab{a}}, \apj, 883, 189, \dodoi{10.3847/1538-4357/ab371a}

\bibitem[{{Sebastian} {et~al.}(2019{\natexlab{b}}){Sebastian}, {Kharb},
  {O'Dea}, {Colbert}, \& {Baum}}]{Seb19}
---. 2019{\natexlab{b}}, \apj, 883, 189, \dodoi{10.3847/1538-4357/ab371a}

\bibitem[{{Sebokolodi} {et~al.}(2020){Sebokolodi}, {Perley}, {Eilek},
  {Carilli}, {Smirnov}, {Laing}, {Greisen}, \& {Wise}}]{Seb20}
{Sebokolodi}, M. L.~L., {Perley}, R., {Eilek}, J., {et~al.} 2020, \apj, 903,
  36, \dodoi{10.3847/1538-4357/abb80e}

\bibitem[{{Sedgwick} {et~al.}(2021){Sedgwick}, {Baldry}, {James}, {Kaviraj}, \&
  {Martin}}]{Sed21}
{Sedgwick}, T.~M., {Baldry}, I.~K., {James}, P.~A., {Kaviraj}, S., \& {Martin},
  G. 2021, arXiv e-prints, arXiv:2106.13812, \dodoi{10.48550/arXiv.2106.13812}

\bibitem[{{Shi} {et~al.}(2021){Shi}, {Li}, {Yuan}, \& {Zhu}}]{Shi21}
{Shi}, F., {Li}, Z., {Yuan}, F., \& {Zhu}, B. 2021, Nature Astronomy, 5, 928,
  \dodoi{10.1038/s41550-021-01394-0}

\bibitem[{{Shi} {et~al.}(2025){Shi}, {Yuan}, {Tombesi}, \& {XIe}}]{Shi25}
{Shi}, F., {Yuan}, F., {Tombesi}, F., \& {XIe}, F.-g. 2025, arXiv e-prints,
  arXiv:2502.08138, \dodoi{10.48550/arXiv.2502.08138}

\bibitem[{{Silpa} {et~al.}(2022){Silpa}, {Kharb}, {Harrison}, {Girdhar},
  {Mukherjee}, {Mainieri}, \& {Jarvis}}]{Sil22}
{Silpa}, S., {Kharb}, P., {Harrison}, C.~M., {et~al.} 2022, \mnras, 513, 4208,
  \dodoi{10.1093/mnras/stac1044}

\bibitem[{{Silpa} {et~al.}(2021{\natexlab{a}}){Silpa}, {Kharb}, {Harrison},
  {Ho}, {Jarvis}, {Ishwara-Chandra}, \& {Sebastian}}]{Silpa21b}
---. 2021{\natexlab{a}}, \mnras, 507, 991, \dodoi{10.1093/mnras/stab1870}

\bibitem[{{Silpa} {et~al.}(2023){Silpa}, {Kharb}, {Ho}, \& {Harrison}}]{Sil23}
{Silpa}, S., {Kharb}, P., {Ho}, L.~C., \& {Harrison}, C.~M. 2023, \apj, 958,
  47, \dodoi{10.3847/1538-4357/acf7c9}

\bibitem[{{Silpa} {et~al.}(2021{\natexlab{b}}){Silpa}, {Kharb}, {O'Dea},
  {Baum}, {Sebastian}, {Mukherjee}, \& {Harrison}}]{Silpa21}
{Silpa}, S., {Kharb}, P., {O'Dea}, C.~P., {et~al.} 2021{\natexlab{b}}, \mnras,
  507, 2550, \dodoi{10.1093/mnras/stab2110}

\bibitem[{{Slee} {et~al.}(2001){Slee}, {Roy}, {Murgia}, {Andernach}, \&
  {Ehle}}]{Sle01}
{Slee}, O.~B., {Roy}, A.~L., {Murgia}, M., {Andernach}, H., \& {Ehle}, M. 2001,
  \aj, 122, 1172, \dodoi{10.1086/322105}

\bibitem[{{Stawarz} {et~al.}(2006){Stawarz}, {Aharonian}, {Kataoka},
  {Ostrowski}, {Siemiginowska}, \& {Sikora}}]{Sta06}
{Stawarz}, {\L}., {Aharonian}, F., {Kataoka}, J., {et~al.} 2006, \mnras, 370,
  981, \dodoi{10.1111/j.1365-2966.2006.10525.x}

\bibitem[{{Su} {et~al.}(2025){Su}, {Natarajan}, {Cho}, {Narayan}, {Hopkins},
  {Angl{\'e}s-Alc{\'a}zar}, \& {Prather}}]{Su25}
{Su}, K.-Y., {Natarajan}, P., {Cho}, H., {et~al.} 2025, \apjl, 981, L33,
  \dodoi{10.3847/2041-8213/adb7dd}

\bibitem[{{Tan} \& {Blackman}(2005)}]{Tan05}
{Tan}, J.~C., \& {Blackman}, E.~G. 2005, \mnras, 362, 983,
  \dodoi{10.1111/j.1365-2966.2005.09364.x}

\bibitem[{{Tingay} {et~al.}(2013){Tingay}, {Goeke}, {Bowman}, {Emrich}, {Ord},
  {Mitchell}, {Morales}, {Booler}, {Crosse}, {Wayth}, {Lonsdale}, {Tremblay},
  {Pallot}, {Colegate}, {Wicenec}, {Kudryavtseva}, {Arcus}, {Barnes},
  {Bernardi}, {Briggs}, {Burns}, {Bunton}, {Cappallo}, {Corey}, {Deshpande},
  {Desouza}, {Gaensler}, {Greenhill}, {Hall}, {Hazelton}, {Herne}, {Hewitt},
  {Johnston-Hollitt}, {Kaplan}, {Kasper}, {Kincaid}, {Koenig}, {Kratzenberg},
  {Lynch}, {Mckinley}, {Mcwhirter}, {Morgan}, {Oberoi}, {Pathikulangara},
  {Prabu}, {Remillard}, {Rogers}, {Roshi}, {Salah}, {Sault}, {Udaya-Shankar},
  {Schlagenhaufer}, {Srivani}, {Stevens}, {Subrahmanyan}, {Waterson},
  {Webster}, {Whitney}, {Williams}, {Williams}, \& {Wyithe}}]{Tin13}
{Tingay}, S.~J., {Goeke}, R., {Bowman}, J.~D., {et~al.} 2013, \pasa, 30, e007,
  \dodoi{10.1017/pasa.2012.007}

\bibitem[{{Tombesi} {et~al.}(2013){Tombesi}, {Cappi}, {Reeves}, {Nemmen},
  {Braito}, {Gaspari}, \& {Reynolds}}]{Tom13}
{Tombesi}, F., {Cappi}, M., {Reeves}, J.~N., {et~al.} 2013, \mnras, 430, 1102,
  \dodoi{10.1093/mnras/sts692}

\bibitem[{{Turner} {et~al.}(2018{\natexlab{a}}){Turner}, {Rogers}, {Shabala},
  \& {Krause}}]{Tur18c}
{Turner}, R.~J., {Rogers}, J.~G., {Shabala}, S.~S., \& {Krause}, M. G.~H.
  2018{\natexlab{a}}, \mnras, 473, 4179, \dodoi{10.1093/mnras/stx2591}

\bibitem[{{Turner} {et~al.}(2018{\natexlab{b}}){Turner}, {Shabala}, \&
  {Krause}}]{Tur18a}
{Turner}, R.~J., {Shabala}, S.~S., \& {Krause}, M. G.~H. 2018{\natexlab{b}},
  \mnras, 474, 3361, \dodoi{10.1093/mnras/stx2947}

\bibitem[{{Turner} {et~al.}(2018{\natexlab{c}}){Turner}, {Shabala}, \&
  {Krause}}]{Tur18b}
---. 2018{\natexlab{c}}, \mnras, 474, 3361, \dodoi{10.1093/mnras/stx2947}

\bibitem[{{Wang} {et~al.}(2013){Wang}, {Nowak}, {Markoff}, {Baganoff},
  {Nayakshin}, {Yuan}, {Cuadra}, {Davis}, {Dexter}, {Fabian}, {Grosso},
  {Haggard}, {Houck}, {Ji}, {Li}, {Neilsen}, {Porquet}, {Ripple}, \&
  {Shcherbakov}}]{Wang13}
{Wang}, Q.~D., {Nowak}, M.~A., {Markoff}, S.~B., {et~al.} 2013, Science, 341,
  981, \dodoi{10.1126/science.1240755}

\bibitem[{{Wayth} {et~al.}(2018){Wayth}, {Tingay}, {Trott}, {Emrich},
  {Johnston-Hollitt}, {McKinley}, {Gaensler}, {Beardsley}, {Booler}, {Crosse},
  {Franzen}, {Horsley}, {Kaplan}, {Kenney}, {Morales}, {Pallot}, {Sleap},
  {Steele}, {Walker}, {Williams}, {Wu}, {Cairns}, {Filipovic}, {Johnston},
  {Murphy}, {Quinn}, {Staveley-Smith}, {Webster}, \& {Wyithe}}]{Wayth18}
{Wayth}, R.~B., {Tingay}, S.~J., {Trott}, C.~M., {et~al.} 2018, \pasa, 35,
  e033, \dodoi{10.1017/pasa.2018.37}

\bibitem[{{Weinberger} {et~al.}(2017){Weinberger}, {Springel}, {Hernquist},
  {Pillepich}, {Marinacci}, {Pakmor}, {Nelson}, {Genel}, {Vogelsberger},
  {Naiman}, \& {Torrey}}]{Weinberger17}
{Weinberger}, R., {Springel}, V., {Hernquist}, L., {et~al.} 2017, \mnras, 465,
  3291, \dodoi{10.1093/mnras/stw2944}

\bibitem[{{Werner} {et~al.}(2010){Werner}, {Simionescu}, {Million}, {Allen},
  {Nulsen}, {von der Linden}, {Hansen}, {B{\"o}hringer}, {Churazov}, {Fabian},
  {Forman}, {Jones}, {Sanders}, \& {Taylor}}]{Wer10}
{Werner}, N., {Simionescu}, A., {Million}, E.~T., {et~al.} 2010, \mnras, 407,
  2063, \dodoi{10.1111/j.1365-2966.2010.16755.x}

\bibitem[{{Wu} {et~al.}(2020){Wu}, {Wu}, {Feng}, {Lu}, \& {Fan}}]{Wu20}
{Wu}, L.-H., {Wu}, Q.-W., {Feng}, J.-C., {Lu}, R.-S., \& {Fan}, X.-L. 2020,
  Research in Astronomy and Astrophysics, 20, 122,
  \dodoi{10.1088/1674-4527/20/8/122}

\bibitem[{{Wu} {et~al.}(2013){Wu}, {Cao}, {Ho}, \& {Wang}}]{Wu13}
{Wu}, Q., {Cao}, X., {Ho}, L.~C., \& {Wang}, D.-X. 2013, \apj, 770, 31,
  \dodoi{10.1088/0004-637X/770/1/31}

\bibitem[{{Wykes} {et~al.}(2015){Wykes}, {Hardcastle}, {Karakas}, \&
  {Vink}}]{Wyk15}
{Wykes}, S., {Hardcastle}, M.~J., {Karakas}, A.~I., \& {Vink}, J.~S. 2015,
  \mnras, 447, 1001, \dodoi{10.1093/mnras/stu2440}

\bibitem[{{Wylezalek} \& {Zakamska}(2016)}]{Wy16}
{Wylezalek}, D., \& {Zakamska}, N.~L. 2016, \mnras, 461, 3724,
  \dodoi{10.1093/mnras/stw1557}

\bibitem[{{Xie} {et~al.}(2023){Xie}, {Narayan}, \& {Yuan}}]{Xie23}
{Xie}, F.-G., {Narayan}, R., \& {Yuan}, F. 2023, \apj, 942, 20,
  \dodoi{10.3847/1538-4357/aca534}

\bibitem[{{Xie} \& {Yuan}(2012)}]{Xie12}
{Xie}, F.-G., \& {Yuan}, F. 2012, \mnras, 427, 1580,
  \dodoi{10.1111/j.1365-2966.2012.22030.x}

\bibitem[{{Yang} {et~al.}(2021){Yang}, {Yuan}, {Yuan}, \& {White}}]{Yang21}
{Yang}, H., {Yuan}, F., {Yuan}, Y.-F., \& {White}, C.~J. 2021, \apj, 914, 131,
  \dodoi{10.3847/1538-4357/abfe63}

\bibitem[{{Yang} {et~al.}(2024){Yang}, {Yuan}, {Li}, {Mizuno}, {Guo}, {Lu},
  {Ho}, {Lin}, {Zdziarski}, \& {Wang}}]{Yang24}
{Yang}, H., {Yuan}, F., {Li}, H., {et~al.} 2024, Science Advances, 10,
  eadn3544, \dodoi{10.1126/sciadv.adn3544}

\bibitem[{{Yuan} {et~al.}(2012{\natexlab{a}}){Yuan}, {Bu}, \& {Wu}}]{Yuan12b}
{Yuan}, F., {Bu}, D., \& {Wu}, M. 2012{\natexlab{a}}, \apj, 761, 130,
  \dodoi{10.1088/0004-637X/761/2/130}

\bibitem[{{Yuan} {et~al.}(2015){Yuan}, {Gan}, {Narayan}, {Sadowski}, {Bu}, \&
  {Bai}}]{Yuan15}
{Yuan}, F., {Gan}, Z., {Narayan}, R., {et~al.} 2015, \apj, 804, 101,
  \dodoi{10.1088/0004-637X/804/2/101}

\bibitem[{{Yuan} \& {Narayan}(2014)}]{yuan14}
{Yuan}, F., \& {Narayan}, R. 2014, \araa, 52, 529,
  \dodoi{10.1146/annurev-astro-082812-141003}

\bibitem[{{Yuan} {et~al.}(2022{\natexlab{a}}){Yuan}, {Wang}, \&
  {Yang}}]{Yuan22}
{Yuan}, F., {Wang}, H., \& {Yang}, H. 2022{\natexlab{a}}, \apj, 924, 124,
  \dodoi{10.3847/1538-4357/ac4714}

\bibitem[{{Yuan} {et~al.}(2022{\natexlab{b}}){Yuan}, {Wang}, \&
  {Yang}}]{Yuan22a}
---. 2022{\natexlab{b}}, \apj, 924, 124, \dodoi{10.3847/1538-4357/ac4714}

\bibitem[{{Yuan} {et~al.}(2012{\natexlab{b}}){Yuan}, {Wu}, \& {Bu}}]{Yuan12a}
{Yuan}, F., {Wu}, M., \& {Bu}, D. 2012{\natexlab{b}}, \apj, 761, 129,
  \dodoi{10.1088/0004-637X/761/2/129}

\bibitem[{{Yuan} {et~al.}(2018){Yuan}, {Yoon}, {Li}, {Gan}, {Ho}, \&
  {Guo}}]{Yuan18}
{Yuan}, F., {Yoon}, D., {Li}, Y.-P., {et~al.} 2018, \apj, 857, 121,
  \dodoi{10.3847/1538-4357/aab8f8}

\bibitem[{{Yue} {et~al.}(2025){Yue}, {Duncan}, {Best}, {Arnaudova}, {Morabito},
  {Petley}, {R{\"o}ttgering}, {Shenoy}, \& {Smith}}]{Yue25}
{Yue}, B.~H., {Duncan}, K.~J., {Best}, P.~N., {et~al.} 2025, \mnras, 537, 858,
  \dodoi{10.1093/mnras/staf077}

\bibitem[{{Zhang} {et~al.}(2025){Zhang}, {Xia}, {Ji}, {Yuan}, {Guo}, {Zhang},
  {Zhu}, {Di}, {He}, {Su}, \& {Zou}}]{Zhang25}
{Zhang}, H., {Xia}, H., {Ji}, S., {et~al.} 2025, arXiv e-prints,
  arXiv:2504.06342, \dodoi{10.48550/arXiv.2504.06342}

\bibitem[{{Zhu} {et~al.}(2023){Zhu}, {Yuan}, {Ji}, {Peng}, \& {Ho}}]{Zhu23}
{Zhu}, B., {Yuan}, F., {Ji}, S., {Peng}, Y., \& {Ho}, L.~C. 2023, \mnras, 525,
  4840, \dodoi{10.1093/mnras/stad2640}

\bibitem[{{Zubovas} {et~al.}(2011){Zubovas}, {King}, \& {Nayakshin}}]{Zub11}
{Zubovas}, K., {King}, A.~R., \& {Nayakshin}, S. 2011, \mnras, 415, L21,
  \dodoi{10.1111/j.1745-3933.2011.01070.x}

\end{thebibliography}
\end{document}